\documentclass[journal]{IEEEtran}
\usepackage[hidelinks, breaklinks=true]{hyperref}

\ifCLASSINFOpdf
  \usepackage[pdftex]{graphicx}
\else
  \usepackage[dvips]{graphicx}
\fi
\usepackage{amsmath}
\usepackage{amssymb}
\usepackage{amsthm}
\usepackage{xfrac}
\usepackage[capitalize]{cleveref}
\usepackage{siunitx}
\usepackage{hhline}
\usepackage{makecell}
\usepackage{multirow}

\usepackage{pgf}
\usepackage{pgfplots}
\usepackage{tikz}
\pgfplotsset{compat=newest}
\usetikzlibrary{pgfplots.groupplots}
\usetikzlibrary{pgfplots.dateplot}
\usepackage[percent]{overpic}
\usepackage{booktabs}

\usepackage{float}
\usepackage{diagbox}
\usepackage[hypcap, labelformat=simple]{subcaption}

\usepackage{import}

\renewcommand{\textrightarrow}{$\rightarrow$}

\usepackage{layouts}

\newcommand*{\Package}[1]{\textsc{#1}}

\begin{document}
%
\title{\makebox[0pt]{PlenoptiCam v1.0: A light-field imaging framework}}
%
%
%

\author{Christopher~Hahne~\IEEEmembership{Member,~IEEE,} and  %
        Amar~Aggoun \\
        University of Wolverhampton, School of Mathematics and Computer Science \\
        Wulfruna Street, WV1 1LY Wolverhampton, United Kingdom \\
        \texttt{\{c.hahne, a.aggoun\}@wlv.ac.uk}%
}

\newcommand*{\defeq}{\stackrel{\text{def}}{=}}

\maketitle

\begin{abstract}
Light-field cameras play a vital role for rich \mbox{3-D} information retrieval in narrow range depth sensing applications. 
The key obstacle in composing light-fields from exposures taken by a plenoptic camera is to computationally calibrate, align and rearrange four-dimensional image data. %
Several attempts have been proposed to enhance the overall image quality by tailoring pipelines dedicated to particular plenoptic cameras and improving the consistency across viewpoints at the expense of high computational loads. %
The framework presented herein advances prior outcomes thanks to its novel micro image scale-space analysis for generic camera calibration independent of the lens specifications and its parallax-invariant, cost-effective viewpoint color equalization from optimal transport theory. 
%
Artifacts from the sensor and micro lens grid are compensated in an innovative way to enable superior quality in sub-aperture image extraction, computational refocusing and Scheimpflug rendering with sub-sampling capabilities.
Benchmark comparisons using established image metrics suggest that our proposed pipeline outperforms state-of-the-art tool chains in the majority of cases. Results from a Wasserstein distance further show that our color transfer outdoes the existing transport methods.
%
Our algorithms are released under an open-source license, offer cross-platform compatibility with few dependencies and different user interfaces. This makes the reproduction of results and  experimentation with plenoptic camera technology convenient for peer researchers, developers, photographers, data scientists and others working in this field.


\end{abstract}


%
\IEEEpeerreviewmaketitle

\section{Introduction}
\label{sec:1}
%
%
%
%
\IEEEPARstart{T}{here} is a growing body of literature in the areas of experimental photography~\cite{NGLEV, Hahne:OPEX:16}, medical imaging~\cite{prevedel:2014:simultaneous, Li439315, Bedard:17, Palmer:18} and machine learning~\cite{Srinivasan:ICCV:2017} recognizing capabilities offered by light-fields. %
\subsection{Background}
The probably most influential light-field model in computer graphics was devised by Levoy and Hanrahan~\cite{LEVHAN} who described a light-field to be a collection of ray vectors piercing through two planes stacked behind one another. The intersections of ray vectors at the two planes make up four Cartesian coordinates, which is why light-fields are often referred to as four-dimensional (\mbox{4-D}). In their much celebrated paper, Levoy and Hanrahan demonstrate that the two planes serve as the spatial and angular image domain providing two different light-field representations that can be transferred into each other. Such light-field transformation corresponds to changing the sequential order of the two planes. 
%
\par
Capturing a light-field from a monocular lens is achieved with a single sensor stacked behind an aperture grid~\cite{AW} such as a Micro Lens Array (MLA).
%
%
This optical setup is known as a plenoptic camera and can be thought of as accommodating an array of consistently spaced virtual cameras located at the aperture plane~\cite{Hahne:IJCV:18}. %
As opposed to light-fields from a camera array, captures from the plenoptic camera need to undergo additional processing to be represented as such~\cite{Cho:2013, DANSCAL}. %
%
\par
\begin{figure*}[!t]
	\centering
	\resizebox{0.734\textwidth}{!}{\includegraphics{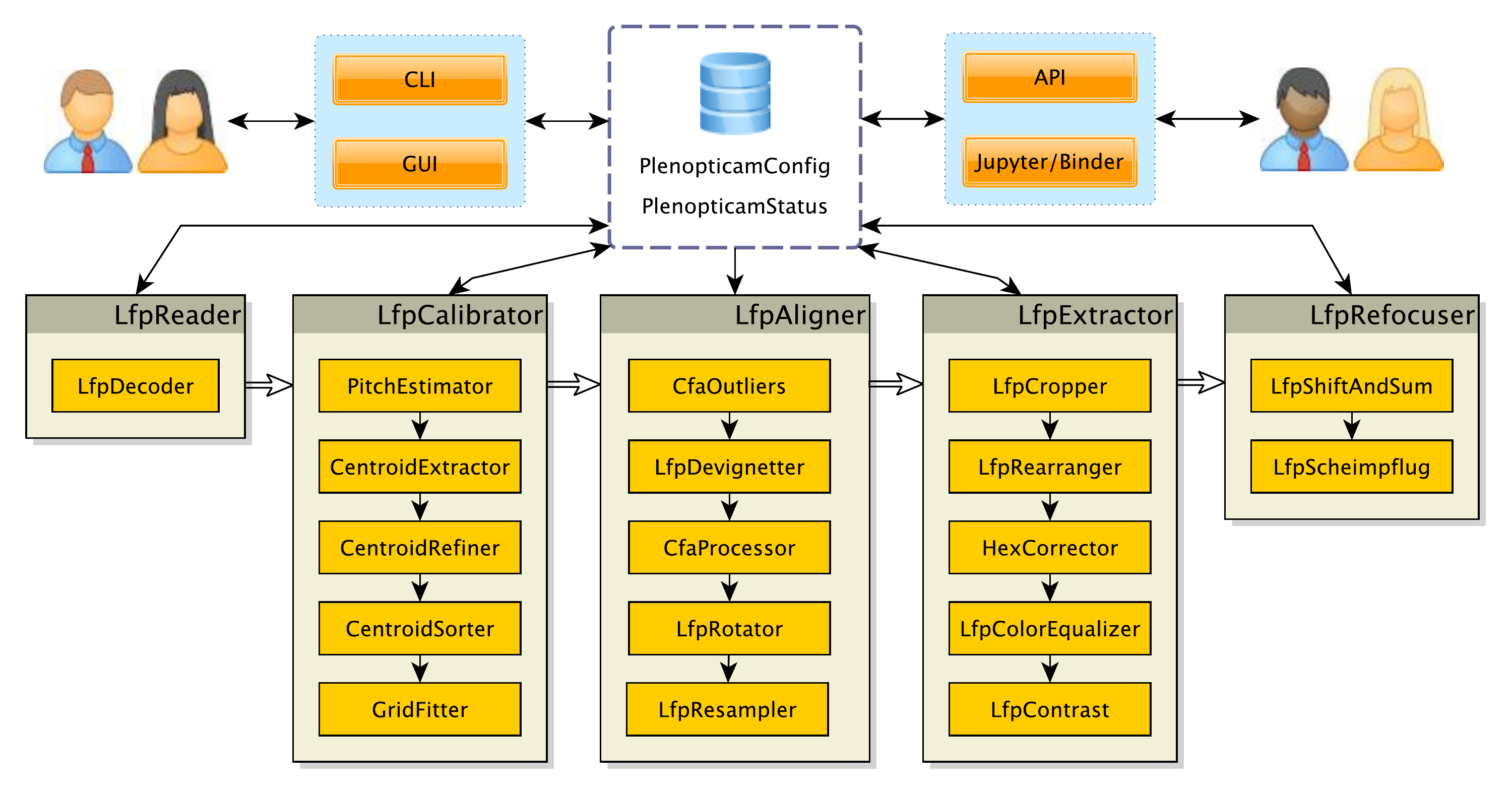}}
	\caption{\textbf{Package architecture} with gray blocks representing modules whereas orange blocks indicate essential top-level classes.\label{fig:arch} }
\end{figure*}
%
%
Until today, the landscape of software tools treating plenoptic content has been characterized by heterogeneity. %
One of the most mature and influential software applications was released by camera manufacturer Lytro in 2012. Lytro's image processing pipeline remains closed-source and is not publicly maintained since the company shut down business in 2018. In the earlier days of Lytro's lifetime, independent programmers developed binary file decoders by reverse-engineering Lytro's file formats~\cite{lfp:nirav, lfp:behnam}. These tools, however, are unable to perform light-field rendering functionalities such as refocusing. Several scientists published methods on how to algorithmically calibrate and decode Lytro's plenoptic camera within Matlab~\cite{Cho:2013, DANSCAL, bok:eccv14, Pendu:2020}. These methods are based on the given metadata while concentrating on the micro image center detection, \mbox{4-D} rearrangement and rectification of radial lens distortions. %
More recently, research has taken the direction to successfully recover physical information at light-field boundaries~\cite{David:2017, Matysiak:2018}. Although these studies revealed convincing results, light-field color consistency as proposed by Matysiak {\it et al.}~\cite{Matysiak:2018} requires high computational resources. %
\par%
%
%
\subsection{Novelties}
While consolidating preliminary decoding procedures, we introduce novel calibration, resampling and refocusing techniques. An overview of our contributions is listed below: %
\begin{enumerate}
\item Scale space analysis of micro images is conducted early in the calibration pipeline to support plenoptic cameras with arbitrary sensor and MLA dimensions and, unlike other methods in the field,  handle footage from custom-built prototypes as well as Lytro cameras. %
\item Centroid grid fitting is employed using the Levenberg-Marquardt optimization to reduce least-squares errors of detected micro lens centers globally and determine the centroid spacing and projective mapping simultaneously.
\item De-vignetting is based on a 4-D white image least-squares fit model to suppress noise propagation as it occurs in a conventional white image division.
\item To counteract color variances across sub-aperture views, we propose a novel, rapid and effective illumination channel correction scheme that outperforms the previous color distribution transport techniques~\cite{Pitie:2007, Perrot:2016}. 
\item For accurate angular sampling, we offer micro image resampling followed by a hexagonal artifact removal. %
\item Taking advantage of the popular computational refocusing, we pave the way for its Scheimpflug equivalent.
\end{enumerate}
These innovative algorithms are bundled as a tool coined \Package{PlenoptiCam} to accomplish outstanding results, which are validated in this study with standard image metrics. 
\subsection{Scope}
Equipped with generic calibration and novel rendering routines, \Package{PlenoptiCam} may build the foundation for future work on light-field image algorithms. Research goals are more easily attained as image scientists can adapt code and focus on their individual idea. 
In general, all kinds of plenoptic cameras are covered by \Package{PlenoptiCam} while raw buffer conversion is fully supported for the Lytro Illum. This framework yet serves as a starting point for plenoptic~2.0 and Raytrix~\cite{RAYTRIX} images, which share calibration and resampling requirements.
%
Note that \Package{PlenoptiCam} can be used in conjunction with the geometry tool \Package{PlenoptiSign}~\cite{Plenoptisign:2019} to pinpoint metric object positions in a light-field captured by a plenoptic camera. \par
\subsection{Structure of the Paper}
The organisation of this paper starts with an overview of the module architecture, which can be regarded as a roadmap for Section~\ref{sec:3}. Subsections contained in Section~\ref{sec:3} give insightful details on novel algorithmic aspects with respect to plenoptic image calibration, sub-aperture processing and refocusing. A benchmark comparison of results rendered by \Package{PlenoptiCam} and other tool chains is carried out in Section~\ref{sec:4}. 
Section~\ref{sec:5} draws conclusions, while reflecting on the framework's potential impact and sketching out ideas for future work. %
\par
%
\section{Module Functionalities}
\label{sec:3}
An architectural scheme of the processing pipeline is depicted in Fig.~\ref{fig:arch}. Objects of a \Package{PlenopticamConfig} and \Package{PlenopticamStatus} class are thought to be singletons and shared across modules. The processing direction of input image data is from left to right (on module-level) and top to bottom (on class-level) as indicated by the arrows.
The diagram in Fig.~\ref{fig:arch} can be regarded as a roadmap for this section in which the core functionality of each module is presented. 
%
%
%
\subsection{LfpReader}
\label{sec:3:1}
The \Package{LfpReader} module supports standard image file types ({\it tiff, bmp, jpg, png}) and the more specific raw Lytro file type decoding ({\it lfp, lfr, raw}) for Bayer image composition according to the findings by Nirav Patel~\cite{lfp:nirav}. For raw data from a Lytro camera, this module exports an image as a {\it tiff} file in Bayer representation as well as a {\it json} file containing corresponding metadata which is used for gamma correction. Other image file types are expected to be in sRGB space for gamma handling.
\subsection{LfpCalibrator}
\label{sec:3:2}
\begin{figure}[H]
	\centering
	\resizebox{\columnwidth}{!}{\includegraphics{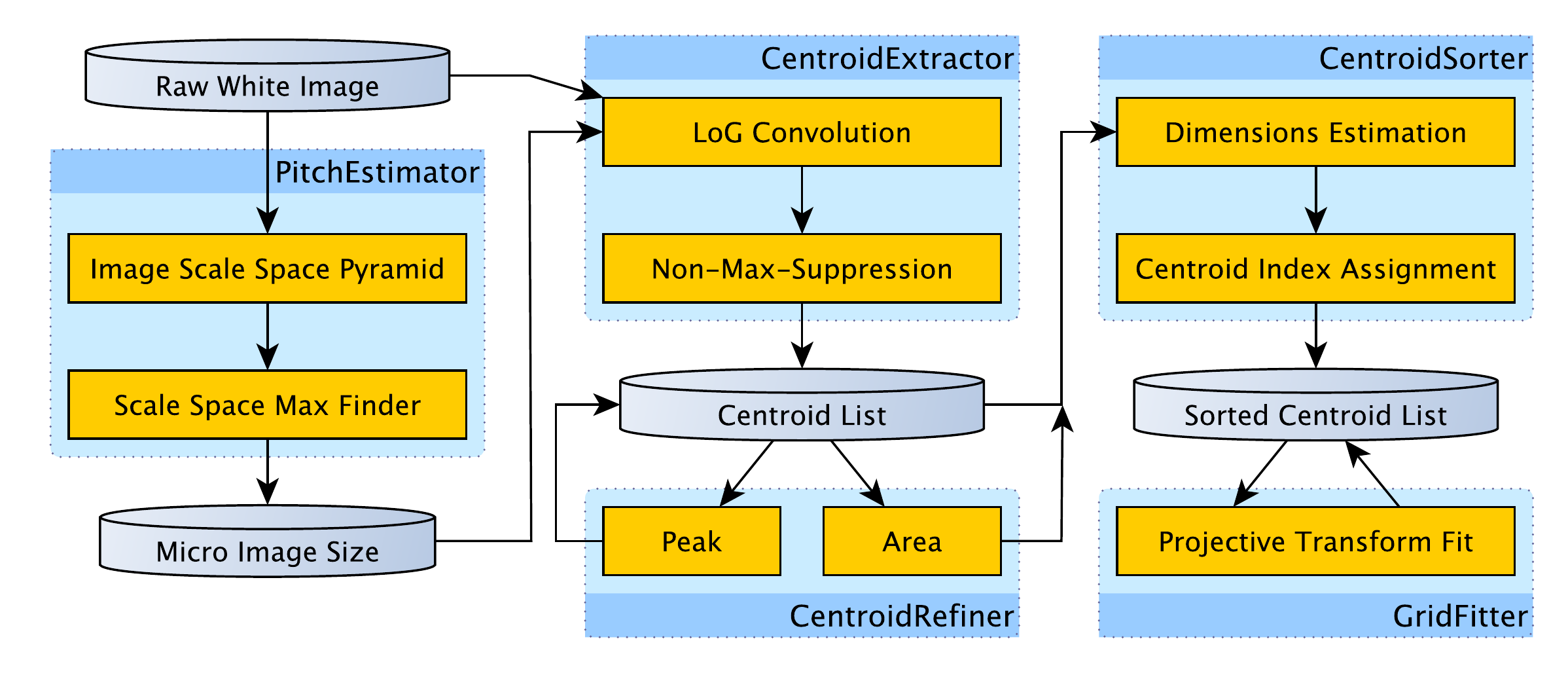}}
	\caption{\textbf{Calibration pipeline}\label{fig:calib_pipe}}
\end{figure}

The fundamental problem we aim to solve when calibrating a plenoptic camera is to register geometric properties of the \mbox{4-D} micro image representation. This enables a light-field to be rearranged as if captured by multiple cameras with consistent spacing~\cite{Hahne:IJCV:18}. %
\Package{PlenoptiCam} introduces a novel calibration pipeline with a sequence of steps shown in Fig.~\ref{fig:calib_pipe}. %
The entire calibration procedure can be sub-divided into pitch estimation, centroid extraction along with refinement, sorting and grid fitting. To let the pipeline cover multiple types of plenoptic cameras with varying lens and sensor specifications, a blob detector is employed as a first step for analysis of the micro image size. %
\subsubsection{PitchEstimator}
\label{sec:3:2:1}
Based on a white calibration image $I_w(\mathbf{x})$ where $\mathbf{x}=\begin{bmatrix}k & l\end{bmatrix}^\intercal \in \mathbb{N}^2$ consists of two spatial coordinates $k, l$ across arbitrary micro images, we adapt the classical scale space theory~\cite{lindeberg:scale-space} using half-octave pyramids~\cite{Crowley:2002} denoted by $P(\nu, \mathbf{x})$ and built via
\begin{align}
\forall\nu, \, \, P(\nu+1, \mathbf{x}) = \mathcal{D}_2\left(P(\nu, \mathbf{x})\ast \nabla^2 G(\sigma, \mathbf{x})\right)
\label{eq:scale_space}
\end{align}
with $\{\nu \in \mathbb{N} \mid \nu < \log_{2}\left(\operatorname{min}(K, L)\right)\}$ while $K, L$ are spatial resolutions of $I_w(\mathbf{x})$. %
Downsampling is represented by $\mathcal{D}_2(\cdot)$ considering the half-octave requirement 
and $\nabla^2G(\sigma, \mathbf{x})$ is the Laplacian of Gaussian (LoG) convolution kernel, known as the Mexican hat, which is approximated by the Difference of Gaussians. Each pyramid level at index $\nu$ is then a downscaled LoG-filtered version of the previous level where $P(0, \mathbf{x})=I_w(\mathbf{x}) \ast \nabla^2 G(\sigma, \mathbf{x})$ serves as the initial scale. 
Let $\nu^{\star}$ be the level exhibiting the maximum intensity across all $\nu$ given by
\begin{align}
\nu^{\star} = \underset{\nu}{\operatorname{arg\,max}}\left\{\underset{\mathbf{x}}{\operatorname{max}} \, P\left(\nu, \mathbf{x}\right)\right\}
\end{align}
we infer $\sigma^\star$ as the detected blob radius at the initial scale by
\begin{align}
\sigma^{\star} = 2^{\sfrac{\nu^{\star}}{2}} \sqrt{2^{\operatorname{mod}(\nu^\star, 2)}} 
\label{eq:lap_scale}
\end{align}
where the base-2 terms account for halved resolutions and the half-octave representation across pyramid levels, respectively. 
It may be an intuitive observation that a diameter $M$ of a micro image proportionally scales with the blob radius $\sigma^\star$. A proof for this relationship is provided in Appendix~\ref{subsec:app:scale_max}.
Section~\ref{sec:calib} demonstrates that our automatic detection of the micro image size $M$ is an important feature for micro image registration and subsequent light-field alignment processes, which enable \Package{PlenoptiCam} to cope with a variety of different MLA and objective lens specifications. \par
\subsubsection{CentroidExtractor}
\label{sec:3:2:2}
Initial approximation of Micro Image Centers (MICs) is a key task in plenoptic image registration and has been subject of existing research. 
An early method developed by Dansereau~{\it et al.}~\cite{DANSCAL} convolves a white image with a kernel of fixed size dedicated to a Lytro camera and subsequently analyzes local peaks. Cho~{\it et al.}~\cite{Cho:2013} and similarly Liang~{\it et al.}~\cite{Liang:2016} iteratively apply a morphological erosion operator to white images until a pattern of isolated micro image regions is obtained. Although image erosion facilitates rough centroid detection from intensity maxima, the iterative nature of this approach leaves space for optimization with regard to speed and robustness. %
Unlike the previous detection methods, we identify MICs as local extrema in a LoG-convoluted image $I_{c}(\mathbf{x})=I_w(\mathbf{x}) \ast \nabla^2 G(\sigma^\star, \mathbf{x})$ by using Non-Maximum-Suppression~(NMS) in the form of the $3 \times 3$ neighborhood scan. Here the kernel $\nabla^2 G(\sigma^\star, \mathbf{x})$ scales with $M$ and 
helps carve out micro image peaks as it is responsive to the detected feature size.
We employ Pham's NMS method~\cite{NMS} and denote an MIC by $\mathbf{c}_n=\left[k_n,l_n\right]^\intercal \in \mathbb{Z}^{2}$. 
\subsubsection{CentroidRefiner}
\label{sec:3:2:2:5}
It is important to note that the preliminary \Package{CentroidExtractor} yields centroids located at integer sensor coordinates. Several studies have shown that integer coordinate precision induces artifacts when decomposing a raw plenoptic image to the sub-aperture image representation of a light-field~\cite{Cho:2013,DANSCAL, Liang:2016}. For accurate light-field decomposition, we refine centroids with sub-pixel precision by taking pixel intensities into account that belong to the same micro image region $\mathcal{R}_n$, which is typically in a range of $M \times M$ size. To obtain sub-pixel precise centroids
$\bar{{\mathbf{c}}}_n=\begin{bmatrix}\bar{k}_n &\bar{l}_n\end{bmatrix}^\intercal \, \in \, \mathbb{R}^{2}$, we perform coordinate refinement by
\begin{align}
\bar{k}_n = \sum_{(k,l) \, \in \, \mathcal{R}_n} \frac{I_{c}(k, l)\cdot k}{\sum I_{c}(k, l)} 
\quad \text{and} \quad
\bar{l}_n = \sum_{(k,l) \, \in \, \mathcal{R}_n} \frac{I_{c}(k, l)\cdot l}{\sum I_{c}(k, l)} 
\label{eq:peak}
\end{align}
where $I_c(k, l)$ represents a LoG-convoluted white image. As an alternative, we compute area centroids from a binary micro image after thresholding with the 75th percentile of $I_c(k, l)$ which then simplifies Eq.~(\ref{eq:peak}) to
\begin{align}
\bar{k}_n = \frac{1}{|\mathcal{R}_n|} \sum_{(k,l) \, \in \, \mathcal{R}_n} k
\quad \text{and} \quad
\bar{l}_n = \frac{1}{|\mathcal{R}_n|} \sum_{(k,l) \, \in \, \mathcal{R}_n} l
\label{eq:area}
\end{align}
where $|\cdot|$ denotes the cardinality \if which in this case\fi providing the total number of elements within a micro image region $\mathcal{R}_n$ above the threshold value. By default, \Package{PlenoptiCam} uses the latter approach given in Eq.~(\ref{eq:area}) as it proves to be the more generic solution for white images suffering from noise or saturation. Figures~\ref{fig:calib_res} and \ref{fig:regress} show examples of detected centroids using this method. \par
\subsubsection{CentroidSorter}
\label{sec:3:2:3}
At this point, only little is known about the dimensions and geometric micro lens arrangement. To enable generic calibration for different camera models with custom MLAs and arbitrary lens numbers, it is mandatory to examine such fundamental properties. Let~$\mathbf{C}~=~\{\bar{{\mathbf{c}}}_n~|~n~\in~\mathbb{N}\}$ be a finite and unordered set of centroids $\bar{{\mathbf{c}}}_n = \begin{bmatrix}\bar{k}_n & \bar{l}_n\end{bmatrix}^\intercal$. %
%
For a first centroid spacing approximation, we assume the ratio of sensor dimensions $K$ and $L$ matches the aspect ratio of the MLA giving $H=\sqrt{|{\mathbf{C}}| \times \sfrac{L}{K}}$ for the horizontal micro lens resolution with $|{\mathbf{C}}|$ as the total number of centroids. The centroid spacing is estimated via $\sfrac{L}{H}$ and used to form a set of neighbours $\mathbf{S}$ from an arbitrary centroid $\mathbf{\bar{c}}_r \in \mathbf{C}$ via
\begin{align}
\mathbf{S} =
\{
\bar{{\mathbf{c}}}_n-\mathbf{\bar{c}}_r \mid \bar{{\mathbf{c}}}_n \in \mathbf{C} \land \sfrac{L}{2H}< \left\|\mathbf{\bar{c}}_n-\mathbf{\bar{c}}_r\right\|_2<\sfrac{3L}{2H}
\}
\label{eq:mla_struct_a}
\end{align}
where $\left\| \cdot \right\|_2$ denotes the $\ell^2$ norm. 
The MLA packing geometry $\mathcal{P}$ is determined by analyzing angles $\alpha_n$ of $\bar{{\mathbf{c}}}_n \in \mathbf{S}$ as follows
\begin{align}
\mathcal{P} := \begin{cases}
\text{hexagonal},
& \text{if } \forall \bar{\mathbf{c}}_n \in \mathbf{S}, \lfloor\alpha_n\sfrac{12}{\pi}\rceil \, \in \{1, 3\}
\\
\text{rectangular}, & \text{if } \forall \bar{\mathbf{c}}_n \in \mathbf{S}, \lfloor\alpha_n\sfrac{12}{\pi}\rceil \, \in \{0,3\}
\end{cases}
\label{eq:mla_struct_b}
\end{align}
where $\lfloor\cdot\rceil$ is the nearest integer operation and $\alpha_n$ is given by
\begin{align}
	\alpha_n = \arccos\left(\frac{\left(\mathbf{1}^\intercal \bar{\mathbf{c}}_n\right)}
	{\sqrt{\left(\mathbf{1}^\intercal\mathbf{1}\right)\left(\bar{\mathbf{c}}_n^\intercal\bar{\mathbf{c}}_n\right)}}\right)
\end{align}
while $\mathbf{1}=\begin{bmatrix}1 & 1\end{bmatrix}^\intercal\in\mathbb{R}^2$ acts as a reference vector at $45\si{\degree}$. \par
Rearranging plenoptic micro images to a sub-aperture light-field requires centroids to be indexed in 2-D since their relative positions act as spatial coordinates in the sub-aperture image domain. However, the order of centroids within the array remains ambiguous. Thus, we seek a procedure that assigns indices $j\in\{1, 2, \ldots, J\}$ and $h\in\{1, 2, \ldots, H\}$ to each micro image centroid $\bar{{\mathbf{c}}}_{j,h}$. %
The proposed sort procedure begins with the search for the most upper left centroid $\bar{{\mathbf{c}}}_{1,1}$ with $j=1$, $h=1$ which is found by the minimum Euclidean distance to the image origin as given by 
\begin{align}
\bar{{\mathbf{c}}}_{1,1} = \bar{\mathbf{c}}_{n^{\star}}, \quad \text{where} \quad n^{\star} \in \underset{1\le n \le|\mathbf{C}|}{\mathrm{arg\,min}} \big( \left\| \bar{\mathbf{c}}_n \right\|_2 \big)
\end{align}
On the basis of an indexed centroid $\bar{{\mathbf{c}}}_{j,h}$, we search for its spatial neighbor by iterating through $\mathbf{C}$ until finding a candidate $\bar{{\mathbf{c}}}_n = \begin{bmatrix}\bar{k}_n & \bar{l}_n\end{bmatrix}^\intercal$ as seen in
\begin{align}
	\bar{{\mathbf{c}}}_{j+1,h} = \{\bar{{\mathbf{c}}}_n \mid \bar{{\mathbf{c}}}_n \in \mathbf{C} \land \varphi(\bar{{\mathbf{c}}}_{j, h}, \bar{{\mathbf{c}}}_n)\}
\end{align}
that satisfies a boundary condition $\varphi(\cdot) \in \{0, 1\}$ given by
\begin{equation}
\begin{aligned}
%
\varphi(\bar{{\mathbf{c}}}_a, \bar{{\mathbf{c}}}_b) ={} &\bar{k}_{a}+\Upsilon_1 < \bar{k}_b < \bar{k}_{a}+\Upsilon_2 \, \, \land \\ &\bar{l}_{a}+\Upsilon_3 < \bar{l}_b < \bar{l}_{a}+\Upsilon_4
 \label{eq:cond}
\end{aligned}
\end{equation}
where $\mathbf{\Upsilon}=\mathbf{e}\sfrac{M}{2}$ with $\mathbf{e}=\begin{bmatrix}e_1 & e_2 & e_3 & e_4\end{bmatrix}^\intercal$ is a boundary scale vector considering earlier identified MLA properties. Note that horizontal and vertical coordinates in Eq.~(\ref{eq:cond}) may be swapped to switch between search directions. An exemplary result showing indexed centroids is depicted in Fig.~\ref{fig:calib_res}. \par
\subsubsection{GridFitter}
\label{sec:3:2:4}
At this stage of the calibration pipeline, centroids rely on intensity distributions, which may be affected by a broad range of irregularities arising from the MLA-sensor compound. %
Dansereau~{\it et al.}~\cite{DANSCAL} compress centroid information by forming a consistently spaced grid based on provided metadata. Usage of Delaunay triangulation as proposed by Cho~{\it et al.}~\cite{Cho:2013} suits hexagonal arrangements, but lacks to compensate for grid inconsistencies. Instead, we elaborate on a least-squares (LSQ) regression as potentially intended in a patent filed by Lytro~\cite{Liang:2016}. \par %
Based on previously determined micro lens numbers $(J,H)$ and geometric packing $\mathcal{P}$, we tailor a grid model function $G(\cdot)$ to produce an ordered array of points $\tilde{\mathbf{g}}_{j,h}=G(J,H;\mathcal{P})$ where $\tilde{\mathbf{g}}_{j,h} = \begin{bmatrix}\tilde{k}_{j,h} & \tilde{l}_{j,h} & \tilde{z}_{j,h}\end{bmatrix}^\intercal \in \mathbb{R}^3$ consists of consistently spaced and normalized spatial centroid coordinates $\tilde{k}_{j,h}$, $\tilde{l}_{j,h}$ and $\tilde{z}_{j,h}$ that may be homogenized, i.e., $\tilde{z}_{j,h}=1$. The grid generation is followed by a projective transformation using $\mathbf{P} \in \mathbb{R}^{3\times 3}$ that yields a 3-vector centroid $\tilde{\mathbf{c}}'_{j,h}=\mathbf{P}\tilde{\mathbf{g}}_{j,h}$, which in matrix notation is written as
\begin{align}
\begin{bmatrix}
\tilde{k}'_{j,h} \\
\tilde{l}'_{j,h} \\
\tilde{z}'_{j,h} \\
\end{bmatrix}
=
\begin{bmatrix}
p_1 & p_2 & p_3 \\
p_4 & p_5 & p_6 \\
p_7 & p_8 & 1 \\
\end{bmatrix}
\begin{bmatrix}
\tilde{k}_{j,h} \\
\tilde{l}_{j,h} \\
\tilde{z}_{j,h} \\
\end{bmatrix}
\end{align}
covering spatial center offsets, grid scales and tilts about three axes~\cite{Hartley2004}. A centroid estimate $\hat{{\mathbf{c}}}_{j,h}=\begin{bmatrix}\hat{k}_{j,h} & \hat{l}_{j,h} & 1\end{bmatrix}^\intercal \in \mathbb{R}^{3}$ is then obtained by
$\hat{{\mathbf{c}}}_{j,h}=\mathbf{P}\tilde{\mathbf{g}}_{j,h}/\tilde{z}'_{j,h}$. %
%
\par
To determine an optimal $\mathbf{P}^*$, we employ a distance metric in $\mathbf{F}_{j,h}$ comparing all measured $\bar{{\mathbf{c}}}_{j,h}$ and generated $\hat{\mathbf{c}}_{j,h}$ by
\begin{align}
\mathbf{F}_{j,h}=\lVert \bar{{\mathbf{c}}}_{j,h}-\hat{{\mathbf{c}}}_{j,h}\rVert_2 + \beta R\left(\bar{{\mathbf{c}}}_{j,h}, \hat{{\mathbf{c}}}_{j,h}, M\right) , \quad \forall j, h  \label{eq:obj_fun} 
\end{align}
with a regularization term $R(\cdot)$ and adjustable weight $\beta$.
\par %
Here, the regularizer penalizes false centroid shifts caused by asymmetric vignetting at off-center micro images by
\begin{align}
R\left(\bar{{\mathbf{c}}}_{j,h}, \hat{{\mathbf{c}}}_{j,h}, M\right) = \begin{cases}
0, & \text{if}\ \bar{\mathbf{d}}_{j,h} + M/\hat{M} < 0 \\
\sum_{(\bar{k}, \bar{l})}\bar{\mathbf{d}}_{j,h}, & \text{otherwise}
\end{cases}
\end{align}
with $\bar{\mathbf{d}}_{j,h}=|\bar{{\mathbf{c}}}_{j,h}-(p_3, p_6)|-|\hat{{\mathbf{c}}}_{j,h}-(p_3, p_6)|$ as an auxiliary distance measure and $\hat{M}\approx 20$ as a micro image size divider. %
\par
Let $\mathbf{P} \in \mathbb{R}^{3\times 3} \rightarrow \mathbf{p} \in \mathbb{R}^{9}$ be reshaped to a parameter vector $\mathbf{p}=\begin{bmatrix}p_1 & p_2 & p_3 & p_4 & p_5 & p_6 & p_7 & p_8 & 1 \end{bmatrix}^\intercal$
and similarly $\mathbf{F}_{j,h} \in \mathbb{R}^{J\times H} \rightarrow \mathbf{f} \in \mathbb{R}^{|{\mathbf{C}}|}$ be flattened to a vector-valued cost function $\mathbf{f}=\begin{bmatrix} f_1 & f_2 & \dots & f_{|{\mathbf{C}}|}\end{bmatrix}^\intercal$, the objective function is
\begin{align}
\underset{\mathbf{p}}{\operatorname{arg\,min}} \sum_{n=1}^{|{\mathbf{C}}|} f_n
\end{align}
 for which we employ the Levenberg-Marquardt (LM) step by
\begin{align}
\mathbf{p}_{k+1}=\mathbf{p}_k-(\mathbf{J}^\intercal\mathbf{J}+\mu\mathbf{D}^\intercal\mathbf{D})^{-1}\mathbf{J}^\intercal\mathbf{f} \label{eq:lma_step}
\end{align}
where $\mathbf{J}$ is the Jacobian, $\mathbf{J}^\intercal\mathbf{J}$ approximates the Hessian with $^\intercal$ as the matrix transpose and $\mathbf{J}^\intercal\mathbf{f}$ acts as the gradient~\cite{LMA:1978}. Here, a diagonal matrix $\mathbf{D}$ with an adaptive damping term $\mu$ allows for fast LM convergence. As opposed to an analytically derived Jacobian $\mathbf{J}$, a multi-variate numerical approximate $\mathbf{\tilde{J}}(\mathbf{p})~\approx~\mathbf{J}$ is employed. The iterative update procedure proceeds until a convergence condition is met. Successful estimates $\mathbf{P}^\star$ are fed into the projective matrix for a final centroid assignment $\hat{{\mathbf{c}}}_{j,h}^\star=\mathbf{P}^\star\tilde{\mathbf{g}}_{j,h}/\tilde{z}'_{j,h}$. Enhancements of the herein proposed LSQ grid regression are examined in Section~\ref{sec:calib}.
%
\subsection{LfpAligner}
\label{sec:3:3}
\subsubsection{CfaOutliers}
\label{sec:3:3:1}
Often, hot and dead pixels arise from electrical response variations in the sensor hardware and become noticeable as intensity outliers. %
In contrast to other pipelines~\cite{DANSCAL, Matysiak:2018}, we detect and rectify outliers prior to demosaicing channels from the Color Filter Array (CFA). The reasoning behind our decision is that false intensities are propagated to adjacent pixels during demosaicing, which may pass unnoticed by a detection at a later processing stage. To identify outliers, we regard each of the four Bayer channels as a two-dimensional (2-D) grey scale image $I_B(\mathbf{x})$ and analyze the difference from its Median-filtered version given by
\begin{align}
I_R(\mathbf{x}) = I_B(\mathbf{x}) - \mathcal{M}\big(I_B(\mathbf{x})\big)
\end{align}
where $I_R(\mathbf{x})$ is the reference image used for further analysis and $\mathcal{M}(\cdot)$ denotes the Median filter operator. From $I_R(\mathbf{x})$, we obtain the arithmetic mean $\bar{I}_R$ and the standard deviation $\sigma_R$ as local statistical measures from a sliding window of $2n+1$ size to replace potential outliers $I_B(\mathbf{x})$ as follows
\begin{align}
I_B(\mathbf{x}) = 
\begin{cases}
\mathcal{M}\big(I_B(\mathbf{x})\big),
& \text{if } I_B(\mathbf{x}) > \bar{I}_R + 4 \sigma_R
\\
I_B(\mathbf{x}), & \text{otherwise}
\end{cases}
\end{align}
where the if clause statement helps detect intensity outliers. Once a condition is fulfilled, nearby intensities in the range of $2n+1$ are used to replace an outlier. The selection is further constrained by only accepting a small number $n$ of candidates in a $n^2\times n^2$ window. Without this constraint, many pixels of a saturated image area would be falsely detected as outliers. 
\subsubsection{LfpDevignetter}
\label{sec:3:3:2}
In optical imaging, vignetting occurs at non-paraxial image areas as a result of either mechanical blocking of light or illumination fall-off from the cosine-fourth law, which arise from the Lambertian reflectance, pupil size reduction and the distance-dependant inverse square law~\cite{Vignetting:16}. Plenoptic images suffer from vignetting in similar ways whereas the appearance is given in \mbox{4-D} representation. So far, micro image vignetting is treated by a pixel-wise division with a normalized white image~\cite{DANSCAL, Matysiak:2018}. \Package{PlenoptiCam} uses this method by default, as it works well for white images not severely suffering from noise. 
To combat potential noise, \Package{LfpDevignetter} offers alternative \mbox{4-D} de-vignetting based on LSQ fitting inspired by classical 2-D image processing that has not yet been applied to plenoptic images. We adapt the classical procedure by iterating through each micro image and dividing it with normalized LSQ fit values to prevent noise propagation during the light-field de-vignetting.
\par %
The intensity surface of a white micro image is given as $I_m(u,v)$ and approximated by a multivariate polynomial regression. For 2\textsuperscript{nd} order polynomials, this fit function writes
\begin{align}
	I_m(u,v) = w_1 + w_2 u + w_3 v + \ldots + w_7 u^2v^2 \label{eq:vign:poly}
\end{align}
with $\mathbf{w}= \begin{bmatrix} w_1 & w_2 & \dots & w_7 \end{bmatrix}^\intercal$ as the regression coefficients. Provided the \mbox{2-D} micro image coordinate indices $u$ and $v$, this is translated to matrix form by
\begin{align}
\mathbf{A} = 
\begin{bmatrix}
1 & u_1 & v_1 & \cdots & u_1^2 v_1^2 \\
1 & u_2 & v_2 & \cdots & u_1^2 v_2^2 \\
\vdots  & \vdots & \vdots  & \ddots & \vdots  \\
1 & u_M & v_M & \cdots & u_M^2 v_M^2
\end{bmatrix}
, \,
\mathbf{b} =
\begin{bmatrix} 
I_m(u_1, v_1) \\ 
I_m(u_1, v_2) \\
\vdots \\ 
I_m(u_M, v_M) 
\end{bmatrix}
\label{eq:vign:vander}
\end{align}
with $\mathbf{A}$ as the Vandermonde matrix and $\mathbf{b}$ containing micro image intensities. %
With $\mathbf{A}$ generally being non-square, the equation system is solved via the pseudo-inverse~$^+$~given by
\begin{align}
\mathbf{A}^+ = \left(\mathbf{A}^\intercal \, \mathbf{A}\right)^{-1} \, \mathbf{A}^\intercal \label{eq:moore_penrose}
\end{align}
so that we obtain fit values $\mathbf{w}$ for each micro image by
\begin{align}
\mathbf{w} &= \mathbf{A}^+ \mathbf{b}\label{eq:vign:pseudo_inv}
\end{align}
After estimation of weight coefficients, we divide each micro image by its fitted white image counterpart. 
The effectiveness of the above method is demonstrated in Section~\ref{subsec:vign}.
\subsubsection{CfaProcessor}
The \Package{CfaProcessor} class is dedicated to raw sensor images taking care of debayering using Menon's algorithm~\cite{Menon2007c}, white balancing and color correction. Previous findings made with regards to highlight processing~\cite{Matysiak:2018} were adopted as they yield enhanced sub-aperture image quality.
\subsubsection{LfpRotator}
\label{sec:3:3:3}
At the assembling stage of a plenoptic camera, a micro lens grid is ideally placed so that its tilt angles are in line with that of a sensor array. 
In the real world, however, this grid may likely be displaced with respect to the pixel grid. 
To counteract rotations about the $z$ axis, our approach exploits the fact that images exposing aberrations tend to be aberration-free along their central axes. This suggests that a centroid row close to the image center forms a line, which may be a reliable indicator for the MLA rotation angle. Such a central row is obtained by  $\hat{\mathbf{c}}^\star_{j,\delta}=\begin{bmatrix}\bar{k}^\star_{j,\delta} & \bar{l}^\star_{j,\delta}\end{bmatrix}^\intercal\in \mathbb{R}^2$ with $\delta=\lfloor\sfrac{(H-1)}{2}\rceil$ and $H$ as the total number of micro lenses in the vertical direction. Applying LSQ regression on $\hat{\mathbf{c}}^\star_{j,\delta}$ similar to~\cref{eq:obj_fun,eq:lma_step} yields an angle $\theta_z$. Centroid and image rotation is accomplished by successively multiplying coordinates with a rotation matrix $\mathring{\mathbf{R}}_{z}$ and a translation matrix $\mathbf{T}$ giving a rotationally aligned centroid set $\hat{\mathbf{c}}_{j,h}' = \mathbf{T}^{-1} \, \mathring{\mathbf{R}}_{z} \, \mathbf{T} \, \hat{\mathbf{c}}^\star_{j,h}$ for further processing. 
The \Package{LfpRotator} is left optional as rotational alignments are covered in the more general resampling stage.
%
%
\subsubsection{LfpResampler}
\label{sec:3:3:4}
Micro image centroids $\hat{\mathbf{c}}^\star_{j,h} \in \mathbb{R}^2$ are likely to be found at sub-pixel positions, as Section~\ref{sec:3:2} reveals. Taking pixel intensities from nearest integer coordinates leads to errors and thus image artifacts~\cite{Cho:2013}. It is mandatory to preserve fractional digits of centroids to accurately decompose sub-aperture images. Therefore, we provide two competitive alignment schemes, which retain geometric properties via 
\begin{enumerate}
\item aligning the entire light-field by a single transformation 
\item resampling each micro image $I_m(u, v)$ individually\if a method that differs from existing approaches\fi 
\end{enumerate}
\par
For global alignment, the projective matrix $\mathbf{P}^\star \in \mathbb{R}^{3\times 3}$ from the \Package{GridFitter} in Section~\ref{sec:3:2:4} is used along with another projection matrix $\mathbf{P}_g$ representing a desired target grid $\tilde{\mathbf{g}}_{j,h}$ at consistently spaced pixel coordinates. The ideal global transfer matrix is then $\mathbf{P}_t=\mathbf{P}_g\left(\mathbf{P}^\star\right)^{-1}$ having 8 degrees of freedom. Similar to\if a method proposed by Dansereau {\it et al.}\fi~\cite{DANSCAL}, our transfer matrix $\mathbf{P}_t$ dictates a global light-field transformation interpolating all micro images such that their centroids exhibit consistent spacing and coincide with actual pixel centers afterwards. Note that this procedure accounts for rotational MLA deviations and facilitates decomposition into sub-aperture images at a later stage.
\par
\begin{figure}[H]
	\centering
	\begin{minipage}[t]{.49\linewidth}
		\centering
		\resizebox{.8\textwidth}{!}{\includegraphics{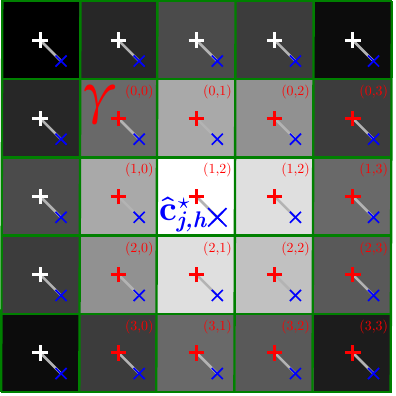}}
		\subcaption{\label{fig:intpolscheme:a}}
	\end{minipage}
	\begin{minipage}[t]{.49\linewidth}
		\centering
		\resizebox{.8\textwidth}{!}{\includegraphics[width=.2625\linewidth]{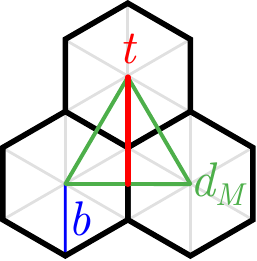}}
		\subcaption{\label{fig:intpolscheme:b}}
	\end{minipage}
	\caption{\textbf{Interpolation schemes} with \subref{fig:intpolscheme:a} centroid $\hat{\mathbf{c}}^\star_{j,h}$ in a micro image and \subref{fig:intpolscheme:b} hexagonal grid geometry with $t$ as the height and $d_M$ as the side length of an equilateral triangle.}
\end{figure}

As an alternative, resampling is conducted locally by \mbox{2-D} interpolation of each micro image so that its central pixel and detected centroid match after spatial shifting. Figure~\ref{fig:intpolscheme:a} depicts local resampling showing a centroid $\hat{\mathbf{c}}^\star_{j,h}$ in a micro image with weighting coefficients $\gamma$ from surrounding pixels. 
%
\setlength{\columnsep}{20pt}%
Following the detection of a hexagonal MLA using Eq.~(\ref{eq:mla_struct_b}), \Package{LfpResampler} takes care of the conversion to a rectangular grid, which exploits the fact that three hexagonally arranged micro image centroids span an equilateral triangle with equal distance to its barycenter as seen in Fig.~\ref{fig:intpolscheme:b}. Considering this geometry, a rectangular micro image grid can be obtained by averaging every set of three adjacent micro images. %
Figure~\ref{fig:intpolscheme:b} illustrates this with $d_M=b\sqrt{3}$ as the spacing of two centroids and the side length of a triangle whereas $t=\sfrac{3b}{2}$ is its height in a rectangular grid. One may note that $\sfrac{d_M}{t} =\sqrt{3}/\sfrac{3}{2}$ becomes $\sfrac{2}{\sqrt{3}}$ after rearranging. From this, it follows that the sampling density along $t$ is $\sfrac{2}{\sqrt{3}}$ times higher in relation to $d_M$. Therefore, upsampling the less dense spatial dimension by $\sfrac{2}{\sqrt{3}}$ will achieve consistency in the sampling density. \par
In local resampling, rectangular grid conversion is accomplished by de-interleaving and elongating the dimension orthogonal to $t$, which corresponds to the horizontal direction in Lytro cameras. This shift and stretch alignment breaks~down to translating every other coordinate vector by $\sfrac{d_M}{2}$ and simultaneously upsampling the micro image number by $\sfrac{2}{\sqrt{3}}$. The light-field array is reshaped prior to the spatial interpolation as consecutive pixels are angular neighbors on the sensor.%
%
%
%
\subsection{LfpExtractor}
\label{sec:3:4}
%
%
\subsubsection{LfpRearranger}
\label{sec:3:4:1}
Let $E_{fs}\left[s_j,u_{c+i}\right]$ be a spatio-angular \mbox{2-D} slice of an aligned 4-D micro image array, the rearrangement to a sub-aperture slice $E_i\left[s_j\right]$ at view position $i$ reads
\begin{align}
E_{i}\left[s_j\right] = E_{f_s}\left[s_j \, , \, u_{c+i}\right] 
\label{eq:vpExtract}
\end{align}
where micro image pixels at $u_{c+i}$ are consecutively collected and relocated in a new vector $E_i\left[s_j\right]$. The preliminary resampling alignment enables each centroid to be represented by $c = (M-1)/2$ with an odd micro image pixel diameter $M$ so that $\operatorname{mod}(M, 2) = 1$. A change in index $i \in \{-c, \ldots, c\}$ controls the relative view location in the light-field whereas index $j \in \{1, 2, \ldots, J \}$ iterates through micro lenses as the spatial domain. Note that Eq.~\eqref{eq:vpExtract} refers to one direction, which is equally applied orthogonal to it for complete rendering. \par %
%
%
\subsubsection{HexCorrector}
\label{sec:3:4:2}
As described in the \Package{LfpResampler}, the elongation scheme in local resampling implies shifting every other row by half the centroid spacing to form a consistent rectangular lattice. However, this technique causes a shortcoming becoming visible as zipper-like artifacts along straight object edges in sub-aperture images (e.g., see Fig.~\ref{fig:hexart}). \par
%
Peers addressed this issue via barycentric interpolation~\cite{Cho:2013}, posterior demosaicing~\cite{Seifi:14} and depth-guided resampling~\cite{Yongwei:19}. We tackle this artifact after the fact by the identification of affected pixels.
For the detection, we compose two auxiliary images $E_i[\hat{s}_j]$ and $E_i[\check{s}_j]$ by de-interlacing, i.e. taking every other row, of a sub-aperture image $E_i[s_j]$ where $E_i[\hat{s}_j]$ are the unshifted pixel vectors and $E_i[\check{s}_j]$ the shifted counterparts. A pixel-wise subtraction yields local variances $\tilde{V}_i[\check{s}_j]$ given by
\begin{align}
\tilde{V}_i[\check{s}_j] = E_i[\check{s}_j] - E_i[\hat{s}_j]
\end{align}
which contain strong responses at edges parallel to the shift and stretch direction. To neglect real object edges, we further subtract the magnitude of the partial derivative $\partial_{\hat{s}} E_i[\hat{s}_j]$ by 
\begin{align}
\bar{V}_i[\check{s}_j] = \lvert \tilde{V}_i[\check{s}_j] \rvert - \lvert \partial_{\hat{s}} E_i[\hat{s}_j] \rvert
\end{align}
where it is assumed that $\bar{V}_i[\check{s}_j]$ exhibits peaks for potential candidates. Noisy responses are eliminated via threshold $\tau$ by
\begin{align}
V_i[\check{s}_j] = 
\begin{cases}
1, & \text{if } \bar{V}_i[\check{s}_j] > \tau \\
0, & \text{otherwise}
\end{cases}
\end{align}
To further exclude false positives along the shift direction, we reject candidates in $V_i[\check{s}_j]$ not being part of a consecutive sequence of minimum length 4. %
In doing so, treated areas have a sufficient size to be visually recognized. The remainders in $V_i[\check{s}_j]$ are then used to make substitutions in $E_i[s_j]$ by
\begin{align}
E_i[s_j] = 
\begin{cases}
\frac{1}{\mathcal{|R|}} \sum_{h\in \mathcal{R}}^{} E_i[t_{h}], & \text{if } V_i[\check{s}_j] = 1 \\
E_i[s_j], & \text{otherwise}
\end{cases}
\end{align}
for each $j \in \{1, 2, \ldots,  J \}$ while $t_{h}$ is the orthogonal counterpart of $s_j$ in a 2-D region $\mathcal{R}$. The results are seen in Fig.~\ref{fig:hexart}. %
\subsubsection{LfpColorEqualizer}
\label{sec:3:4:3}
Lens components generally expose a gradual intensity decline toward image edges caused by the cosine-fourth law~\cite{Vignetting:16}. With a micro image fully covered by the sensor, this illumination fall-off is spread across a relatively small group of Bayer pattern pixels merged during demosaicing. Thus, visible intensity variances in off-axis sub-aperture views arise from micro image vignetting that we aim to rectify. %
Here, the goal is to propagate trusted intensities from paraxial image areas that expose no severe image aberrations to sub-aperture images located at the edge of a light-field. It can thus be used in addition to de-vignetting by utilizing the information redundancy within the light-field. A recent study employed Gaussian Mixture Models (GMMs) in conjunction with disparity correspondences computed at a preceding stage to reduce color inconsistencies~\cite{Matysiak:2018}. However, the authors note that their procedure requires a high computational load (240 minutes per light-field) due to the many steps involved. \par
To overcome this limitation, we employ linear mappings based on probability distributions. Histogram Matching~(HM) may be a starting point since it is invariant of the texture, preserves parallax information while requiring relatively little computational effort as opposed to iterative methods~\cite{Matysiak:2018}. %
Besides using HM, we extend a recent advancement in image color transfer based on the Monge-Kantorovich-Linearization~(MKL). Piti{\'e} and Kokaram~\cite{Pitie:2007} introduced MKL to the field of image processing to facilitate automatic color grading in media production. As of now, MKL has not been applied in the context of light-field color transfer and appears to not have been combined with channel-wise HM. Due to MKL using Multi-Variate Gaussian Distributions~(MVGDs) in combination with HM, we expect our novel HM-MKL-HM compound to outperform a stand-alone HM and Piti{\'e}'s pure MKL in terms of accuracy while largely reducing the computational complexity imposed by Matysiak's method~\cite{Matysiak:2018}. %
Mathematical details are presented hereafter. \par
\begin{figure*}
	\centering
	\includegraphics[width=.8\linewidth]{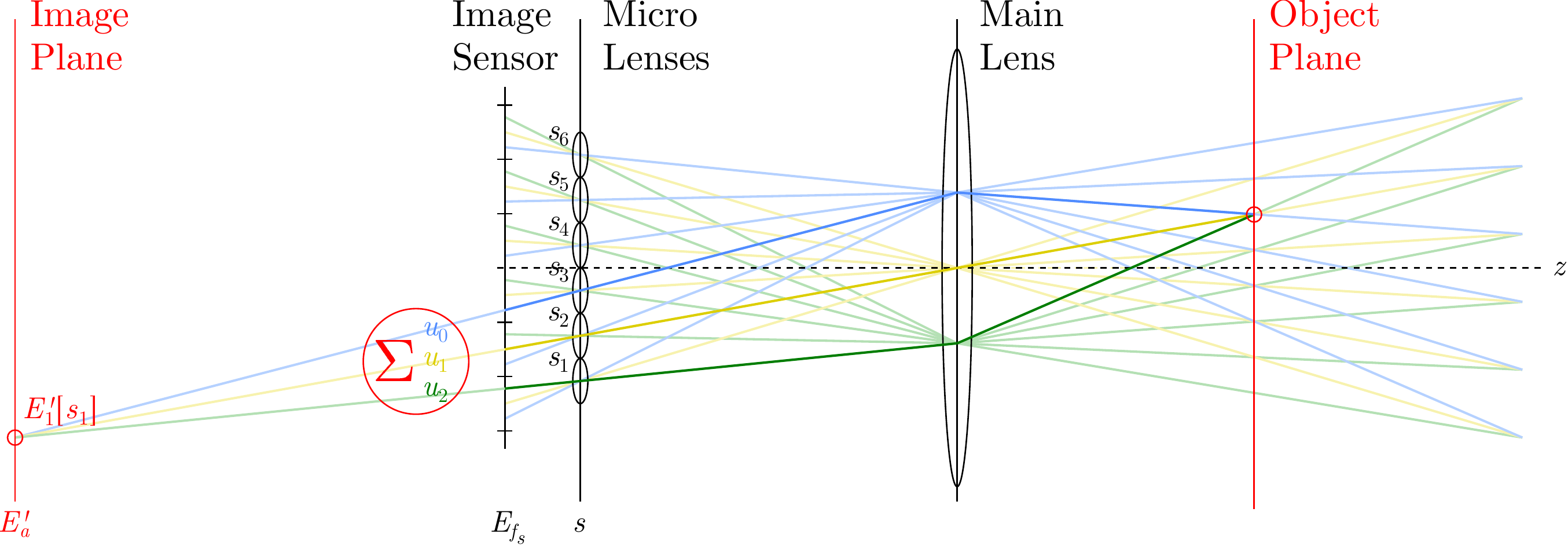}
	\caption{\textbf{Refocusing concept} with chief ray intersections indicating object and image plane for a recovered point $E'_1[s_1]$.\label{fig:refo_model}}
\end{figure*}
%
%
Let $\mathbf{r}\in \mathbb{R}^{JH}$ be single color channel intensities of an image $E_{i}\left[s_j\right]$, its Probability Density Function~(PDF) is given by
\begin{align}
f(k,\mathbf{r}) = \frac{\eta\left(k,\mathbf{r}\right)}{JH} \,  , 
\quad \forall k \in \{1, 2, \ldots, \mathcal{L}\}
\label{eq:pdf}
\end{align}
where $\eta(k , \, \cdot \, )$ yields the number of pixels with intensity level $k$ while $\mathcal{L}$ is the maximum level~\cite{gonzalez2017digital}. The histogram is normalized by $JH$ as the total pixel count of the image. From this we compute $F(k, \mathbf{r})$ as the Cumulative Density Function (CDF), which is obtained using an auxiliary index $\epsilon$ via
\begin{align}
F(k, \mathbf{r}) = \sum_{\epsilon=1}^{k} \frac{\eta\left(\epsilon,\mathbf{r}\right)}{JH} \, ,
\quad \forall k \in \{1, 2, \ldots, \mathcal{L}\}
\label{eq:cdf}
\end{align}
To match a source $F(k,\mathbf{r})$ with a target $G(k,\mathbf{z}) \defeq F(k,\mathbf{z})$, we perform a mapping with $T : F(k,\mathbf{r}) \rightarrow G(k,\mathbf{z})$ that yields
\begin{align}
\mathbf{r} \leftarrow T\left(F(k,\mathbf{r})\right)
\end{align}
where each $F(k,\mathbf{r})$ gets assigned a new value $G(k,\mathbf{z})$ from the probability map $T$ implemented as a discrete lookup table.  \par
While a channel-wise HM is effective, it fails to transfer colors at a level satisfying our visual perception. It is thus our goal to determine an optimal PDF transport. 
Due to the stochastic nature of intensity distributions $(r, g, b)$, we regard each source $\mathbf{R}=\begin{bmatrix}\mathbf{r}^{(r)} & \mathbf{r}^{(g)} & \mathbf{r}^{(b)}\end{bmatrix}^\intercal$ and target $\mathbf{Z}=\begin{bmatrix}\mathbf{z}^{(r)} & \mathbf{z}^{(g)} & \mathbf{z}^{(b)}\end{bmatrix}^\intercal$ as a correlated channel MVGD $\mathcal{N}(\cdot)$ given by
\begin{align}
\mathcal{N}(\mathbf{R};\boldsymbol\mu_r, \mathbf{\Sigma}_r) = \frac{\exp\left(-\frac{1}{2} (\mathbf{R}-\boldsymbol\mu_r)^\intercal \mathbf{\Sigma}^{-1}_r (\mathbf{R}-\boldsymbol\mu_r) \right)}{\sqrt{(2\pi)^{\text{rank}\left(\mathbf{\Sigma}_r\right)} |\mathbf{\Sigma}_r |}} 
\label{eq:mvgd}
\end{align}
where $\mathbf{\Sigma}_{r} \in \mathbb{R}^{3\times 3}$ and $\mathbf{\Sigma}_{z} \in \mathbb{R}^{3\times 3}$ denote covariance matrices with $\mathbf{\boldsymbol\mu}_r \in \mathbb{R}^{3\times 1}$ and $\mathbf{\boldsymbol\mu}_z \in \mathbb{R}^{3\times 1}$ as the mean vectors of \mbox{$\mathbf{R} \in \mathbb{R}^{3\times JH}$} and $\mathbf{Z} \in \mathbb{R}^{3\times N}$. 
A desired transfer $\hat{t}(\mathbf{R})$ requires MVGDs to be 
$\mathcal{N}(\mathbf{Z}; \boldsymbol\mu_z, \mathbf{\Sigma}_z) \propto \mathcal{N}(\hat{t}(\mathbf{R}); \boldsymbol\mu_z, \mathbf{\Sigma}_z)$ so that a result $\hat{t}(\mathbf{R})$ is substituted for a $\mathbf{Z}$ when setting
\begin{align}
(\mathbf{Z} - \boldsymbol\mu_z)^\intercal \mathbf{\Sigma}^{-1}_z \left(\hat{t}(\mathbf{R}) - \boldsymbol\mu_z\right) = (\mathbf{R} - \boldsymbol\mu_r)^\intercal \mathbf{\Sigma}^{-1}_r(\mathbf{R} - \boldsymbol\mu_r)
\label{}
\end{align}
after dropping the constant leading terms. Using the pseudo-inverse $^+$ as in Eq.~(\ref{eq:moore_penrose}), we arrange the above equation to
\begin{align}
\hat{t}(\mathbf{R})-\boldsymbol\mu_z = \left((\mathbf{Z} - \boldsymbol\mu_z)^\intercal \mathbf{\Sigma}^{-1}_z\right)^{+}(\mathbf{R} - \boldsymbol\mu_r)^\intercal \mathbf{\Sigma}^{-1}_r(\mathbf{R} - \boldsymbol\mu_r)
\label{eq:mvgd:rearrange}
\end{align}
As we seek a compact transfer matrix, we define $\mathbf{M}$ to be
\begin{align}
\mathbf{M} &= \left((\mathbf{Z} - \boldsymbol\mu_z)^\intercal \mathbf{\Sigma}^{-1}_z\right)^+ (\mathbf{R} - \boldsymbol\mu_r)^\intercal \mathbf{\Sigma}^{-1}_r \label{eq:analytical_transfer}
\end{align}
so that after substitution and rearranging, Eq.~(\ref{eq:mvgd:rearrange}) becomes
\begin{align}
\hat{t}(\mathbf{R}) &= \mathbf{M}(\mathbf{R} - \boldsymbol\mu_r) + \boldsymbol\mu_z
\end{align}
%
for the forward MVGD transfer. The determination of $\mathbf{M}$ is key for an optimal transport. Piti{\'e} and Kokaram employed MKL with $\mathbf{M}^\intercal \mathbf{\Sigma}^{-1}_z \mathbf{M} = \mathbf{\Sigma}^{-1}_r$ which is given by
\begin{align}
\mathbf{M} =
\mathbf{\Sigma} _{r}^{\sfrac{-1}{2}}\big{(}\mathbf{\Sigma}_{r}^{\sfrac{1}{2}}\mathbf{\Sigma}_{z}^{\frac{}{}}\mathbf{\Sigma}_{r}^{\sfrac{1}{2}}\big{)}^{\sfrac{1}{2}} \mathbf{\Sigma}_{r}^{\sfrac{-1}{2}} 
\label{eq:mkl}
\end{align}
One may note that the authors elaborated on a variety of solutions for~$\mathbf{M}$ (e.g., Cholesky factorization) among which MKL proved to be the most successful in terms of accuracy~\cite{Pitie2008bookchapter}. We complement this concept by providing an analytical solution in Eq.~(\ref{eq:analytical_transfer}) and combining it with HM in a sequential order (e.g., HM-MVGD-HM). Taking the central light-field image as a target $\mathbf{Z}$, we iterate through the \mbox{2-D} angular light-field where each angular view is a source $\mathbf{R}$. \par
After the color transfer, intensities undergo an automatic dynamic range alignment using the lower and upper histogram percentiles (0.005 and 99.9) of the central light-field image. This is followed by a gamma correction according to the sRGB standard. An evaluation of the herein proposed methods is carried out in Section~\ref{sec:4}.
\subsection{LfpRefocuser}
\label{sec:3:5}
The \Package{LfpRefocuser} enables computational change of the optical focus in light-field images by offering 3 different mechanisms. The fundamental technique is the conventional \textit{shift-and-sum} method as originally presented by Isaksen~{\it et al.}~\cite{ISAKSEN} who employed a light-field taken by an array of cameras. This concept was transferred to sub-aperture images $E_{i}\left[s_j\right]$ by Ng~\textit{et al.}~\cite{NGLEV}, where a refocused image vector $E'_{a}\left[s_j\right]$ with synthetic focus scale $a$ is given by
\begin{align}
E'_{a}\left[s_j\right] &= \sum_{i=-c}^{c} E_{i}\left[s_{j-a(c+i)}\right] , \quad  a \in \mathbb{Q} \label{eq:refocusingAlt}
\end{align}
for the \mbox{\mbox{1-D}} case leaving out variables of the second spatial domain. A refinement option enables sub-pixel precision in refocusing via upsampling each sub-aperture image before integration. While this numerically increases the spatial resolution, it also allows for sub-sampling the focal depth range. \par
As an alternative, \textit{shift-and-sum} is accomplished based on an aligned micro image array $E_{f_s}\left[ s_j \, , \, u_{c+i}\right]$, which writes
\begin{align}
E'_{a}\left[s_j\right] &= \sum_{i=-c}^{c} E_{f_s}\left[ s_{j+a(c-i)} \, , \, u_{c+i}\right]  , \quad  a \in \mathbb{Q}
\label{eq:refocusing}
\end{align}
and can be thought of as employing an interleaved convolution kernel as shown in our previous work examining FPGA-based refocusing~\cite{Hahne:IEEE:2018}. For an intuitive understanding of computational refocusing, the concept behind Eq.~(\ref{eq:refocusing}) is illustrated in Fig.~\ref{fig:refo_model} with the aid of paraxial optics. Interested readers may note that refocused distances can be pinpointed in space, using this model~\cite{Hahne:OPEX:16, Plenoptisign:2019}. \par
The featured \mbox{\Package{LfpScheimpflug}} class mimics the identically named principle on a computational level. Tilting the refocused plane is achieved by fusing spatial image areas from a monotonically varying synthetic focus parameter $a$. This enables tilted focus renderings along horizontal, vertical and both diagonal image corner-to-corner directions, where focal start and end points rely on provided $a$. Figure~\ref{fig:scheim} depicts a plenoptic-based photograph exhibiting the Scheimpflug effect. 
\section{Results}
\label{sec:4}
\subsection{Calibration}
\label{sec:calib}
\begin{figure}[H]
	\centering
	\begin{minipage}[t]{.49\linewidth}
		\resizebox{\textwidth}{!}{
			\includegraphics{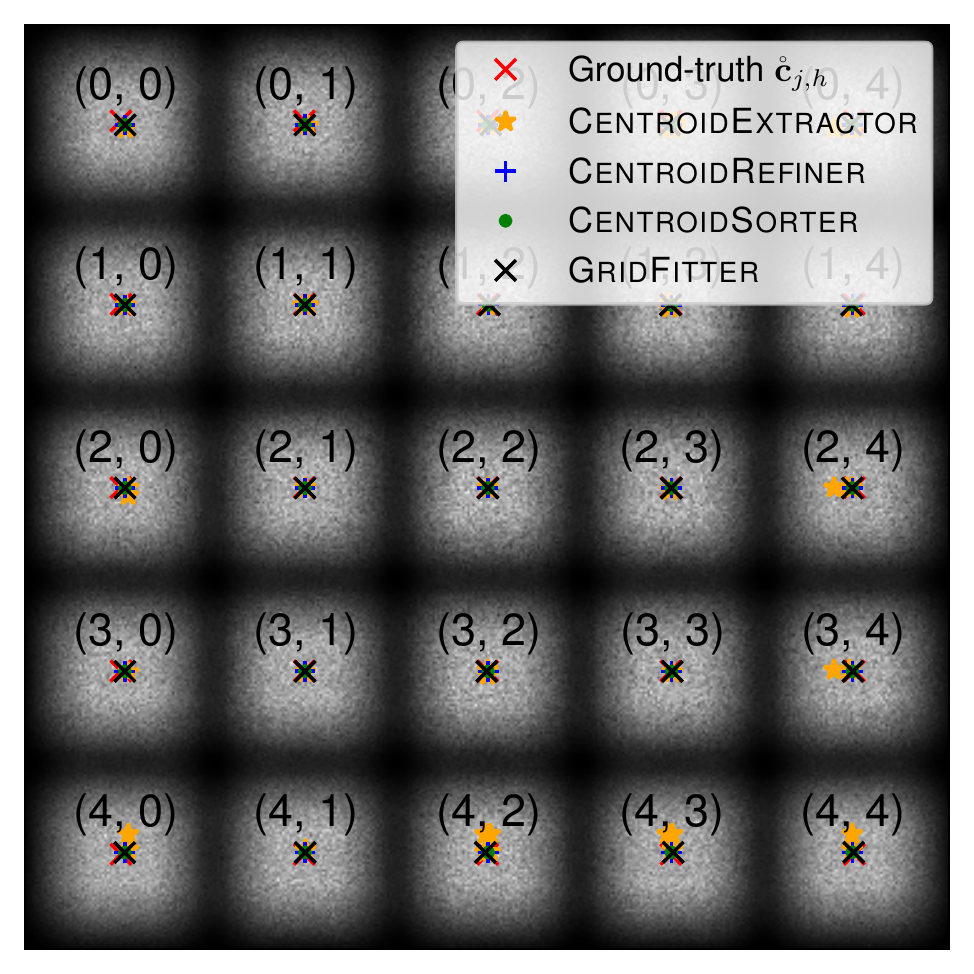}}
		\subcaption{\footnotesize$M=141$, $J=5$,
			$\mathbf{\Theta}=[0\si{\degree}, 0\si{\degree}
			]$
			\label{fig:calib_res:a}}
	\end{minipage}
	\begin{minipage}[t]{.49\linewidth}
		\resizebox{\textwidth}{!}{
			\includegraphics{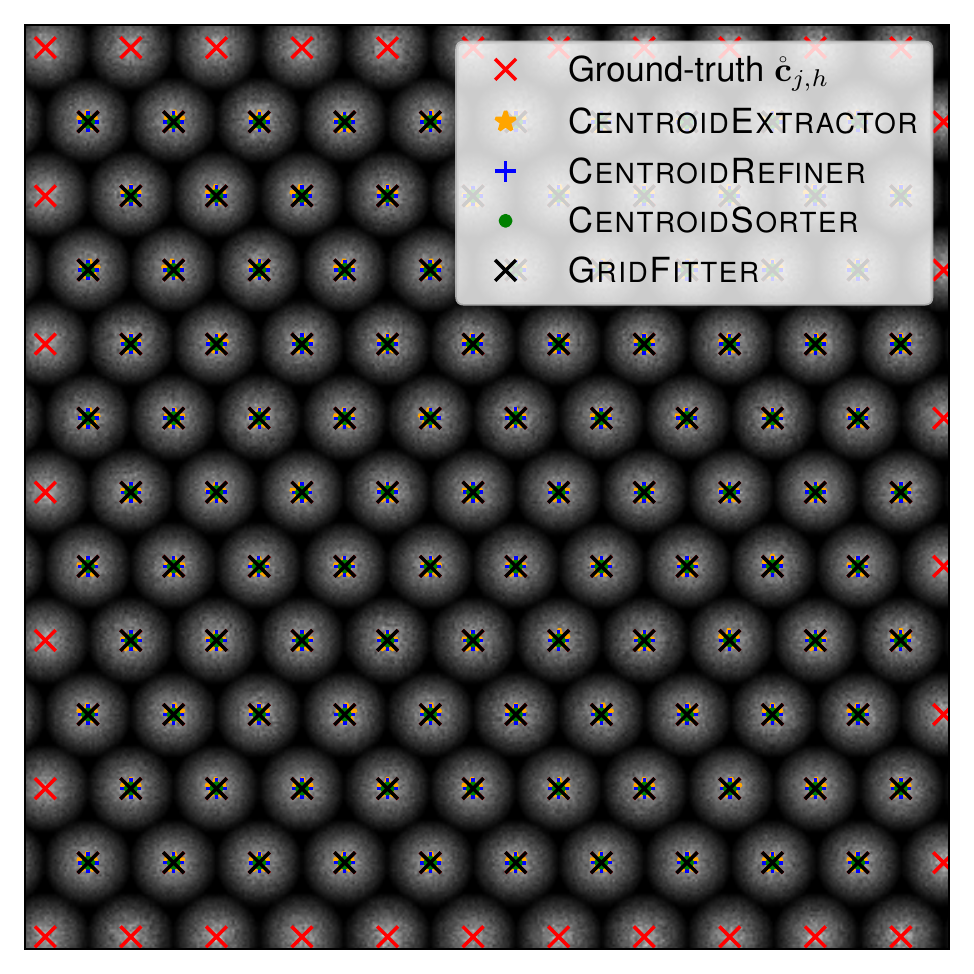}}
		\subcaption{\footnotesize$M=52$, $J=13$, $\mathbf{\Theta}=[0\si{\degree}, 0\si{\degree}
			]$\label{fig:calib_res:b}}
	\end{minipage}
	\\
	\vspace{2pt}
	\begin{minipage}[t]{.49\linewidth}
		\resizebox{\textwidth}{!}{
			\includegraphics{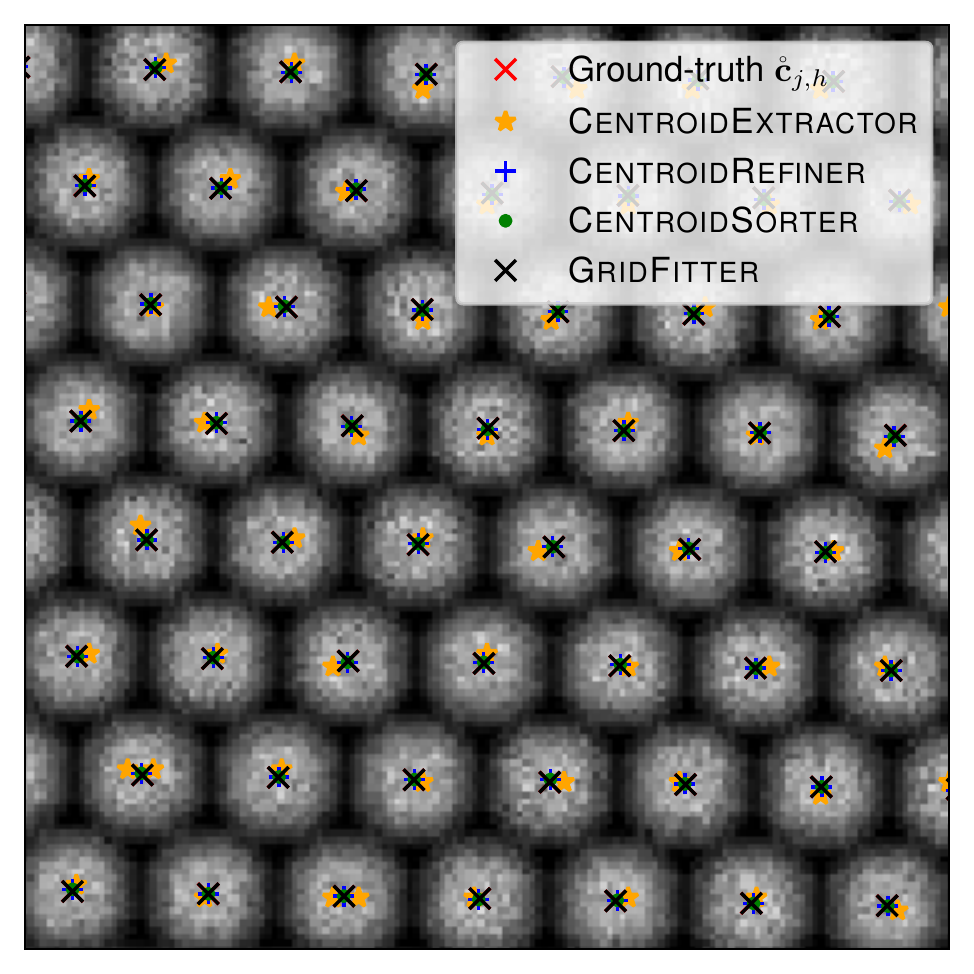}}
		\subcaption{\footnotesize$M=18$, $J=40$, $\mathbf{\Theta}=[-1\si{\degree}, 0\si{\degree}]$\label{fig:calib_res:c}}
	\end{minipage}
	\begin{minipage}[t]{.49\linewidth}
		\resizebox{\textwidth}{!}{
			\includegraphics{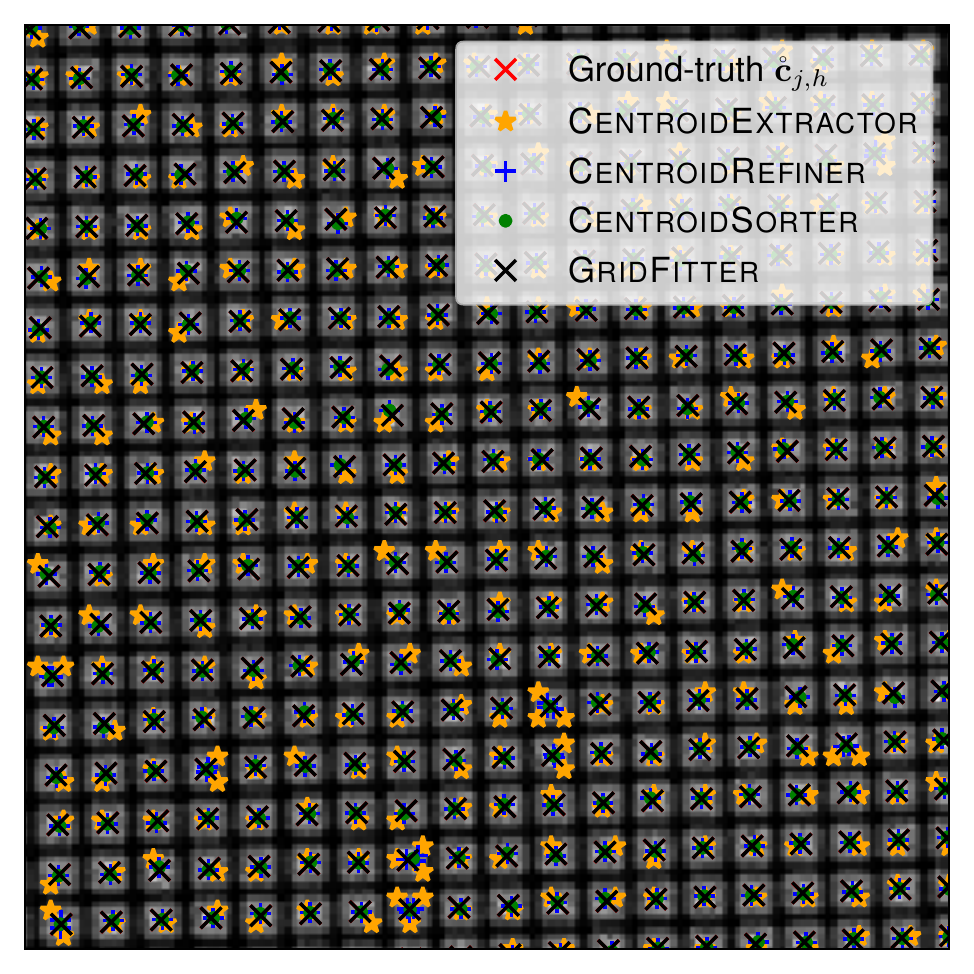}}
		\subcaption{\footnotesize$M=6$, $J=90$, $\mathbf{\Theta}=[2\si{\degree}, 1\si{\degree}
			]$\label{fig:calib_res:d}}
	\end{minipage}
	\caption{\textbf{Auto-calibration results} from synthetic data showing noisy micro images varying in size, tilts $\mathbf{\Theta}=[\theta_z, \theta_x]$ about the $z$-$x$-axes and packing $\mathcal{P}$, which is rectangular in \subref{fig:calib_res:a}; hexagonal in \subref{fig:calib_res:b}; tilted hexagonal in \subref{fig:calib_res:c}; and tilted rectangular in \subref{fig:calib_res:d}. \label{fig:calib_res}}
\end{figure}

\begin{figure}[H]
	\centering
	\includegraphics[width=.905\linewidth]{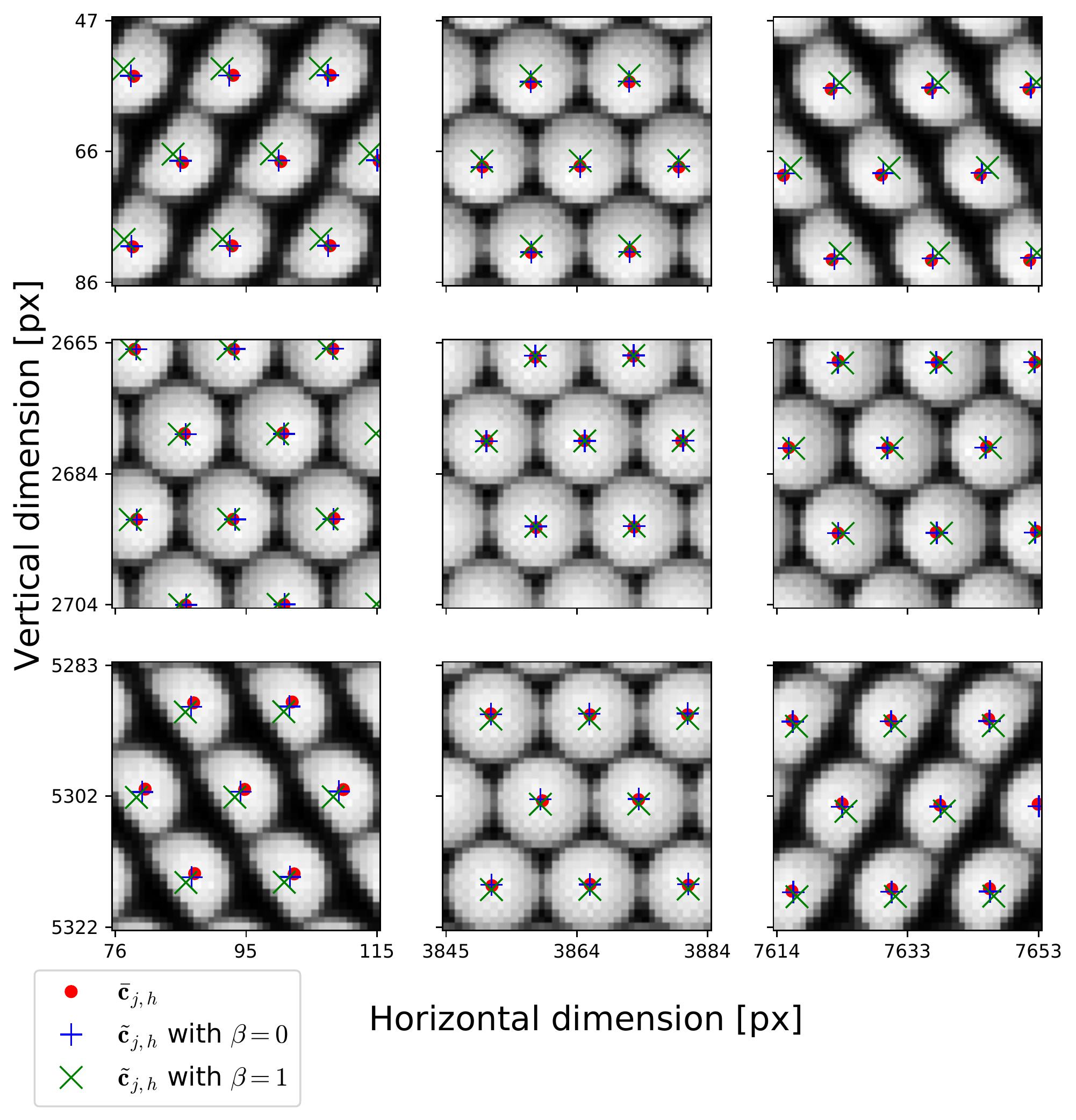}
	\caption[Lytro calibration]{\textbf{Lytro calibration} with centroids from Eq.~(\ref{eq:area}) (blue), LM regression (green) and regularization where $\beta=1$ (red).}
	\label{fig:regress}
\end{figure}

Our proposed generic calibration is tested in different scenarios with synthesized micro images varying in size, number, orientation angles, geometric packing and pixel noise. The validation is carried out through respective ground-truth references. Results are depicted in Fig.~\ref{fig:calib_res} for visual inspection.
To analyze the \Package{PitchEstimator}, scale space maxima along $\nu$ are presented from the examples shown in Fig.~\ref{fig:calib_res}. 
\begin{figure}[H]
	\centering
	\resizebox{\columnwidth}{!}{
		\includegraphics{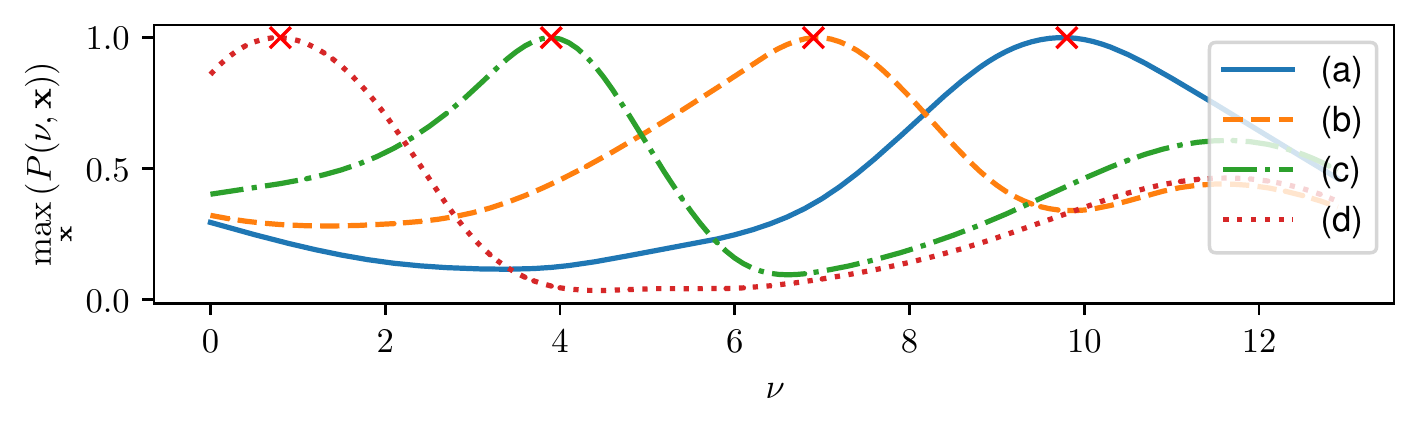}}
	\caption{\textbf{Scale maxima analysis} from \Package{PitchEstimator} with pyramid data from Fig.~\ref{fig:calib_res} where crosses signify respective $\nu^\star$
		\label{fig:scale_max_plot}}
\end{figure}
For a quantitative examination of the centroid accuracy, we use $C$ as a deviation metric in pixel unit given by
\begin{align}
C = \frac{1}{JH} \sum_{j=1}^{J} \sum_{h=1}^{H} \left\| \mathring{\mathbf{c}}_{j,h} - \mathbf{c}_{j,h} \right\|_2	\label{eq:calib_metric}
\end{align}
where $\mathbf{c}_{j,h}$ represents the output of each centroid detection method and $\mathring{\mathbf{c}}_{j,h}$ acts as the ground-truth. Numerical results from Eq.~(\ref{eq:calib_metric}) using examples in Fig.~\ref{fig:calib_res} are provided in Table~\ref{tab:calib_metric}. %
Figure~\ref{fig:calib_res:a} suggests that approximates from the \Package{CentroidExtractor} appear significantly off with regards to the ground-truth while subsequent refinement stages enhance the accuracy. 
Closer inspection reveals that the \Package{CentroidExtractor} may yield several centroids for a micro image suffering from noise. This is a side effect of NMS which would have been of greater concern if white images were not convolved with $\nabla^2G(\sigma^\star, \mathbf{x})$ prior to NMS, as this proves to cancel out false maxima.
Candidates passing through \Package{LfpExtractor} thus tend to be close to the ground-truth. %
\begin{table}[H]
	\small
	\centering
	\caption{Centroid deviation $C$ in pixel unit}
	\label{tab:calib_metric}
	\renewcommand{\arraystretch}{1}
	\resizebox{\linewidth}{!}{
		\begin{tabular}{
				|c
				|c
				|c
				|c
				|c
				|c
				|c
				|c
				|	
			}
			\hline
			\multirow{3}{*}{
				\begin{minipage}{.2in}
					\centering
					Fig.\\ \ref{fig:calib_res}
				\end{minipage}}
				& \textit{Stage 1} & \multicolumn{2}{c|}{\textit{Stage 2}} & \textit{Stage 3} & \textit{Stage 4} & \multicolumn{2}{c|}{\textit{Benchmark comparison}} \\
			\cline{2-8}
			 & \multirow{2}{*}{
			 	\begin{minipage}{.4in}
			 		\centering
			 		\Package{Cen.}\\
			 		\Package{Extr}.
			 	\end{minipage}} 
		 		& \multicolumn{2}{c|}{
		 			\begin{minipage}{.8in}
		 				\centering
		 				\Package{Cen.Refiner}
	 				\end{minipage}}
		 		& 			 	
		 		\multirow{2}{*}{
		 		\begin{minipage}{.4in}
				 	\centering
				 	\Package{Cen.}\\
				 	\Package{Sorter}
		 		\end{minipage}}
	 			& \multirow{2}{*}{
	 			\begin{minipage}{.4in}
	 				\centering
	 				\Package{Grid}\\
	 				\Package{Fitter}
	 			\end{minipage}}
 				&
 				\multirow{2}{*}{
 				\begin{minipage}{.4in}
 					\centering
 					\cite{DANSCAL}
 				\end{minipage}}
 				& \multirow{2}{*}{
 				\begin{minipage}{.4in}
 					\centering
 					\cite{SchambachPuenteLeon}
 				\end{minipage}}
 				\\
			\cline{3-4}
			 & & \begin{minipage}{.4in}\centering\textit{peak}\end{minipage} & \begin{minipage}{.4in}\centering\textit{area}\end{minipage} & & & & \\
			\hline
			\subref{fig:calib_res:a} & 6.171 & 2.984 & 1.941 & 1.849 & \textbf{1.845} & n.a. & n.a. \\
			\subref{fig:calib_res:b} & 0.854 & 0.237 & 0.124 & 0.124 & \textbf{0.027} & 0.381 & n.a. \\
			\subref{fig:calib_res:c} & 1.741 & 0.667 & 0.167 & 0.166 & \textbf{0.010} & 0.261 & 1.121\\
			\subref{fig:calib_res:d} & 1.162 & 0.419 & 0.311 & 0.297 & \textbf{0.007} & n.a. & n.a. \\
			\hline	
		\end{tabular}
	}
\end{table}
From Table~\ref{tab:calib_metric}, it follows that the \Package{CentroidRefiner} uses Eq.~(\ref{eq:area}) as opposed to Eq.~(\ref{eq:peak}), with the latter propagating more imprecise coordinates to subsequent procedures. %
The \Package{CentroidSorter} yields better results, because it merges centroids belonging to the same micro image. %
Improvements of the \Package{GridFitter} arise from a large number of data points in the global centroid regression, which outperforms state-of-the-art methods~\cite{Dansereau:2015, SchambachPuenteLeon}. 
Due to the absence of vignetting in Fig.~\ref{fig:calib_res}, regularization was omitted, but is shown in Fig.~\ref{fig:regress}.
On closer inspection of Fig.~\ref{fig:regress}, one may note that mechanical vignetting causes off-center micro images to be of non-radially symmetric shape, shifting detected centers away from their actual physical counterparts. This phenomenon has been recognized and addressed by peers~\cite{pitts2013compensating, Liang:2016, mignarddebise, SchambachPuenteLeon}. 
To work against this displacement in a cost-efficient manner, we have introduced a regularization term $\beta R(\cdot)$ to the LM regression in Eq.~(\ref{eq:obj_fun}) for penalization of corrupted centroid peaks. Despite the promising \Package{GridFitter} results, its usage is left optional to allow for calibration of inconsistent lattice spacings. %
%
\subsection{Sub-aperture images}
\begin{figure*}[ht]
	\centering
	\resizebox{\textwidth}{!}{
		\includegraphics[width=1\linewidth]{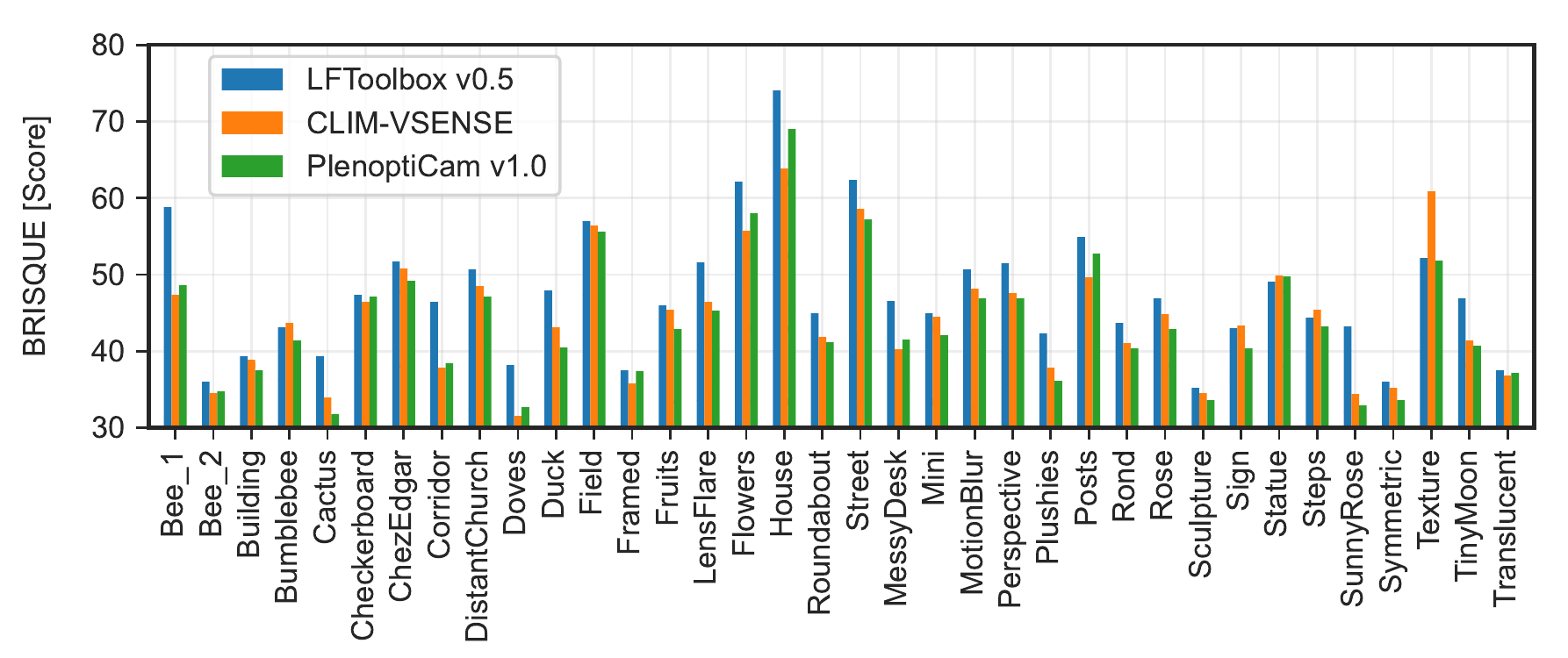}
	}
	\caption{\textbf{Analysis of central sub-aperture images} from~\cite{illum:dataset} along the horizontal axis and BRISQUE metric scores along the vertical axis where lower values signify higher quality. Our tool chain outperforms others in 24 out of 36 images in total.}
	\label{fig:brisque}
\end{figure*}
\begin{figure}[H]
	\centering
	\begin{minipage}[t]{.321\linewidth}
		\begin{minipage}[t]{\linewidth}
			\centering
			\Package{LFToolbox\\ v0.5}~\cite{DANSCAL}
			\vspace{.5em}
		\end{minipage}
		\vspace{2pt}
		\begin{minipage}[t]{\linewidth}
			\resizebox{\textwidth}{!}{
				\includegraphics{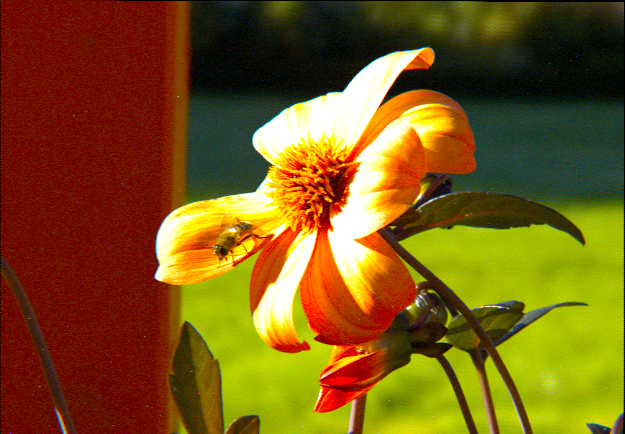}
			}
			\subcaption{\label{fig:compare:sub:a}}
		\end{minipage}
		\vspace{2pt}
		\begin{minipage}[t]{\linewidth}
			\resizebox{\textwidth}{!}{
				\includegraphics{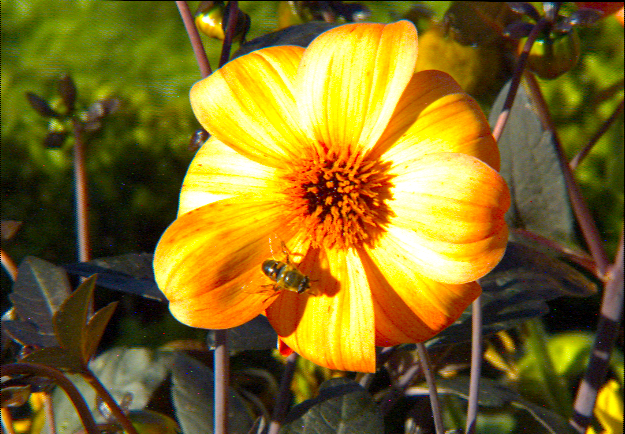}
			}
			\subcaption{\label{fig:compare:sub:b}}
		\end{minipage}
		\vspace{2pt}
		\begin{minipage}[t]{\linewidth}
			\resizebox{\textwidth}{!}{
				\includegraphics{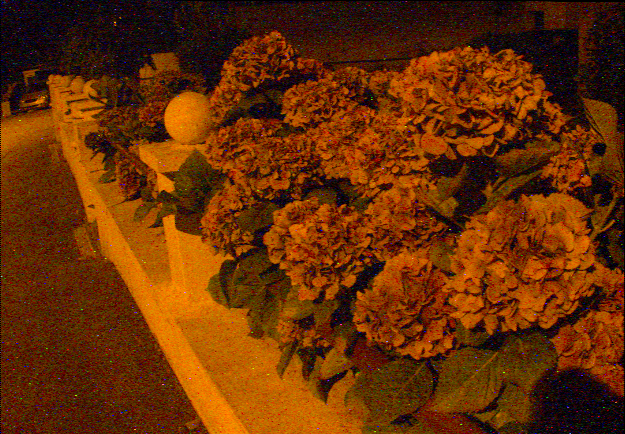}
			}
			\subcaption{\label{fig:compare:sub:c}}
		\end{minipage}
		\vspace{2pt}
		\begin{minipage}[t]{\linewidth}
			\resizebox{\textwidth}{!}{
				\includegraphics{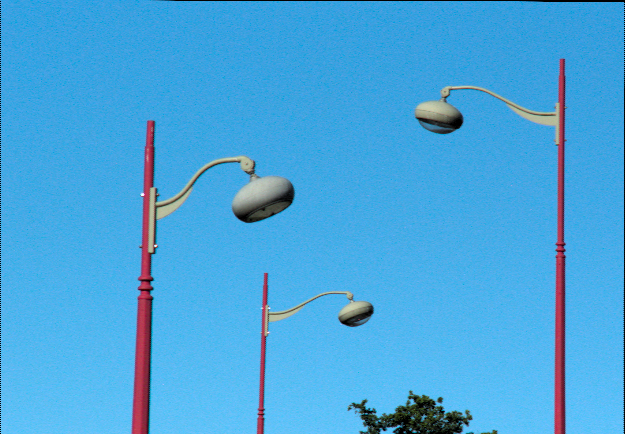}
			}
			\subcaption{\label{fig:compare:sub:d}}
		\end{minipage}
	\end{minipage}
	\hfill
	\begin{minipage}[t]{.321\linewidth}
		\begin{minipage}[t]{\linewidth}
			\centering
			\Package{CLIM-\\VSENSE}~\cite{Matysiak:2018}
			\vspace{.5em}
		\end{minipage}
		\vspace{2pt}
		\begin{minipage}[t]{\linewidth}
			\resizebox{\textwidth}{!}{
				\includegraphics{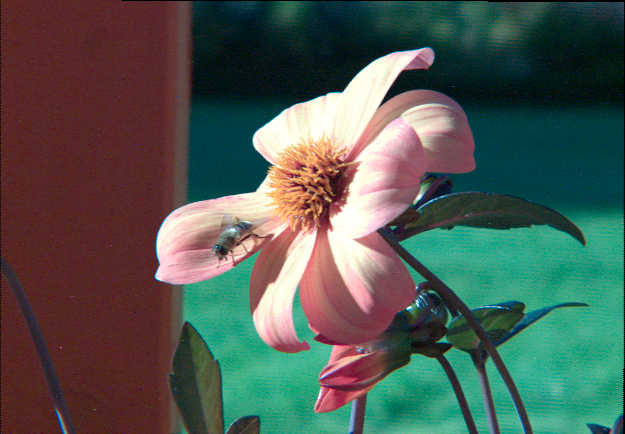}
			}
			\subcaption{\label{fig:compare:sub:e}}
		\end{minipage}
		\vspace{2pt}
		\begin{minipage}[t]{\linewidth}
			\resizebox{\textwidth}{!}{
				\includegraphics{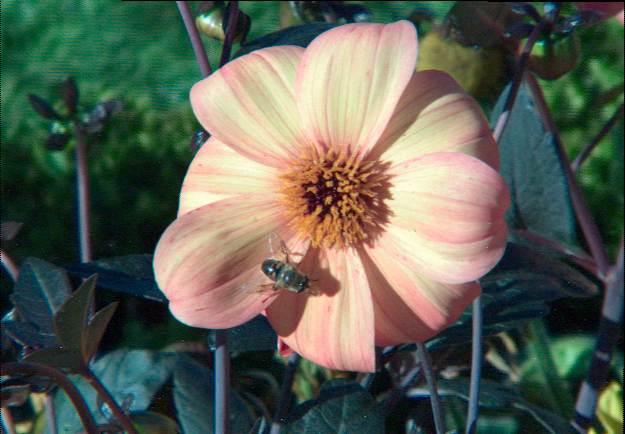}
			}
			\subcaption{\label{fig:compare:sub:f}}
		\end{minipage}
		\vspace{2pt}
		\begin{minipage}[t]{\linewidth}
			\resizebox{\textwidth}{!}{
				\includegraphics{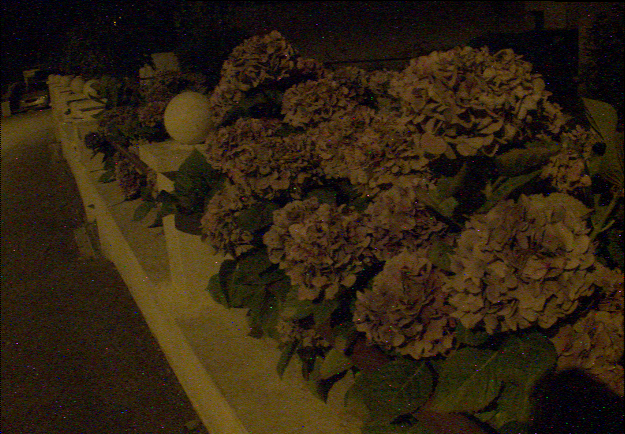}
			}
			\subcaption{\label{fig:compare:sub:g}}
		\end{minipage}
		\vspace{2pt}
		\begin{minipage}[t]{\linewidth}
			\resizebox{\textwidth}{!}{
				\includegraphics{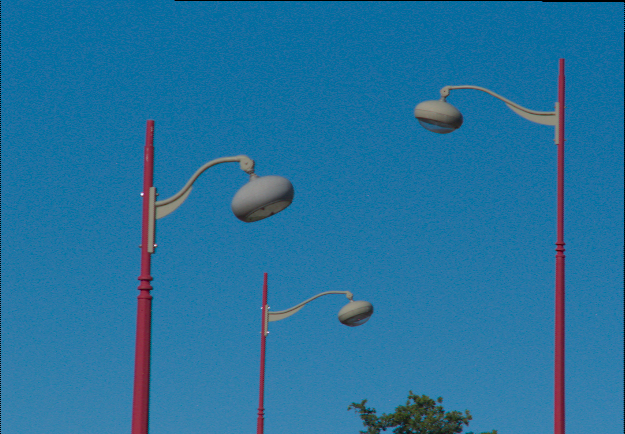}
			}
			\subcaption{\label{fig:compare:sub:h}}
		\end{minipage}
	\end{minipage}
	\hfill
	\begin{minipage}[t]{.321\linewidth}
		\begin{minipage}[t]{\linewidth}
			\centering
			\Package{PlenoptiCam\\ v1.0}
			\vspace{.64em}
		\end{minipage}
		\vspace{2pt}
		\begin{minipage}[t]{\linewidth}
			\resizebox{\textwidth}{!}{
				\includegraphics{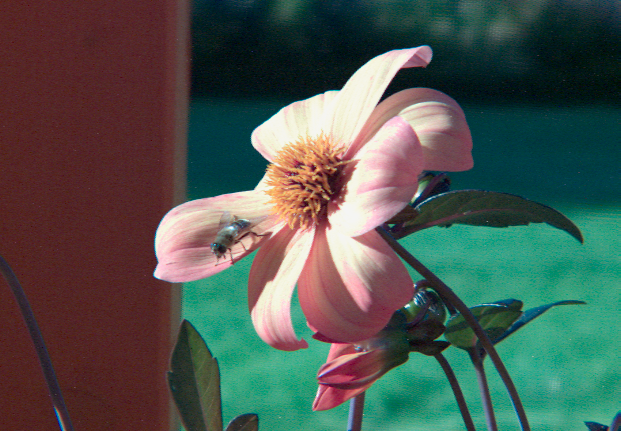}
			}
			\subcaption{\label{fig:compare:sub:i}}
		\end{minipage}
		\vspace{2.1pt}
		\begin{minipage}[t]{\linewidth}
			\resizebox{\textwidth}{!}{
				\includegraphics{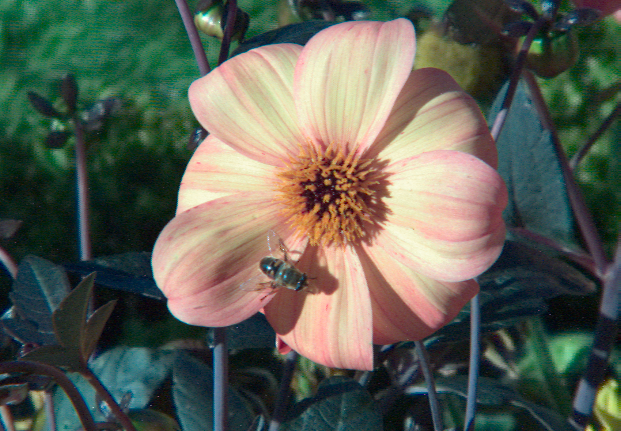}
			}
			\subcaption{\label{fig:compare:sub:j}}
		\end{minipage}
		\vspace{2.1pt}
		\begin{minipage}[t]{\linewidth}
			\resizebox{\textwidth}{!}{
				\includegraphics{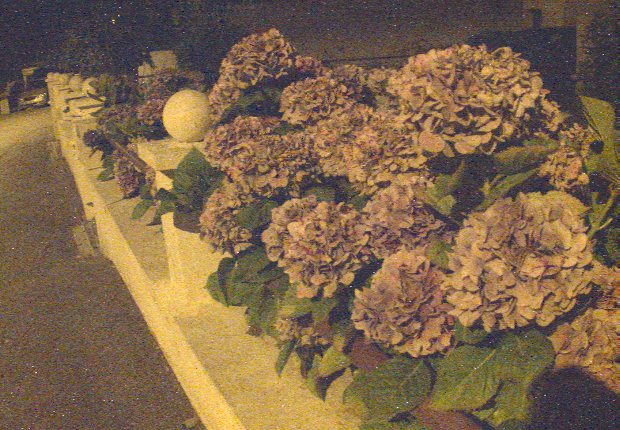}
			}
			\subcaption{\label{fig:compare:sub:k}}
		\end{minipage}
		\vspace{2.1pt}
		\begin{minipage}[t]{\linewidth}
			\resizebox{\textwidth}{!}{
				\includegraphics{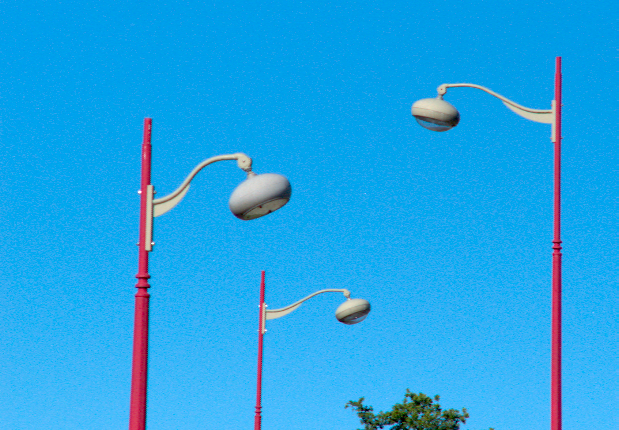}
			}
			\subcaption{\label{fig:compare:sub:l}}
		\end{minipage}
	\end{minipage}
	\caption{\textbf{Central sub-aperture images} from the IRISA dataset~\cite{illum:dataset} rendered in \subref{fig:compare:sub:a} to \subref{fig:compare:sub:d} by \Package{LFToolbox~v0.5}~\cite{DANSCAL}, in \subref{fig:compare:sub:e} to \subref{fig:compare:sub:h} by \Package{CLIM-VSENSE}~\cite{Matysiak:2018} and ours in \subref{fig:compare:sub:i} to \subref{fig:compare:sub:l}. \label{fig:compare}}
\end{figure}

The central view of a light-field generally suffers the least of aberrations and thus, exposes best image quality. Figure~\ref{fig:compare} depicts decoded central views of light-field photographs from the available IRISA dataset~\cite{illum:dataset} rendered by state-of-the-art plenoptic imaging pipelines for comparison. 
Closer inspection reveals that central views from \Package{LFToolbox~v0.5} appear brighter, however, fail to preserve bright image details.
%
For a quantitative assessment of the different pipelines, we seek a metric serving as an objective measure. Since central sub-aperture images lack ground-truth references, we employ a widely accepted no-reference-based technique named {\it Blind Reference-less Image Spatial Quality Evaluator} (BRISQUE)~\cite{Mittal2012NoReferenceIQ} from the {\it pybrisque} implementation~\cite{gh:pybrisque}. Scores from the BRISQUE metric are presented in Fig.~\ref{fig:brisque}. Light-field denoising as borrowed by~\cite{Matysiak:2018} was left out in the evaluation as it is considered equally effective for each pipeline and thus, not crucial for the decomposition. For our benchmark comparison, all pipelines are set to render with de-vignetting, color and sRGB options. \par
%
\begin{figure}[H]
	\begin{minipage}[t]{\linewidth}
		\begin{minipage}{.49\linewidth}
			\centering
			\Package{Lytro~Power Tools v1.0}
			\vspace{8pt}
		\end{minipage}
		\hfill
		\begin{minipage}{.49\linewidth}
			\centering
			\Package{PlenoptiCam v1.0}
			\vspace{8pt}
		\end{minipage}
	\end{minipage}
	\begin{minipage}[t]{\linewidth}
		\begin{minipage}{.49\linewidth}
			\begin{overpic}[width=\textwidth]{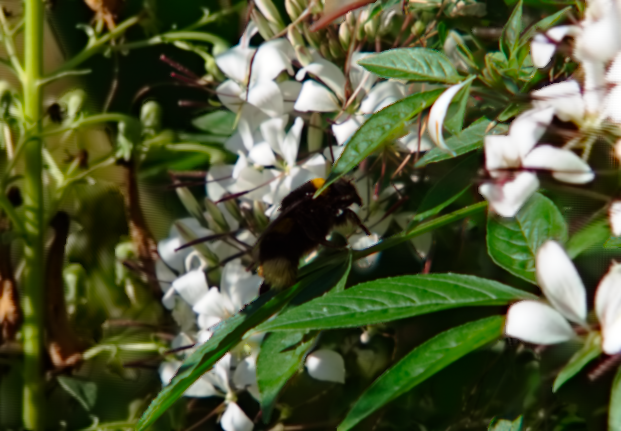}
				\put(-2.9,0){
					\begin{tikzpicture}[
					green!40!black,
					colored/.style={fill=#1!20,draw=#1!50!black!50}
					]
					\node[anchor=south west,inner sep=0] (image) at (0,0) {
						\includegraphics[clip, trim=340 178 218 213, width=.4\textwidth]{img/sai_comp/lytro_power_tools_621x431/Bumblebee.png}
					};
					\begin{scope}[x={(image.south east)},y={(image.north west)}]
					\draw[red,ultra thick] (0,0) rectangle (1,1);	
					\end{scope}
					\begin{scope}
					\path (1.75,0) coordinate (p1) {};
					\path (0,1.13) coordinate (p2) {};
					\path (2.8,1.2) coordinate (p3) {};
					\path (2.4,1.5) coordinate (p4) {};
					\draw[dotted, red, line width=.4mm](p1.east)--(p3.east)(p2.west)--(p4.west);
					\end{scope}
					\end{tikzpicture}
				}
				\put(51,27){
					\begin{tikzpicture}
					\begin{scope}
					\draw[red,ultra thick] (0,0) rectangle (.43225,.3);	
					\end{scope}
					\end{tikzpicture}
				}
			\end{overpic}
			\subcaption{\label{fig:lpt_comp:sub:a}}
		\end{minipage}
		\hfill
		\begin{minipage}{.49\linewidth}
			\begin{overpic}[width=\textwidth]{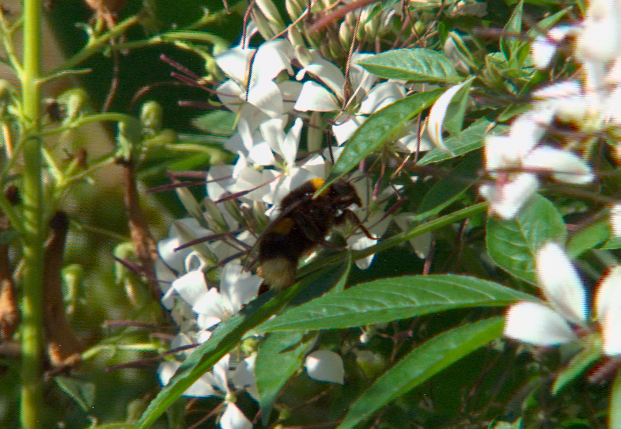}
				\put(-2.9,0){
					\begin{tikzpicture}
					\node[anchor=south west,inner sep=0] (image) at (0,0) {
						\includegraphics[clip, trim=340 175 220 215, width=.4\textwidth]{img/sai_comp/thumb_collection/Bumblebee.png}
					};
					\begin{scope}[x={(image.south east)},y={(image.north west)}]
					\draw[red,ultra thick] (0,0) rectangle (1,1);	
					\end{scope}
					\begin{scope}
					\path (1.75,0) coordinate (p1) {};
					\path (0,1.13) coordinate (p2) {};
					\path (2.8,1.2) coordinate (p3) {};
					\path (2.4,1.5) coordinate (p4) {};
					\draw[dotted, red, line width=.4mm](p1.east)--(p3.east)(p2.west)--(p4.west);
					\end{scope}
					\end{tikzpicture}
				}
				\put(51,27){
					\begin{tikzpicture}
					\begin{scope}
					\draw[red,ultra thick] (0,0) rectangle (.43225,.3);	
					\end{scope}
					\end{tikzpicture}
				}
			\end{overpic}
			\subcaption{\label{fig:lpt_comp:sub:b}}
		\end{minipage}
		\vspace{5pt}
	\end{minipage}
	\begin{minipage}[t]{\linewidth}
		\begin{minipage}{.49\linewidth}
			\begin{overpic}[width=\textwidth]{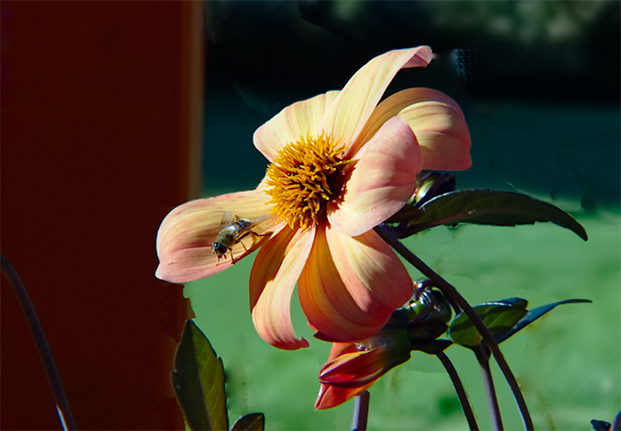}
				\put(55.6,0){					
					\begin{tikzpicture}
					\node[anchor=south west,inner sep=0] (image) at (0,0) {
						\includegraphics[clip, trim=150 81.8 400 260, width=.4\textwidth]{img/sai_comp/lpt_v=0.4_marginal/Bee_1_621x431.png}
					};
					\begin{scope}[x={(image.south east)},y={(image.north west)}]
					\draw[red,ultra thick] (0,0) rectangle (1,1);	
					\end{scope}
					\end{tikzpicture}
				}
				\put(20.7,0){
					\begin{tikzpicture}
					\begin{scope}
					\path (1.7,0.57) coordinate (p1) {};
					\path (1.7,1.17) coordinate (p2) {};
					\path (3.3,0) coordinate (p3) {};
					\path (3.3,2.2) coordinate (p4) {};
					\draw[dotted, red, line width=.4mm](p1.east)--(p3.east)(p2.west)--(p4.west);
					\end{scope}
					\end{tikzpicture}
				}
				\put(21,13){
					\begin{tikzpicture}
					\begin{scope}
					\draw[red,ultra thick] (0,0) rectangle (.47,.62);
					\end{scope}
					\end{tikzpicture}
				}
			\end{overpic}
			\subcaption{\label{fig:lpt_comp:sub:c}}
		\end{minipage}
		\hfill
		\begin{minipage}{.49\linewidth}
			\begin{overpic}[width=\textwidth]{img/sai_comp/thumb_collection/Bee_1.png}
				\put(55.6,0){					
					\begin{tikzpicture}
					\node[anchor=south west,inner sep=0] (image) at (0,0) {
						\includegraphics[clip, trim=150 81.8 400 260, width=.4\textwidth]{img/sai_comp/thumb_collection/Bee_1.png}
					};
					\begin{scope}[x={(image.south east)},y={(image.north west)}]
					\draw[red,ultra thick] (0,0) rectangle (1,1);	
					\end{scope}
					\end{tikzpicture}
				}
				\put(20.7,0){
					\begin{tikzpicture}
					\begin{scope}
					\path (1.7,0.57) coordinate (p1) {};
					\path (1.7,1.17) coordinate (p2) {};
					\path (3.3,0) coordinate (p3) {};
					\path (3.3,2.2) coordinate (p4) {};
					\draw[dotted, red, line width=.4mm](p1.east)--(p3.east)(p2.west)--(p4.west);
					\end{scope}
					\end{tikzpicture}
				}
				\put(21,13){
					\begin{tikzpicture}
					\begin{scope}
					\draw[red,ultra thick] (0,0) rectangle (.47,.62);
					\end{scope}
					\end{tikzpicture}
				}
			\end{overpic}
			\subcaption{\label{fig:lpt_comp:sub:d}}
		\end{minipage}
	\end{minipage}
	\caption{\textbf{View comparison with Lytro engine} revealing differences between classical sub-aperture images by our pipeline in \subref{fig:lpt_comp:sub:b}, \subref{fig:lpt_comp:sub:d} and Lytro's results in \subref{fig:lpt_comp:sub:a}, \subref{fig:lpt_comp:sub:c}, a consequence of all-in-focus rendering from depth-based segmentation~\cite{focal_stack_fusion, r4}.\label{fig:lpt_comp}}
\end{figure}
%
The BRISQUE metric is known to be sensitive to noise and blur by local statistical analysis so that an information loss caused by image processing modules would deteriorate the score. Low scores of our pipeline in Fig.~\ref{fig:brisque} justify that on average our decomposition retains the physical information optimally. Our central sub-aperture images outperform the other two pipelines in 24 out of 36 cases of the dataset samples~\cite{illum:dataset}. We achieve a total score of 1567 whereas \Package{CLIM-VSENSE} yields 1595 and the \Package{LFToolbox~v0.5} gets 1702 scores. 
%
The reason for our strong results relies on our proposed extensions to existing pipelines. This includes the treatment of detected hot-pixels prior to Bayer demosaicing to prevent outliers from propagating to adjacent pixels. We further take advantage of Menon's method~\cite{Menon2007c} for debayering. In addition, the \Package{LfpResampler} contributes to lower scores by conducting micro image alignment in an element-wise manner instead of re-mapping the entire plenoptic image as a whole~\cite{Dansereau:2015}. Another reason for the improvements is that \Package{PlenoptiCam} benefits from insights previously made available by peers such as the de-saturation approach~\cite{Matysiak:2018}. Apart from that, our automatic dynamic range alignment based on percentiles of different color spaces tends to be a more generic solution for images of various exposure, which becomes apparent by comparing \subref{fig:compare:sub:g} with \subref{fig:compare:sub:k} in Fig.~\ref{fig:compare}. \par
For comparisons with the Lytro engine, results were computed with \Package{Lytro Desktop v3.4.4} and \Package{Lytro Power Tools v1.0}. Thereby, metrics indicate that the latter yields brighter images with larger field of view and allows for numerical parametrization (e.g., $\lambda$ for focus), making experimental reproduction easier. All Lytro results presented hereafter were therefore generated using \Package{Lytro Power Tools v1.0}.
Figure~\ref{fig:lpt_comp} provides a comparison of light-fields rendered by \Package{Lytro Power Tools v1.0} and \Package{PlenoptiCam v1.0}, respectively. On closer examination of the Lytro results in Figs. \ref{fig:lpt_comp:sub:a} and \ref{fig:lpt_comp:sub:c}, occasional displacements emerge with image regions being removed or unnaturally shifted from their surroundings. %
Since Lytro's code has not been disclosed, the cause for this appearance can only be speculated. A study carried out by scientists formerly affiliated with Lytro suggests that a previously computed depth map is used to fuse a stack of high-resolution refocused frames to an all-in-focus image~\cite{focal_stack_fusion, r4}, which in turn may be a starting point for high-resolution angular view synthesis. 
While this yields higher resolutions, unknown or erroneous disparity is propagated to subsequent stages, i.e. super-resolved views. 
Image region displacements in Figs.~\ref{fig:lpt_comp:sub:a} and \ref{fig:lpt_comp:sub:c} thus might be due to false depth map disparities. Lytro's spatial resolution is numerically higher than the number of present micro lenses, further supporting the assumption of such pipeline design.
%
These observations bring us to the conclusion that the Lytro engine omits direct access to classical sub-aperture images unlike other open-source toolchains~\cite{DANSCAL, Matysiak:2018} including ours. 


Technically, the user can generate sub-aperture images from Lytro's aligned ESLF images by completing their processing manually, including pixel rearrangements, adapting angular view positions and accounting for gamma as well as color correction. Given that those ESLF images are at a different processing stage and require additional resources, %
we compare our sub-aperture images with the scientific toolbox~\cite{DANSCAL} and its extensions~\cite{Matysiak:2018}. Nonetheless, we then contrast our refocusing results against those of the Lytro engine in Section~\ref{sec:3:5}. \par
%
%
For further sub-aperture image analysis, we focus on the \Package{HexCorrector} results where artifacts caused by hexagonal sampling are subject to removal.
\begin{figure}[h]
	\centering
	\begin{minipage}[t]{.2\columnwidth}
		\resizebox{\textwidth}{!}{\includegraphics[trim={182 280 410 110},clip]
			{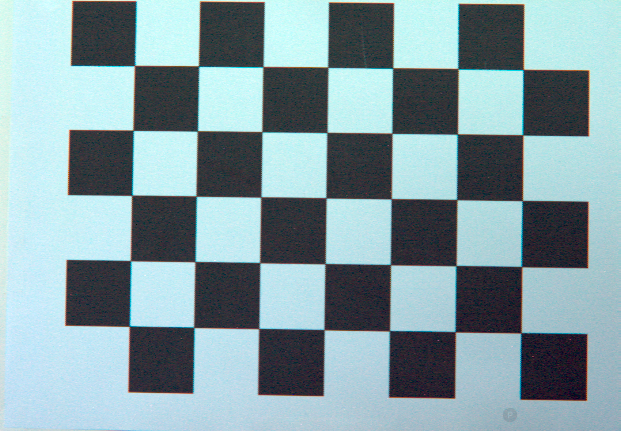}}
		\subcaption{Local
			\label{fig:hexart:sub:a}}
	\end{minipage}
	\begin{minipage}[b]{.042\columnwidth}
		\centering
		\raisebox{1.1cm}{\large\textrightarrow{}}
	\end{minipage}
	\begin{minipage}[t]{.2\columnwidth}
		\resizebox{\textwidth}{!}{\includegraphics[trim={182 280 410 110},clip]
			{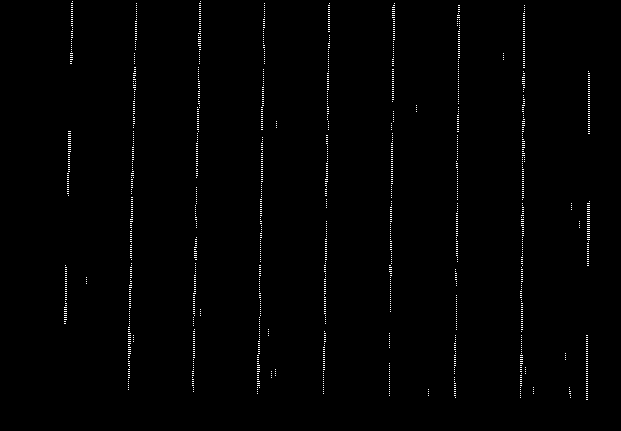}}
		\subcaption{Mask 
			\label{fig:hexart:sub:b}}
	\end{minipage}
	\begin{minipage}[b]{.042\columnwidth}
		\centering
		\raisebox{1.1cm}{\large\textrightarrow{}}
	\end{minipage}
	\begin{minipage}[t]{.2\columnwidth}
		\resizebox{\textwidth}{!}{\includegraphics[trim={182 280 410 110},clip]
			{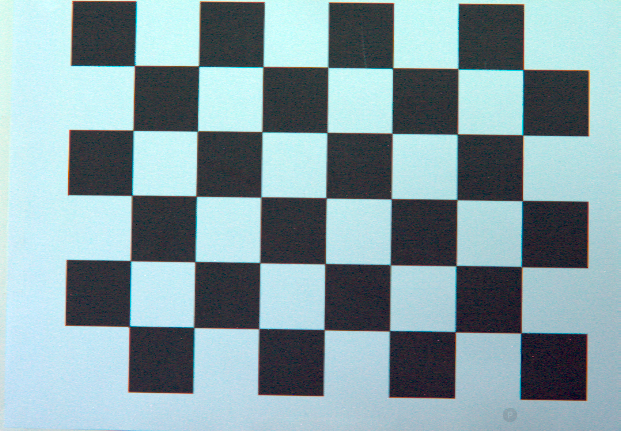}}
		\subcaption{Rectified
			\label{fig:hexart:sub:c}}
	\end{minipage}
	\begin{minipage}[b]{.042\columnwidth}
		\centering
		\hspace{1cm}
	\end{minipage}
	\begin{minipage}[t]{.2\columnwidth}
		\resizebox{\textwidth}{!}{\includegraphics[trim={183.5 281 408.5 111},clip]
			{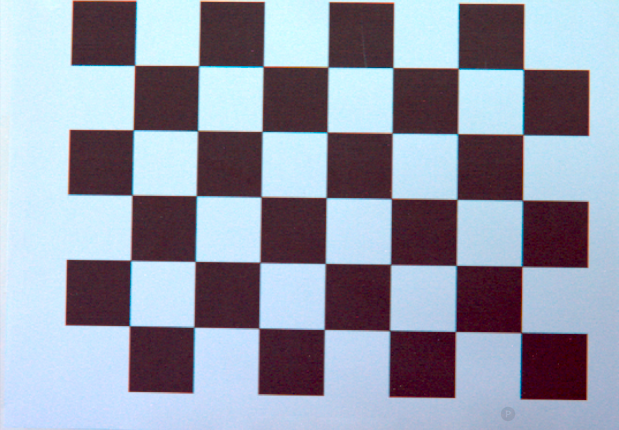}}
		\subcaption{Global
			\label{fig:hexart:sub:g}}
	\end{minipage}
	\\
	\vspace{2pt}
	\begin{minipage}[t]{.2\columnwidth}
		\resizebox{\textwidth}{!}{\includegraphics[trim={505 312 50 53},clip]
			{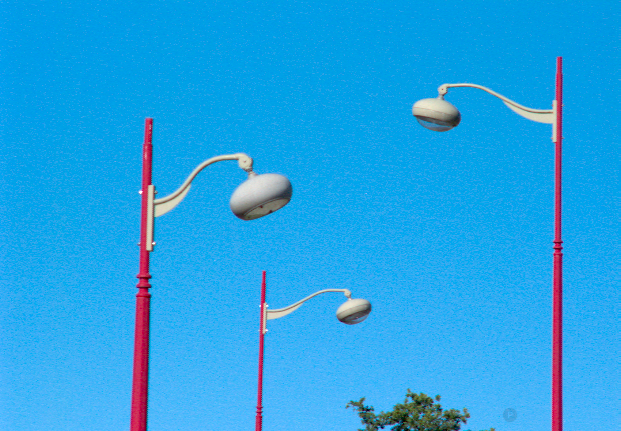}}
		\subcaption{Local
			\label{fig:hexart:sub:d}}
	\end{minipage}
	\begin{minipage}[b]{.042\columnwidth}
		\centering
		\raisebox{.75cm}{\large\textrightarrow{}}
	\end{minipage}
	\begin{minipage}[t]{.2\columnwidth}
		\resizebox{\textwidth}{!}{\includegraphics[trim={505 312 50 53},clip]
			{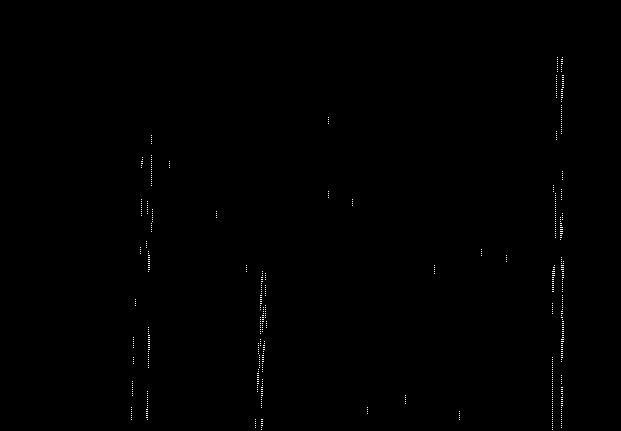}}
		\subcaption{Mask
			\label{fig:hexart:sub:e}}
	\end{minipage}
	\begin{minipage}[b]{.042\columnwidth}
		\centering
		\raisebox{.75cm}{\large\textrightarrow{}}
	\end{minipage}
	\begin{minipage}[t]{.2\columnwidth}
		\resizebox{\textwidth}{!}{\includegraphics[trim={505 312 50 53},clip]
			{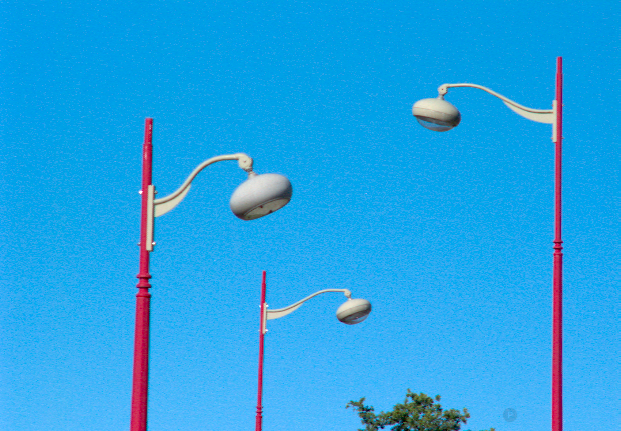}}
		\subcaption{Rectified
			\label{fig:hexart:sub:f}}
	\end{minipage}
	\begin{minipage}[b]{.042\columnwidth}
		\centering
		\hspace{1cm}
	\end{minipage}
	\begin{minipage}[t]{.2\columnwidth}
		\resizebox{\textwidth}{!}{\includegraphics[trim={505 311.5 48 52.5},clip]
			{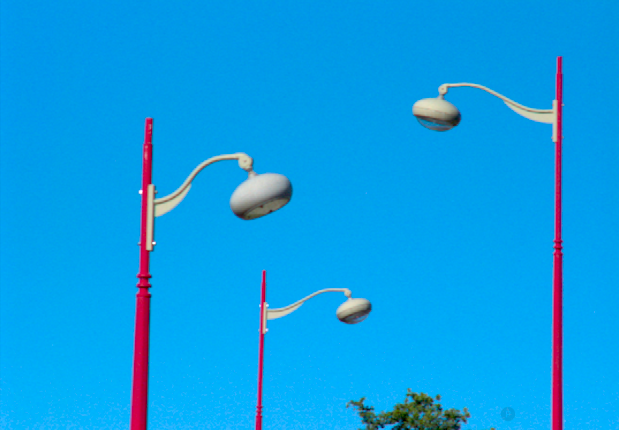}}
		\subcaption{Global
			\label{fig:hexart:sub:h}}
	\end{minipage}
	\caption{\textbf{Fringe artifact reduction results} showing sub-aperture images with local resampling on the left, which exhibit fringes along vertical lines. The binary masks $V_i[\check{s}_j]$ indicate pixels detected by \Package{HexCorrector} that undergo rectification. The right column depicts global resampling results.\label{fig:hexart}}
\end{figure}
Figure~\ref{fig:hexart} shows magnified portions of sub-aperture images with apparent fringe artifacts in Figs.~\ref{fig:hexart:sub:a} and \ref{fig:hexart:sub:d} as well as rectified counterparts exposing smooth edges in Figs.~\ref{fig:hexart:sub:c} and \ref{fig:hexart:sub:f}. Although the efficacy of our treatment is visually notable, this improvement does not present a huge impact on BRISQUE scores. This is likely due to the blind character of the metric and its unawareness of the hexagonal sampling artifact, which can be interpreted as reasonably high spatial frequencies. Another observation made in Fig.~\ref{fig:hexart} is that organic object structures intentionally receive no treatment as they lack straight edges and are thus less affected. \par
%
\begin{figure}[H]
	\centering
	\begin{minipage}[t]{.49\linewidth}
		\resizebox{\textwidth}{!}{\includegraphics{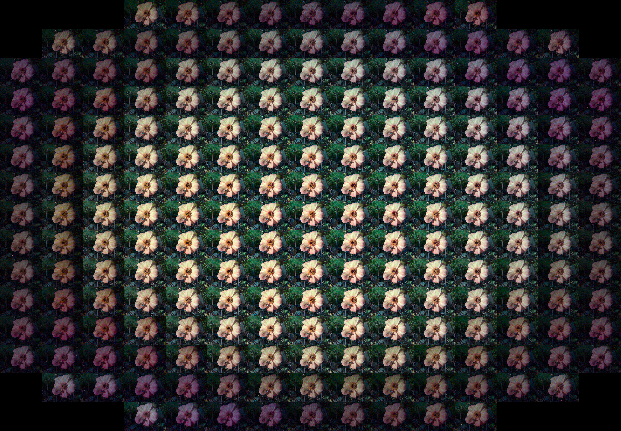}}
		\subcaption{Without treatment \label{fig:coco:sub:a}}
	\end{minipage}
	\begin{minipage}[t]{.49\linewidth}
		\resizebox{\textwidth}{!}{
			\includegraphics
			{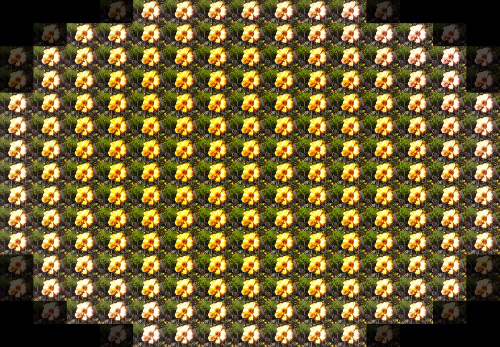}}
		\subcaption{\Package{LFToolbox~v0.5}~\cite{DANSCAL}\label{fig:coco:sub:b}}
	\end{minipage}
	\\
	\vspace{2pt}
	\begin{minipage}[t]{.49\linewidth}
		\resizebox{\textwidth}{!}{
			\includegraphics
			{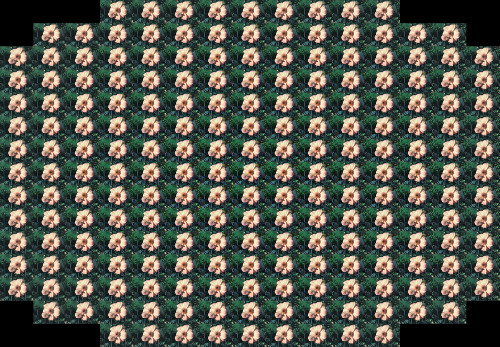}}
		\subcaption{Matysiak~\textit{et al.}~\cite{Matysiak:2018}\label{fig:coco:sub:c}}
	\end{minipage}
	\begin{minipage}[t]{.49\linewidth}
		\resizebox{\textwidth}{!}{\includegraphics{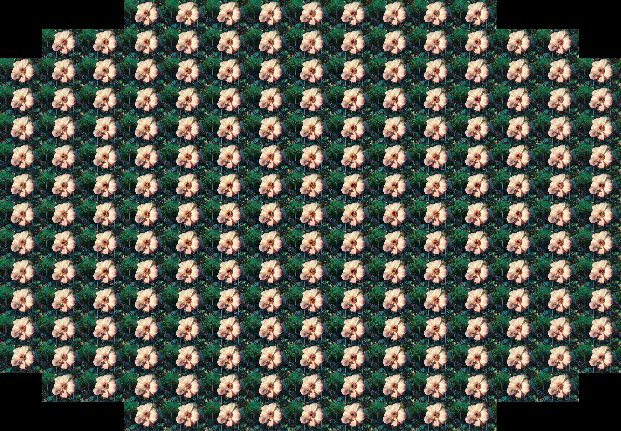}}
		\subcaption{\Package{PlenoptiCam v1.0} \label{fig:coco:sub:d}}
	\end{minipage}
	\caption{\textbf{Color equalization results} with $15 \times 15$ stitched sub-aperture images indicating illumination variances across a decomposed light-field. \subref{fig:coco:sub:a} our result without treatment, \subref{fig:coco:sub:b} Dansereau~\textit{et al.}~\cite{DANSCAL}, \subref{fig:coco:sub:c} Matysiak~\textit{et al.}~~\cite{Matysiak:2018} with 240 mins computation time and \subref{fig:coco:sub:d} our HM-MKL-HM compound with 81 secs computation time. \label{fig:coco}}
\end{figure}
\begin{figure}[H]
	\centering
	\begin{minipage}[t]{.32\linewidth}
		\resizebox{\textwidth}{!}{\includegraphics{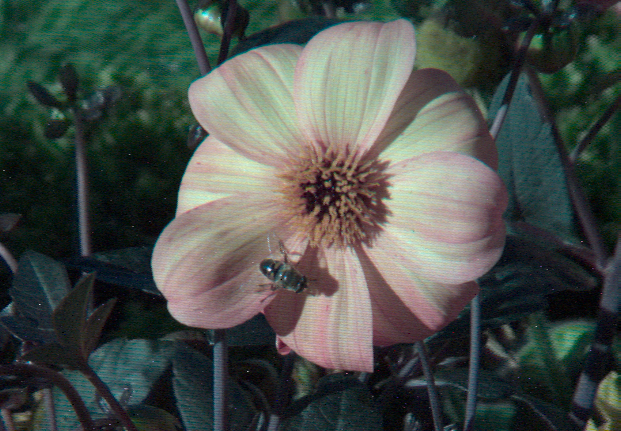}}
		\subcaption{\footnotesize Ours untreated\label{fig:clr_eq:sub:a}}
	\end{minipage}
	\hfill
	\begin{minipage}[t]{.32\linewidth}
		\resizebox{\textwidth}{!}{\includegraphics{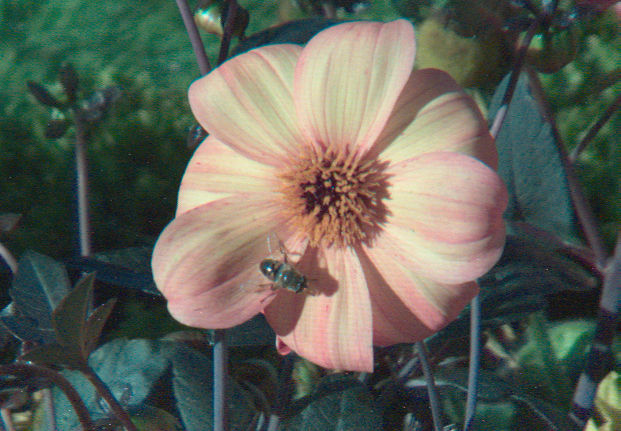}}
		\subcaption{\footnotesize \mbox{MKL as by \cite{Pitie:2007}}\label{fig:clr_eq:sub:b}
		}
	\end{minipage}
	\hfill
	\begin{minipage}[t]{.32\linewidth}
		\resizebox{\textwidth}{!}{\includegraphics{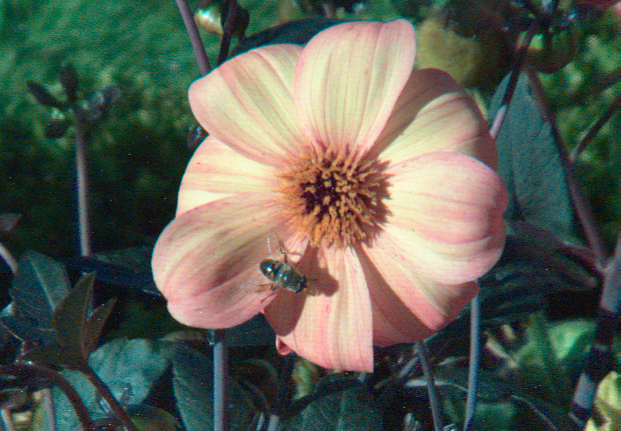}}
		\subcaption{\footnotesize Our \mbox{HM-MKL-HM}\label{fig:clr_eq:sub:c}}
	\end{minipage}
	\\
	\vspace{2pt}
	\begin{minipage}[t]{.32\linewidth}
		\resizebox{\textwidth}{!}{\includegraphics{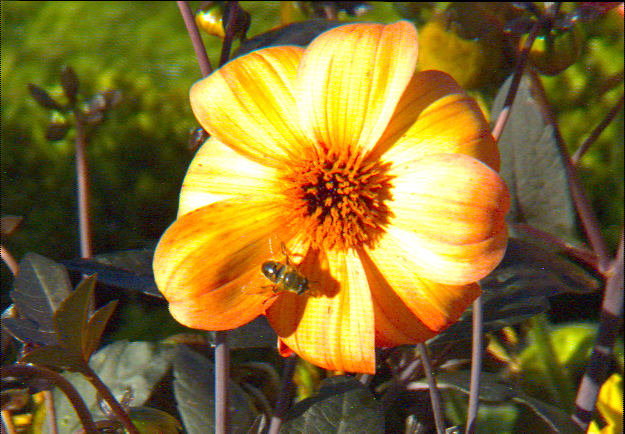}}
		\subcaption{\footnotesize Dansereau {\it et al.}~\cite{DANSCAL}\label{fig:clr_eq:sub:d}}
	\end{minipage}
	\hfill
	\begin{minipage}[t]{.32\linewidth}
		\resizebox{\textwidth}{!}{\includegraphics{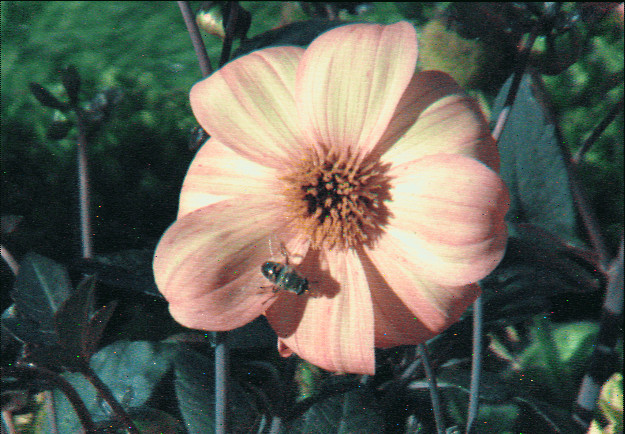}}
		\subcaption{\footnotesize Matysiak {\it et al.}~\cite{Matysiak:2018}\label{fig:clr_eq:sub:e}}
	\end{minipage}
	\hfill
	\begin{minipage}[t]{.32\linewidth}
		\resizebox{\textwidth}{!}{\includegraphics{img/sai_comp/thumb_collection/Bee_2.png}}
		\subcaption{Central view \label{fig:clr_eq:sub:f}}
	\end{minipage}
	\caption{\textbf{Marginal view comparison} of methods combating illumination fall-off at off-axis light-field positions using an exemplary image with $M=15$. The marginal view location is $i=7$, $g=2$ with \subref{fig:clr_eq:sub:a} the untreated image, \subref{fig:clr_eq:sub:b} MKL in Eq. (\ref{eq:mkl}) by Piti{\'e} and Kokaram~\cite{Pitie:2007}, \subref{fig:clr_eq:sub:c} our HM-MKL-HM compound, \subref{fig:clr_eq:sub:d} \Package{LFToolbox~v0.5}~\cite{DANSCAL}, \subref{fig:clr_eq:sub:e} Matysiak~{\it et al.}~\cite{Matysiak:2018} and \subref{fig:clr_eq:sub:f} the central sub-aperture image ($i=7$, $g=7$) as a reference. 
		\label{fig:clr_eq}}
\end{figure}

%
Light-field rays arriving from non-paraxial angles generally suffer from vignetting causing light distributions to be inconsistent among views which we address with the \Package{LfpColorEqualizer}. For visual inspection of the intensity variances, sub-aperture images are stitched together in Fig.~\ref{fig:coco} before and after treatments. A detailed analysis of sub-aperture images at a marginal position ($i=7$, $g=2$) is provided in Fig.~\ref{fig:clr_eq}. \par
\begin{figure*}[ht]
	\centering
	\resizebox{\linewidth}{!}{
		\includegraphics[width=1\linewidth]{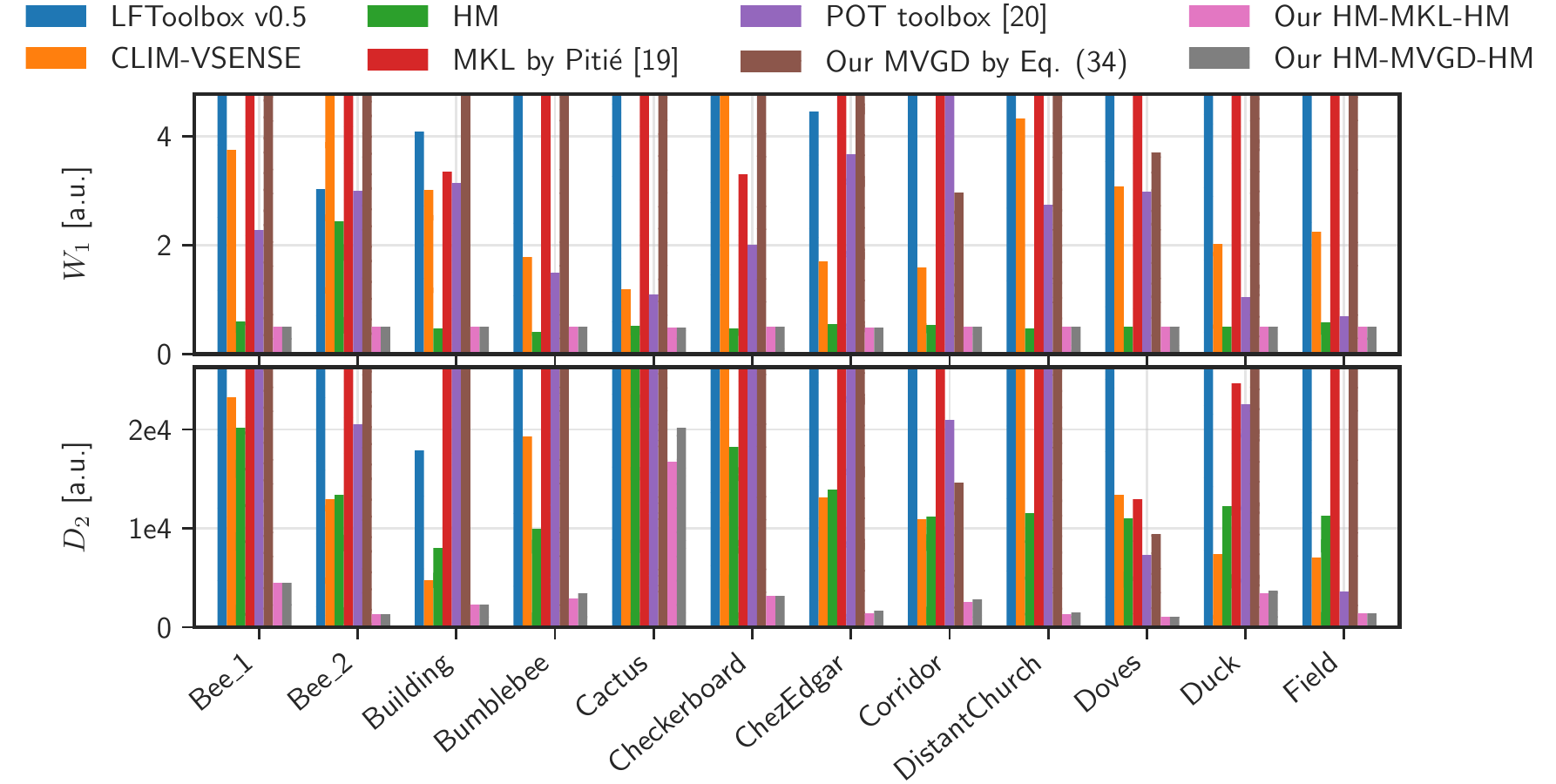}
	}
	\caption{\textbf{Light-field color consistency} from Wasserstein metric $W_1$ and histogram distance $D_2$ where low values indicate high similarity between marginal views $E_g[s_j]$ at $i=7, g=2$ and central views $E_c[s_j]$ from a light-field dataset~\cite{illum:dataset} with $M=15$.}
	\label{fig:marginal}
\end{figure*}

For quantitative assessment of the color transfers we compute normalized CDFs to obtain a Wasserstein metric $W_1$ by
\begin{align}
W_1 = \int_{0}^{\infty} \left| F\left(k,\mathbf{r}^{(g)}\right) - F\left(k,\mathbf{z}^{(g)}\right) \right| \, \mathrm{d}k
\end{align}
where $\mathbf{r} \defeq E_g[s_j]$ represents the marginal and $\mathbf{z} \defeq E_c[s_j]$ the central sub-aperture image serving as the reference. In addition, we employ the average histogram distance $D_2$ by
\begin{align}
D_2 = \left\| f(k,\mathbf{r}) - f(k,\mathbf{z}) \right\|_2
\label{eq:avg_hist}
\end{align}
 with $f(k, \cdot)$ as an all-channel PDF by Eq.~(\ref{eq:pdf}). Color consistency analysis is depicted in Fig.~\ref{fig:marginal} and suggests that our low-cost transport outperforms other methods~\cite{Matysiak:2018, Pitie:2007, Perrot:2016}. \par
%
Table~\ref{tab:comptime} provides computation times for a single Lytro Illum picture at different stages employing one physical processor of an Intel Core i7 @ 2.5 GHz. The extensive computational load imposed by Matysiak's recoloring procedure~\cite{Matysiak:2018} makes it impractical for us to iterate through an entire dataset. 
\begin{table}[H]
	\small
	\centering
	\caption{Computation time comparison where $M=15$}
	\label{tab:comptime}
	\renewcommand{\arraystretch}{1}
	\resizebox{\linewidth}{!}{
		\begin{tabular}{
				|c
				|c
				|c
				|c
				|c
				|
			} %
			\hline
			 \multirow{2}{*}{\textit{Process}} & \footnotesize Dansereau  & \footnotesize Matysiak & \multicolumn{2}{c|}{\multirow{2}{*}{Ours}} \\
			 & \textit{et al.} \cite{DANSCAL} & \textit{et al.} \cite{Matysiak:2018} & \multicolumn{2}{c|}{} \\
			\hline
			Resampling & \multirow{2}{*}{\textbf{70 secs}} & \multirow{2}{*}{125 secs} & local & global \\
			\& Decoding & & & 432 secs & 85 secs\\
			\hline
			Hot Pixels & no support & 100 secs & \multicolumn{2}{c|}{\textbf{52 secs}} \\
			\hline
			Color Eq. & no support & 234 mins & \multicolumn{2}{c|}{\textbf{81 secs}} \\
			\Xhline{2.5\arrayrulewidth}
			\textbf{Total} & 70 secs & 237 mins & 565 secs & 218 secs \\
			\hline	
		\end{tabular}
	}
\end{table}
%
%
%
%
%
%
\subsection{De-Vignetting}
\label{subsec:vign}
\begin{figure}[H]
	\centering
	\begin{minipage}[t]{\linewidth}
		\begin{minipage}{.49\linewidth}
			\begin{overpic}[width=\textwidth]{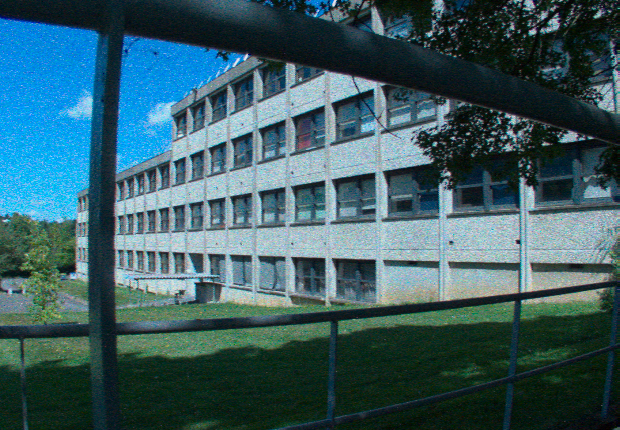}
				\put(55.6,0){					
					\begin{tikzpicture}
					\node[anchor=south west,inner sep=0] (image) at (0,0) {
						\includegraphics[clip, 
						trim=15 181.8 535 160, width=.4\textwidth]{img/vign/build-div-sigd15.png}
					};
					\begin{scope}[x={(image.south east)},y={(image.north west)}]
					\draw[red,ultra thick] (0,0) rectangle (1,1);	
					\end{scope}
					\end{tikzpicture}
				}
				\put(0,0){
					\begin{tikzpicture}
					\begin{scope}
					\path (0,1.22) coordinate (p1) {};
					\path (0,1.88) coordinate (p2) {};
					\path (2.4,0) coordinate (p3) {};
					\path (2.4,2.2) coordinate (p4) {};
					\draw[dotted, red, line width=.4mm](p1.east)--(p3.east)(p2.west)--(p4.west);
					\end{scope}
					\end{tikzpicture}
				}
				\put(0,28.5){
					\begin{tikzpicture}
					\begin{scope}
					\draw[red,ultra thick] (0,0) rectangle (.47,.62);
					\end{scope}
					\end{tikzpicture}
				}
			\end{overpic}
			\subcaption{$74$ \si{\dB} PSNR from division\label{fig:vign:sub:c}}
		\end{minipage}
		\hfill
		\begin{minipage}{.49\linewidth}
			\begin{overpic}[width=\textwidth]{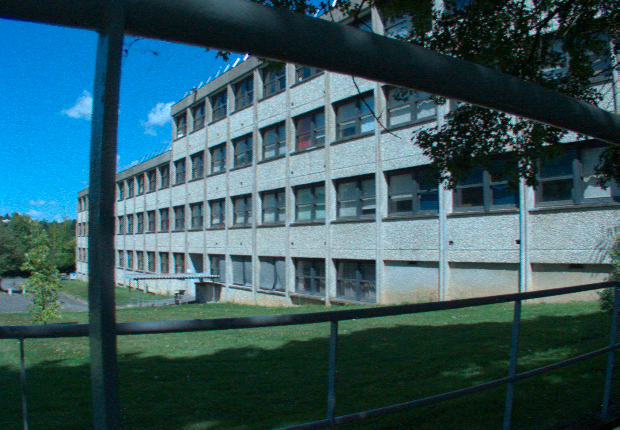}
				\put(55.6,0){					
					\begin{tikzpicture}
					\node[anchor=south west,inner sep=0] (image) at (0,0) {
						\includegraphics[clip, trim=15 181.8 535 160, width=.4\textwidth]{img/vign/build-fit-sigd15.png}
					};
					\begin{scope}[x={(image.south east)},y={(image.north west)}]
					\draw[red,ultra thick] (0,0) rectangle (1,1);	
					\end{scope}
					\end{tikzpicture}
				}
				\put(0,0){
					\begin{tikzpicture}
					\begin{scope}
					\path (0,1.22) coordinate (p1) {};
					\path (0,1.88) coordinate (p2) {};
					\path (2.4,0) coordinate (p3) {};
					\path (2.4,2.2) coordinate (p4) {};
					\draw[dotted, red, line width=.4mm](p1.east)--(p3.east)(p2.west)--(p4.west);
					\end{scope}
					\end{tikzpicture}
				}
				\put(0,28.5){
					\begin{tikzpicture}
					\begin{scope}
					\draw[red,ultra thick] (0,0) rectangle (.47,.62);
					\end{scope}
					\end{tikzpicture}
				}
			\end{overpic}
			\subcaption{$85$ \si{\dB} PSNR from LSQ fit \label{fig:vign:sub:d}}
		\end{minipage}
	\end{minipage}
	\caption{\textbf{De-vignetting} from a white image with Gaussian noise $\sigma_v=0.15$. 
		The noise was absent when de-vignetting the ground-truth. 
		Noise propagates to the light-field during division as seen in \subref{fig:vign:sub:c} and is significantly suppressed in \subref{fig:vign:sub:d} via patch-wise micro image fitting based on \cref{eq:vign:poly,eq:vign:vander,eq:moore_penrose,eq:vign:pseudo_inv}. Our approach gains $\sim10$ \si{\dB} of PSNR compared to the ordinary division as used by~\cite{DANSCAL, Matysiak:2018}. \label{fig:vign}}
\end{figure}
Experimental validation of our LSQ-based de-vignetting in \cref{eq:vign:poly,eq:vign:vander,eq:moore_penrose,eq:vign:pseudo_inv} is assessed using the Peak Signal-to-Noise-Ratio (PSNR) given by
\begin{align}
	\text{PSNR} = 20 \cdot \log_{10}
	\left( 
	\frac{2^8-1}{
		\sqrt{
			\frac{1}{J}\sum_{j=1}^{J} \left(E_T[s_j] - E_c[s_j]\right)^2}
	} 
	\right)
\end{align}
as a metric to analyze its effectiveness. For the ground-truth data $E_T[s_j]$, we use a photograph divided by a white image considered noiseless. This white image is then exposed to additive Gaussian noise with $\sigma_v=0.15$ for de-vignetting by either classical division~\cite{DANSCAL, Matysiak:2018} and our proposed fitting scheme. In doing so, we expect the synthetic noise to propagate during de-vignetting, however, with sufficient suppression (i.e. higher PSNR) for our LSQ approach. The results in Fig.~\ref{fig:vign} show that our method retains the image quality by gaining $\sim10$ \si{\decibel} of dynamic range with respect to noise. This is a consequence of the least-squares fit treating pixel noise as residuals. Although our evaluation contains highly amplified noise, this method is regarded equally beneficial for low-noise images.
\subsection{Refocusing}
\begin{figure}[H]
	\centering
	\begin{minipage}[t]{.32\linewidth}
		\resizebox{\textwidth}{!}{\includegraphics[scale=0.1]{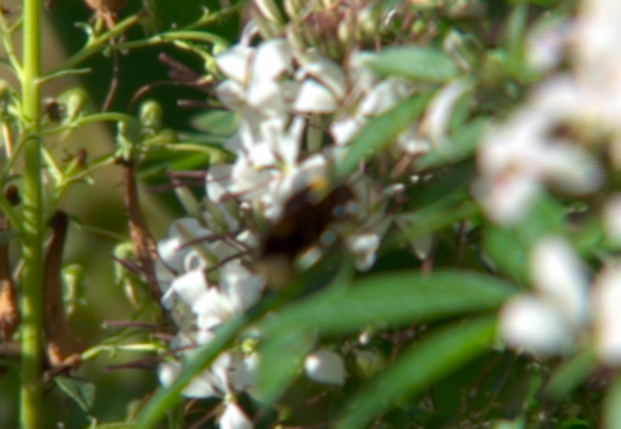}}
		\subcaption{$a=-1.71$\label{fig:refo:sub:a}}
	\end{minipage}
	\begin{minipage}[t]{.32\linewidth}
		\resizebox{\textwidth}{!}{\includegraphics[scale=0.1]{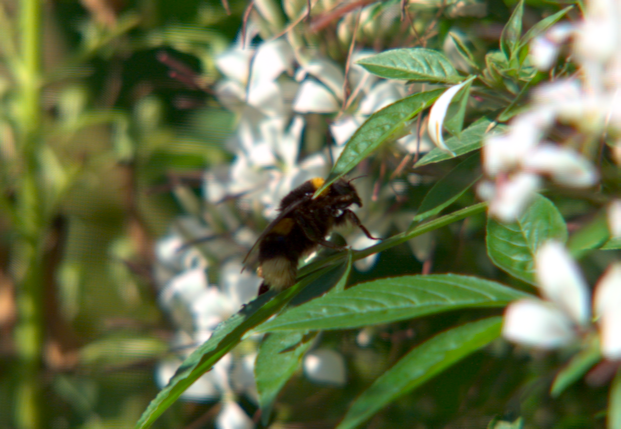}}
		\subcaption{$a=0.57$\label{fig:refo:sub:b}}
	\end{minipage}
	\begin{minipage}[t]{.32\linewidth}
		\resizebox{\textwidth}{!}{\includegraphics[scale=0.1]{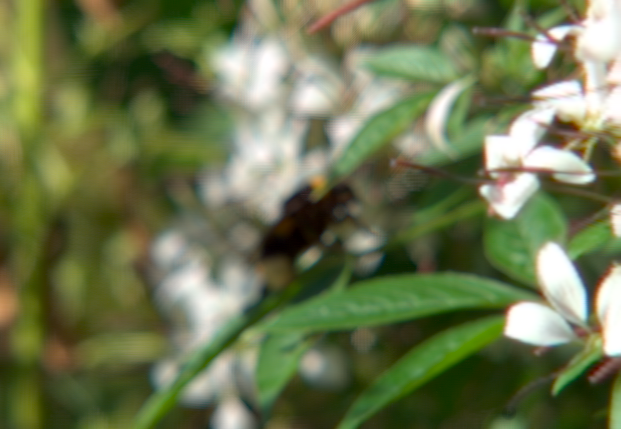}}
		\subcaption{$a=2.43$\label{fig:refo:sub:c}}
	\end{minipage}
	\\
	\vspace{2pt}
	\begin{minipage}[t]{.32\linewidth}
		\resizebox{\textwidth}{!}{\includegraphics[scale=0.1]{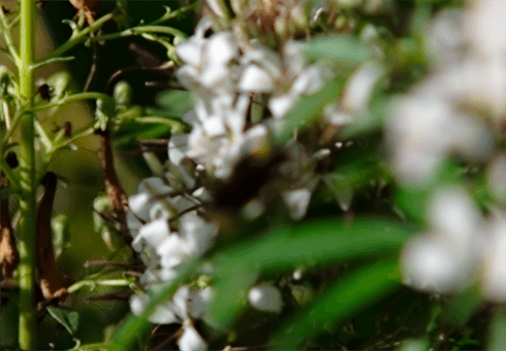}}
		\subcaption{$\lambda=27$\label{fig:refo:sub:d}}
	\end{minipage}
	\begin{minipage}[t]{.32\linewidth}
		\resizebox{\textwidth}{!}{\includegraphics[scale=0.1]{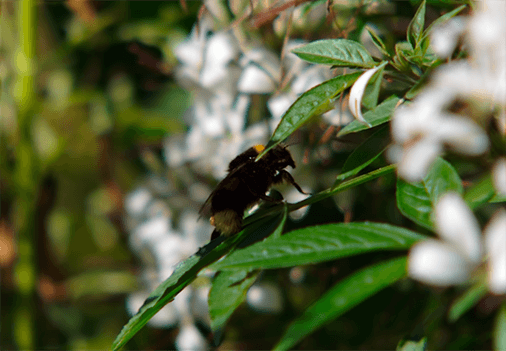}}
		\subcaption{$\lambda=1$\label{fig:refo:sub:e}}
	\end{minipage}
	\begin{minipage}[t]{.32\linewidth}
		\resizebox{\textwidth}{!}{\includegraphics[scale=0.1]{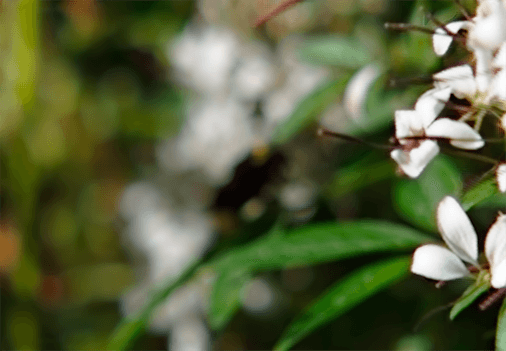}}
		\subcaption{$\lambda=-23$\label{fig:refo:sub:f}}
	\end{minipage}
	\caption{\textbf{Refocused photographs} from \Package{PlenoptiCam v1.0} in the top row with shift and sum parameter $a$ in comparison to results processed by \Package{Lytro Power Tools v1.0} in the bottom row with focus parameter $\lambda$. \label{fig:refo}}
\end{figure}

For the refocusing assessment, we compare the proposed \Package{LfpRefocuser} module against \Package{Lytro Power Tools~v1.0}. Refocused images are depicted in Fig.~\ref{fig:refo}. We choose the {\it Bumblebee.lfp} image from the IRISA dataset~\cite{illum:dataset} as it exhibits organic object structures spread through a wide range of depth. 


%
Quantitative comparison is achieved by analysis of the sharpness and BRISQUE score of local image details. For the sharpness analysis, we follow the Mavridaki and Mezaris~\cite{MAVRIDAKI:2014} approach by transforming cropped versions of a refocused image slice $E''_a\left[s_j \, , \, t_h\right]$ to the Fourier domain using the Discrete Fourier Transformation (DFT) and extracting the magnitude signal $\mathcal{X}\left[\sigma_{\omega} \, , \, \rho_{\psi}\right]$ by
\begin{align}
\mathcal{X}\left[\sigma_{\omega} \, , \, \rho_{\psi}\right] = &\Bigg\lvert \sum_{j=\xi}^{\Xi}\sum_{h=\varpi}^{\Pi} E''_a\left[s_j \, , \, t_h\right] \\ 
&\quad \exp{\left(-2\pi \kappa(j\omega/(\Xi-\xi)+h\psi/(\Pi-\varpi)\right)}\Bigg\rvert \notag
\end{align}
where $\kappa=\sqrt{-1}$ and $|\cdot|$ yields absolute values. %
Given the \mbox{2-D} magnitude signal $\mathcal{X}\left[\sigma_{\omega} \, , \, \rho_{\psi}\right]$, the total energy $TE$ reads
\begin{align}
TE &= \sum_{\omega=1}^{\Omega} \sum_{\psi=1}^{\Psi} \mathcal{X}\left[\sigma_{\omega} \, , \, \rho_{\psi}\right]^2 \,
\end{align}
where $\Omega=\lceil\sfrac{(\Xi-\xi)}{2}\rceil$ and $\Psi=\lceil\sfrac{(\Pi-\varpi)}{2}\rceil$ are borders cropping the first quarter of the unshifted magnitude signal. To isolate the energy of high frequency elements $HE$, we take the power of low frequencies and subtract it from $TE$ reading
\begin{align}
HE &= TE - \sum_{\omega=1}^{Q_H} \sum_{\psi=1}^{Q_V} \mathcal{X}\left[\sigma_{\omega} \, , \, \rho_{\psi}\right]^2 \,
\end{align}
with $Q_H$ and $Q_V$ as scalar limits in the range of $\{1, 2, \ldots, \Omega\}$ and $\{1, 2, \ldots, \Psi\}$ separating low from high frequencies. In our experiments we set the limits to a five hundredths of the cropped image resolution. Finally, the sharpness $S$ is a frequency ratio of refocused image portions obtained by
\begin{align}
S &= \frac{HE}{TE} \label{eq:sharp}
\end{align}
which serves as our blur metric. \par
\begin{figure}[H]
	\centering
	\begin{minipage}[t]{.32\linewidth}
		\centering
		\Package{PlenoptiCam v1.0}
		\vspace{0em}
	\end{minipage}
	\hfill
	\begin{minipage}[t]{.32\linewidth}
		\centering
		\Package{PlenoptiCam v1.0} [refinement]
		\vspace{0em}
	\end{minipage}
	\hfill
	\begin{minipage}[t]{.32\linewidth}
		\centering
		\Package{Lytro Power Tools v1.0}
		\vspace{0em}
	\end{minipage}
	\\
	\vspace{2pt}
	\begin{minipage}[t]{.32\linewidth}
			\resizebox{\textwidth}{!}{\includegraphics{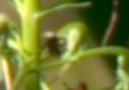}}
			\subcaption{$a=-2$\label{fig:sharp:sub:a}}
		\end{minipage}
	\begin{minipage}[t]{.32\linewidth}
		\resizebox{\textwidth}{!}{\includegraphics{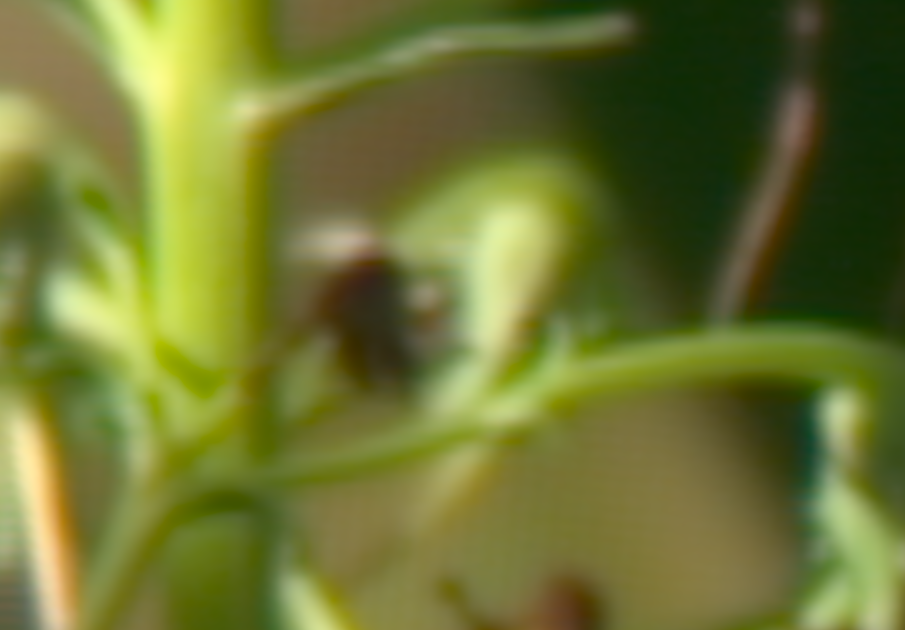}}
		\subcaption{$a=-1.71$\label{fig:sharp:sub:b}}
	\end{minipage}
	\begin{minipage}[t]{.32\linewidth}
		\resizebox{\textwidth}{!}{\includegraphics{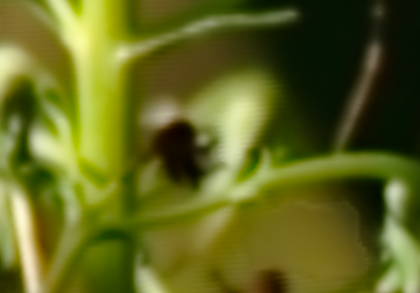}}
		\subcaption{$\lambda=27$\label{fig:sharp:sub:c}}
	\end{minipage}
	\\
	\vspace{2pt}
	\begin{minipage}[t]{.32\linewidth}
		\resizebox{\textwidth}{!}{\includegraphics{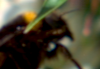}}
		\subcaption{$a=1$\label{fig:sharp:sub:d}}
	\end{minipage}
	\begin{minipage}[t]{.32\linewidth}
		\resizebox{\textwidth}{!}{\includegraphics{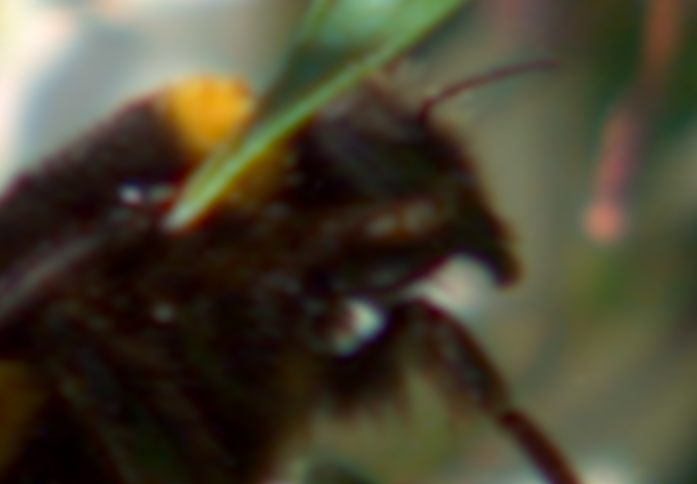}}
		\subcaption{$a=0.57$\label{fig:sharp:sub:e}}
	\end{minipage}
	\begin{minipage}[t]{.32\linewidth}
		\resizebox{\textwidth}{!}{\includegraphics{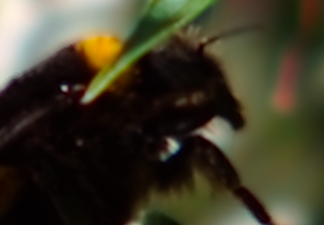}}
		\subcaption{$\lambda=1$\label{fig:sharp:sub:f}}
	\end{minipage}
	\\
	\vspace{2pt}
	\begin{minipage}[t]{.32\linewidth}
		\resizebox{\textwidth}{!}{\includegraphics{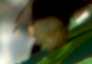}}
		\subcaption{$a=1$\label{fig:sharp:sub:g}}
	\end{minipage}
	\begin{minipage}[t]{.32\linewidth}
		\resizebox{\textwidth}{!}{\includegraphics{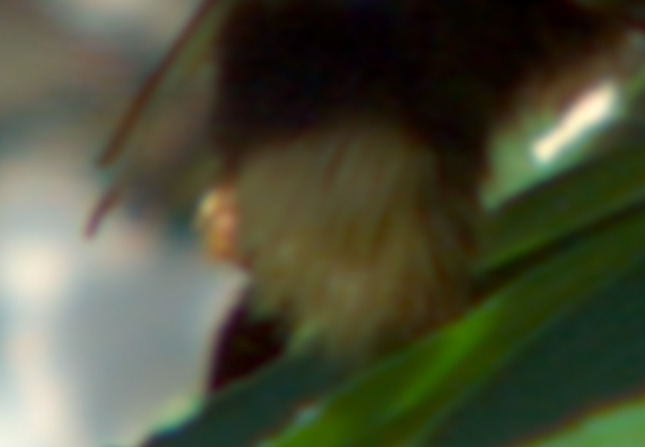}}
		\subcaption{$a=0.57$\label{fig:sharp:sub:h}}
	\end{minipage}
	\begin{minipage}[t]{.32\linewidth}
		\resizebox{\textwidth}{!}{\includegraphics{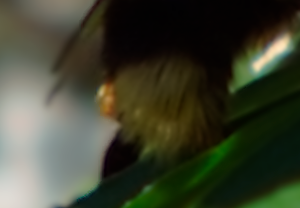}}
		\subcaption{$\lambda=1$\label{fig:sharp:sub:i}}
	\end{minipage}
	\\
	\vspace{2pt}
	\begin{minipage}[t]{.32\linewidth}
		\resizebox{\textwidth}{!}{\includegraphics{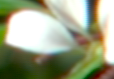}}
		\subcaption{$a=2$\label{fig:sharp:sub:j}}
	\end{minipage}
	\begin{minipage}[t]{.32\linewidth}
		\resizebox{\textwidth}{!}{\includegraphics{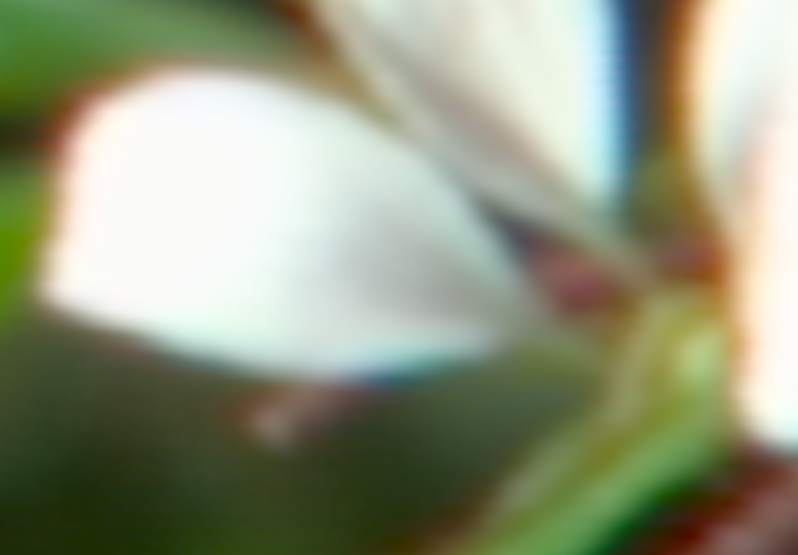}}
		\subcaption{$a=2.43$\label{fig:sharp:sub:k}}
	\end{minipage}
	\begin{minipage}[t]{.32\linewidth}
		\resizebox{\textwidth}{!}{\includegraphics{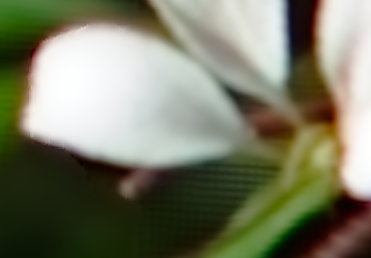}}
		\subcaption{$\lambda=-23$\label{fig:sharp:sub:l}}
	\end{minipage}
	\caption{\textbf{Magnified refocused image tiles} showing texture details which are quantified by our blur metric $S$ in Table~\ref{tab:refo}. The left and middle column contain results from \Package{PlenoptiCam v1.0} whereas \subref{fig:sharp:sub:b}, \subref{fig:sharp:sub:e}, \subref{fig:sharp:sub:h}, \subref{fig:sharp:sub:k} are rendered with sub-pixel precision. The right column is obtained from \Package{Lytro Power Tools v1.0} using focus parameter $\lambda$.\label{fig:sharp}}
\end{figure}

From the cropped view of the Lytro image in Fig.~\ref{fig:sharp:sub:l}, it is seen that there is no gradual decrease in blur between objects at different depth leading to an unnatural appearance of image blur. Images rendered with \Package{LfpRefocuser} do not expose such sudden change in sharpness. It is only up to speculation on how Lytro's refocusing algorithm works. The high resolution and occasional blur artifacts around object edges suggest that Lytro super-resolves images at the first stage and then re-blurs spatial areas guided by a previously computed depth map. This may explain blur artifacts at object boundaries, which rely on the quality of a depth map.
\begin{table}[H]
	\small
	\centering
	\caption{Metric assessment of refocused images}
	\label{tab:refo}
	\resizebox{\columnwidth}{!}{
		\begin{tabular}{
				|c
				|c
				|c
				|c
				|c
				|c
				|c
				|	
			}
			\hline
			\diagbox{Metrics}{Fig.~\ref{fig:sharp}} & \subref{fig:sharp:sub:a} & \subref{fig:sharp:sub:b} & \subref{fig:sharp:sub:c} & \subref{fig:sharp:sub:d} & \subref{fig:sharp:sub:e} & \subref{fig:sharp:sub:f} \\
			\hline
			Sharpness $S$ & 0.034 & 0.040 & \textbf{0.060} & 0.107 & 0.107  & \textbf{0.224} \\
			\hline
			BRISQUE score & 84.29 & 83.51 & \textbf{81.67} & 92.43 & 81.29 & \textbf{77.29} \\
			\hline
			\multicolumn{7}{c}{}\\
			\hline
			\diagbox{Metrics}{Fig.~\ref{fig:sharp}} & \subref{fig:sharp:sub:g} & \subref{fig:sharp:sub:h} & \subref{fig:sharp:sub:i} & \subref{fig:sharp:sub:j} &
			\subref{fig:sharp:sub:k} & \subref{fig:sharp:sub:l} \\
			\hline
			Sharpness $S$ & 0.107 & 0.107  & \textbf{0.125} & 0.022 & 0.027 & \textbf{0.054} \\
			\hline
			BRISQUE score & 92.43 & 81.29 & \textbf{81.19} & 94.30 & 90.56 & \textbf{80.22} \\
			\hline
		\end{tabular}
	}
\end{table}
%
It is therefore questionable whether the measured sharpness in Lytro's refocus algorithm arises from high frequency artifacts. 
By visual inspection, however, it becomes apparent that the magnified views from Lytro expose more image detail at small $\lambda$ value. This observation is backed by respective scores from the BRISQUE metric in Table~\ref{tab:refo}. %
On the contrary, occasional artifact patterns as in Fig.~\ref{fig:sharp:sub:l} may still appeal unpleasant to human visual perception. %
\par
\begin{figure}[H]
	\resizebox{\columnwidth}{!}{\includegraphics{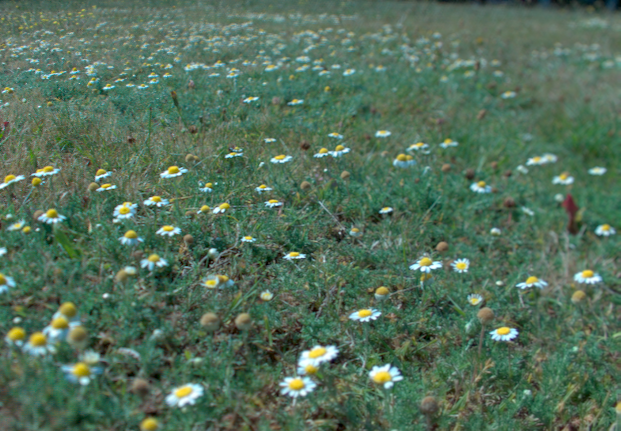}}
	\caption{\textbf{Scheimpflug rendered photograph} from \Package{LfpScheimpflug} class with $a=\{0,1\}$ and sub-pixel precision.\label{fig:scheim}}
\end{figure}

%
%
An exemplary result of the \Package{LfpScheimpflug} rendering engine is depicted in Fig.~\ref{fig:scheim}. Note how image areas in the upper left background expose sharp details, whereas the focus gradually moves toward the lower right foreground. 
This is a consequence of the synthetic focus plane being tilted with respect to the image sensor. \par
\section{Conclusions}
\label{sec:5}
%
Plenoptic image decomposition is a crucial task in light-field rendering for which we propose a  pipeline with outstanding image quality. 
To the best of our knowledge, \Package{PlenoptiCam} is the only framework that allows users to automatically calibrate footage from different types of plenoptic cameras regardless of the micro lens specification. We achieve this by employing blob detection, non-maximum suppression and a generic MLA grid geometry recognition. 
The light-field color equalization stage is taken to a new level by using histogram matching in conjunction with an optimal transport solver to yield convincing quantitative results at reasonable computation times, which outperforms and accelerates previous methods~\cite{Matysiak:2018, Pitie:2007, Perrot:2016}. 
For plenoptic image alignment, we created an alternative element-wise micro image resampling procedure. Moreover, we are first to address fringe artifacts caused by hexagonal micro lens arrangements, which are successfully reduced by our novel identification scheme. An extensive assessment of state-of-the-art sub-aperture image decomposition pipelines has been carried out in this paper with the result of outperforming other tool chains in two thirds of the cases, using a wide range of metrics such as BRISQUE, Wasserstein and histogram distance as well as a blur metric. \par 
These findings and respective implementations are released as an open-source framework made available as a repository on \href{https://github.com/hahnec/plenopticam/}{GitHub}~\cite{gh:pcam} to allow others to participate and contribute. The herein presented framework is intended to disburden newcomers and facilitate first steps in the field of plenoptic imaging. Peers are encouraged to report bugs and actively expand this software. \Package{PlenoptiCam} may lay the groundwork for future algorithm development and testing of plenoptic data. \par
Such extensions may include, but are not limited to, an algorithm development for the focused plenoptic camera as initially discovered by Lumsdaine and Georgiev~\cite{LUMSFULL}. An implementation of a super-resolution technique may be accomplished on the basis of early pioneering work~\cite{Bishop09lightfield, focal_stack_fusion}. 
It should also be feasible to enhance the image quality by combating optical aberrations as shown in previous studies~\cite{Ng:2006, Cohen:14, r4, Liang:2017} or by making use of the Bayer demosaicing similar to what Yu and Yu have revealed~\cite{Yu:2012}. Until now, rectification for lens distortions exceeded the scope as this can be accomplished using traditional computer vision libraries. However, correct disparity estimation requires counteracting distortions in sub-aperture images at a preceding stage. \par
%
%
We believe that our work will be a substantial contribution to the field of plenoptic cameras, not limited to the image processing community. As Dansereau's \Package{LFToolbox} has shown, there is a large group of researchers experimenting with plenoptic cameras demonstrating the demand for an easy-to-use software addressing the special requirements for plenoptic imaging. \Package{PlenoptiCam} may serve this broad user group including image scientists, programmers in the fields of data science, medical and industrial engineering or photography independent of the user's operating system. In particular, \Package{PlenoptiCam} facilitates light-field data preparation for visual machine learning systems such as convolutional neural networks, which currently receive an increasing interest from a variety of scientific and industrial communities. The chosen open source license model enables cost-free usage and code republication with modifications, giving users the opportunity to extend the presented software. %
%

%





\ifCLASSOPTIONcaptionsoff
  \newpage
\fi



\bibliographystyle{IEEEtran}
\bibliography{IEEEabrv,mall-merge.bib}

\begin{thebibliography}{10}
\providecommand{\url}[1]{#1}
\csname url@samestyle\endcsname
\providecommand{\newblock}{\relax}
\providecommand{\bibinfo}[2]{#2}
\providecommand{\BIBentrySTDinterwordspacing}{\spaceskip=0pt\relax}
\providecommand{\BIBentryALTinterwordstretchfactor}{4}
\providecommand{\BIBentryALTinterwordspacing}{\spaceskip=\fontdimen2\font plus
\BIBentryALTinterwordstretchfactor\fontdimen3\font minus
  \fontdimen4\font\relax}
\providecommand{\BIBforeignlanguage}[2]{{%
\expandafter\ifx\csname l@#1\endcsname\relax
\typeout{** WARNING: IEEEtran.bst: No hyphenation pattern has been}%
\typeout{** loaded for the language `#1'. Using the pattern for}%
\typeout{** the default language instead.}%
\else
\language=\csname l@#1\endcsname
\fi
#2}}
\providecommand{\BIBdecl}{\relax}
\BIBdecl

\bibitem{NGLEV}
R.~Ng, M.~Levoy, M.~Br{\`e}dif, G.~Duval, M.~Horowitz, and P.~Hanrahan, ``Light
  field photography with a hand-held plenoptic camera,'' Stanford University,
  Tech. Rep. CTSR 2005-02, 2005.

\bibitem{Hahne:OPEX:16}
\BIBentryALTinterwordspacing
C.~Hahne, A.~Aggoun, V.~Velisavljevic, S.~Fiebig, and M.~Pesch, ``Refocusing
  distance of a standard plenoptic camera,'' \emph{Opt. Express}, vol.~24,
  no.~19, pp. 21\,521--21\,540, September 2016. [Online]. Available:
  \url{http://www.opticsexpress.org/abstract.cfm?URI=oe-24-19-21521}
\BIBentrySTDinterwordspacing

\bibitem{prevedel:2014:simultaneous}
R.~Prevedel, Y.-G. Yoon, M.~Hoffmann, N.~Pak, G.~Wetzstein, S.~Kato,
  T.~Schr{\"o}del, R.~Raskar, M.~Zimmer, E.~S. Boyden \emph{et~al.},
  ``Simultaneous whole-animal 3d imaging of neuronal activity using light-field
  microscopy,'' \emph{Nature methods}, vol.~11, no.~7, p. 727, 2014.

\bibitem{Li439315}
\BIBentryALTinterwordspacing
H.~Li, C.~Guo, D.~Kim-Holzapfel, W.~Li, Y.~Altshuller, B.~Schroeder, W.~Liu,
  Y.~Meng, J.~French, K.-I. Takamaru, M.~Frohman, and S.~Jia, ``Fast,
  volumetric live-cell imaging using high-resolution light-field microscopy,''
  \emph{bioRxiv}, 2018. [Online]. Available:
  \url{https://www.biorxiv.org/content/early/2018/10/10/439315}
\BIBentrySTDinterwordspacing

\bibitem{Bedard:17}
N.~Bedard, T.~Shope, A.~Hoberman, M.~Ann~Haralam, N.~Shaikh, J.~Kovačević,
  N.~Balram, and I.~Tošić, ``Light field otoscope design for 3d in vivo
  imaging of the middle ear,'' \emph{Biomedical Optics Express}, vol.~8, p.
  260, 01 2017.

\bibitem{Palmer:18}
D.~W.~Palmer, T.~Coppin, K.~Rana, D.~G.~Dansereau, M.~Suheimat, M.~Maynard,
  D.~A.~Atchison, J.~Roberts, R.~Crawford, and A.~Jaiprakash, ``Glare-free
  retinal imaging using a portable light field fundus camera,''
  \emph{Biomedical Optics Express}, vol.~9, p. 3178, 07 2018.

\bibitem{Srinivasan:ICCV:2017}
P.~P. Srinivasan, T.~Wang, A.~Sreelal, R.~Ramamoorthi, and R.~Ng, ``Learning to
  synthesize a 4d rgbd light field from a single image,'' in \emph{IEEE
  International Conference on Computer Vision (ICCV)}, October 2017.

\bibitem{LEVHAN}
M.~Levoy and P.~Hanrahan, ``Light field rendering,'' \emph{Proceedings of the
  23rd annual conference on Computer graphics and interactive techniques}, pp.
  31--42, 1996.

\bibitem{AW}
E.~H. Adelson and J.~Y. Wang, ``Single lens stereo with a plenoptic camera,''
  \emph{IEEE Transactions on Pattern Analysis and Machine Intelligence},
  vol.~14, no.~2, pp. 99--106, February 1992.

\bibitem{Hahne:IJCV:18}
\BIBentryALTinterwordspacing
C.~{Hahne}, A.~{Aggoun}, V.~{Velisavljevic}, S.~{Fiebig}, and M.~{Pesch},
  ``Baseline and triangulation geometry in a standard plenoptic camera,''
  \emph{International Journal of Computer Vision}, vol. 126, no.~1, pp. 21--35,
  January 2018. [Online]. Available:
  \url{https://doi.org/10.1007/s11263-017-1036-4}
\BIBentrySTDinterwordspacing

\bibitem{Cho:2013}
D.~Cho, M.~Lee, S.~Kim, and Y.-W. Tai, ``Modeling the calibration pipeline of
  the lytro camera for high quality light-field image reconstruction,'' in
  \emph{IEEE International Conference on Computer Vision (ICCV)}, December
  2013, pp. 3280--3287.

\bibitem{DANSCAL}
D.~G. Dansereau, O.~Pizarro, and S.~B. Williams, ``Decoding, calibration and
  rectification for lenselet-based plenoptic cameras,'' in \emph{IEEE
  International Conference on Computer Vision and Pattern Recognition (CVPR)},
  June 2013, pp. 1027--1034.

\bibitem{lfp:nirav}
N.~Patel, ``lfptools,'' \url{https://github.com/nrpatel/lfptools}, 2011.

\bibitem{lfp:behnam}
B.~Esfahbod, ``python-lfp-reader,''
  \url{https://github.com/behnam/python-lfp-reader}, 2013.

\bibitem{bok:eccv14}
Y.~Bok, H.-G. Jeon, and I.~S. Kweon, ``Geometric calibration of
  micro-lens-based light-field cameras using line features,'' in
  \emph{Proceedings of European Conference on Computer Vision (ECCV)}, 2014.

\bibitem{Pendu:2020}
M.~Le~Pendu and A.~Smolic, ``High resolution light field recovery with fourier
  disparity layer completion, demosaicing, and super-resolution,'' April 2020.

\bibitem{David:2017}
P.~David, M.~Le~Pendu, and C.~Guillemot, ``White lenslet image guided
  demosaicing for plenoptic cameras,'' 10 2017, pp. 1--6.

\bibitem{Matysiak:2018}
\BIBentryALTinterwordspacing
P.~Matysiak, M.~Grogan, M.~L. Pendu, M.~Alain, and A.~Smolic, ``A pipeline for
  lenslet light field quality enhancement,'' \emph{2018 25th IEEE International
  Conference on Image Processing (ICIP)}, October 2018. [Online]. Available:
  \url{http://dx.doi.org/10.1109/ICIP.2018.8451544}
\BIBentrySTDinterwordspacing

\bibitem{Pitie:2007}
F.~Piti{\'e} and A.~Kokaram, ``The linear monge-kantorovitch colour mapping for
  example-based colour transfer,'' \emph{IET Conference Proceedings}, pp.
  23--23(1), January 2007.

\bibitem{Perrot:2016}
M.~Perrot, N.~Courty, R.~Flamary, and A.~Habrard, ``Mapping estimation for
  discrete optimal transport,'' ser. NIPS'16.\hskip 1em plus 0.5em minus
  0.4em\relax Red Hook, NY, USA: Curran Associates Inc., 2016, pp.
  4204–--4212.

\bibitem{RAYTRIX}
\BIBentryALTinterwordspacing
C.~Perwass and L.~Wietzke, ``Single lens {3D}-camera with extended
  depth-of-field,'' in \emph{Human Vision and Electronic Imaging XVII}, vol.
  Proc. SPIE 8291.\hskip 1em plus 0.5em minus 0.4em\relax Raytrix GmbH,
  February 2012. [Online]. Available: \url{http://dx.doi.org/10.1117/12.909882}
\BIBentrySTDinterwordspacing

\bibitem{Plenoptisign:2019}
\BIBentryALTinterwordspacing
C.~Hahne and A.~Aggoun, ``Plenoptisign: An optical design tool for plenoptic
  imaging,'' \emph{SoftwareX}, vol.~10, p. 100259, 2019. [Online]. Available:
  \url{http://www.sciencedirect.com/science/article/pii/S2352711019300159}
\BIBentrySTDinterwordspacing

\bibitem{lindeberg:scale-space}
T.~Lindeberg, \emph{Scale-Space Theory in Computer Vision}.\hskip 1em plus
  0.5em minus 0.4em\relax Springer US, January 1994.

\bibitem{Crowley:2002}
J.~L. Crowley, O.~Riff, and J.~H. Piater, ``Fast computation of characteristic
  scale using a half-octave pyramid,'' in \emph{In: Scale Space 03: 4th
  International Conference on Scale-Space theories in Computer Vision, Isle of
  Skye}, 2002.

\bibitem{Liang:2016}
C.-K. Liang and Z.~Wang, ``Calibration of light-field camera geometry via
  robust fitting,'' U.S. Patent 9 420 276 B2, August 2016.

\bibitem{NMS}
T.~Q. Pham, ``Non-maximum suppression using fewer than two comparisons per
  pixel,'' in \emph{Advanced Concepts for Intelligent Vision Systems},
  J.~Blanc-Talon, D.~Bone, W.~Philips, D.~Popescu, and P.~Scheunders,
  Eds.\hskip 1em plus 0.5em minus 0.4em\relax Berlin, Heidelberg: Springer
  Berlin Heidelberg, 2010, pp. 438--451.

\bibitem{Hartley2004}
R.~I. Hartley and A.~Zisserman, \emph{Multiple View Geometry in Computer
  Vision}, 2nd~ed.\hskip 1em plus 0.5em minus 0.4em\relax Cambridge University
  Press, ISBN: 0521540518, 2004.

\bibitem{LMA:1978}
J.~J. Mor{\'e}, ``The levenberg-marquardt algorithm: Implementation and
  theory,'' in \emph{Numerical Analysis}, G.~A. Watson, Ed.\hskip 1em plus
  0.5em minus 0.4em\relax Berlin, Heidelberg: Springer Berlin Heidelberg, 1978,
  pp. 105--116.

\bibitem{Vignetting:16}
A.~Kordecki, H.~Palus, and A.~Bal, ``Practical vignetting correction method for
  digital camera with measurement of surface luminance distribution,''
  \emph{Signal, Image and Video Processing}, vol.~10, pp. 1417–--1424, 2016.

\bibitem{Menon2007c}
\BIBentryALTinterwordspacing
D.~Menon, S.~Andriani, and G.~Calvagno, ``{Demosaicing With Directional
  Filtering and a posteriori Decision},'' \emph{IEEE Transactions on Image
  Processing}, vol.~16, no.~1, pp. 132--141, January 2007. [Online]. Available:
  \url{http://ieeexplore.ieee.org/document/4032820/}
\BIBentrySTDinterwordspacing

\bibitem{Seifi:14}
M.~Seifi, N.~Sabater, V.~Drazic, and P.~Perez, ``Disparity-guided demosaicking
  of light field images,'' October 2014.

\bibitem{Yongwei:19}
Y.~{Li} and M.~{Sj{\"o}str{\"o}m}, ``Depth-assisted demosaicing for light field
  data in layered object space,'' in \emph{2019 IEEE International Conference
  on Image Processing (ICIP)}, 2019, pp. 3746--3750.

\bibitem{gonzalez2017digital}
R.~Gonzalez and R.~Woods, \emph{Digital Image Processing, Global
  Edition}.\hskip 1em plus 0.5em minus 0.4em\relax Pearson, 2017.

\bibitem{Pitie2008bookchapter}
\BIBentryALTinterwordspacing
F.~Piti{\'e}, A.~Kokaram, and R.~Dahyot, ser. Image Processing Series.\hskip
  1em plus 0.5em minus 0.4em\relax CRC Press, September 2008, ch. Enhancement
  of Digital Photographs Using Color Transfer Techniques, pp. 295--321, 0.
  [Online]. Available: \url{http://dx.doi.org/10.1201/9781420054538.ch11}
\BIBentrySTDinterwordspacing

\bibitem{ISAKSEN}
A.~Isaksen, ``Dynamically reparameterized light fields,'' Master's thesis,
  Electrical Engineering and Computer Science, Massachusetts Institute of
  Technology, November 2000.

\bibitem{Hahne:IEEE:2018}
C.~{Hahne}, A.~{Lumsdaine}, A.~{Aggoun}, and V.~{Velisavljevic}, ``Real-time
  refocusing using an fpga-based standard plenoptic camera,'' \emph{IEEE
  Transactions on Industrial Electronics}, vol.~65, no.~12, pp. 9757--9766,
  December 2018.

\bibitem{SchambachPuenteLeon}
M.~Schambach and F.~{Puente Le{\'o}n}, ``\BIBforeignlanguage{english}{Microlens
  array grid estimation, light field decoding, and calibration},''
  \emph{\BIBforeignlanguage{english}{IEEE transactions on computational
  imaging}}, vol.~6, pp. 591–--603, January 2020.

\bibitem{Dansereau:2015}
\BIBentryALTinterwordspacing
D.~G. Dansereau, O.~Pizarro, and S.~B. Williams, ``Linear volumetric focus for
  light field cameras,'' \emph{ACM Trans. Graph.}, vol.~34, no.~2, pp.
  15:1--15:20, March 2015. [Online]. Available:
  \url{http://doi.acm.org/10.1145/2665074}
\BIBentrySTDinterwordspacing

\bibitem{pitts2013compensating}
C.~Pitts, T.~Knight, C.~Liang, and Y.~Ng, ``Compensating for variation in
  microlens position during light-field image processing,'' U.S. Patent App.
  13/774,971, August 2013.

\bibitem{mignarddebise}
\BIBentryALTinterwordspacing
L.~Mignard-Debise and I.~Ihrke, ``{A Vignetting Model for Light Field Cameras
  with an Application to Light Field Microscopy},'' \emph{{IEEE Transactions on
  Computational Imaging}}, p.~10, Apr. 2019. [Online]. Available:
  \url{https://hal.archives-ouvertes.fr/hal-02263377}
\BIBentrySTDinterwordspacing

\bibitem{illum:dataset}
\BIBentryALTinterwordspacing
R.~C. Daudt and C.~Guillemot. Lytro illum light field dataset. Online;
  accessed: July 2019. [Online]. Available:
  \url{https://www.irisa.fr/temics/demos/IllumDatasetLF/index.html}
\BIBentrySTDinterwordspacing

\bibitem{Mittal2012NoReferenceIQ}
A.~Mittal, A.~K. Moorthy, and A.~Bovik, ``No-reference image quality assessment
  in the spatial domain,'' \emph{IEEE Transactions on Image Processing},
  vol.~21, pp. 4695--4708, 2012.

\bibitem{gh:pybrisque}
Bukalapak, ``Pybrisque,'' \url{https://github.com/bukalapak/pybrisque}, 2020.

\bibitem{focal_stack_fusion}
\BIBentryALTinterwordspacing
C.-K. Liang and R.~Ramamoorthi, ``A light transport framework for lenslet light
  field cameras,'' \emph{ACM Trans. Graph.}, vol.~34, no.~2, March 2015.
  [Online]. Available: \url{https://doi.org/10.1145/2665075}
\BIBentrySTDinterwordspacing

\bibitem{r4}
\BIBentryALTinterwordspacing
L.-Y. Wei, C.-K. Liang, G.~Myhre, C.~Pitts, and K.~Akeley, ``Improving light
  field camera sample design with irregularity and aberration,'' \emph{ACM
  Trans. Graph.}, vol.~34, no.~4, July 2015. [Online]. Available:
  \url{https://doi.org/10.1145/2766885}
\BIBentrySTDinterwordspacing

\bibitem{MAVRIDAKI:2014}
E.~Mavridaki and V.~Mezaris, ``No-reference blur assessment in natural images
  using fourier transform and spatial pyramids,'' in \emph{2014 IEEE
  International Conference on Image Processing (ICIP)}, October 2014, pp.
  566--570.

\bibitem{gh:pcam}
C.~Hahne, ``Plenopticam,'' \url{https://github.com/hahnec/plenopticam}, 2020.

\bibitem{LUMSFULL}
A.~Lumsdaine and T.~Georgiev, ``Full resolution lightfield rendering,'' Adobe
  Systems, Inc., Tech. Rep. No. 1, January 2008.

\bibitem{Bishop09lightfield}
T.~E. Bishop, S.~Zanetti, and P.~Favaro, ``Light field superresolution,'' in
  \emph{IEEE International Conference on Computational Photography (ICCP)},
  2009.

\bibitem{Ng:2006}
\BIBentryALTinterwordspacing
R.~Ng and P.~Hanrahan, ``{Digital correction of lens aberrations in light field
  photography},'' in \emph{International Optical Design Conference 2006}, G.~G.
  Gregory, J.~M. Howard, and R.~J. Koshel, Eds., vol. 6342, International
  Society for Optics and Photonics.\hskip 1em plus 0.5em minus 0.4em\relax
  SPIE, 2006, pp. 441--454. [Online]. Available:
  \url{https://doi.org/10.1117/12.692290}
\BIBentrySTDinterwordspacing

\bibitem{Cohen:14}
\BIBentryALTinterwordspacing
N.~Cohen, S.~Yang, A.~Andalman, M.~Broxton, L.~Grosenick, K.~Deisseroth,
  M.~Horowitz, and M.~Levoy, ``Enhancing the performance of the light field
  microscope using wavefront coding,'' \emph{Opt. Express}, vol.~22, no.~20,
  pp. 24\,817--24\,839, October 2014. [Online]. Available:
  \url{http://www.opticsexpress.org/abstract.cfm?URI=oe-22-20-24817}
\BIBentrySTDinterwordspacing

\bibitem{Liang:2017}
C.-K. Liang, C.~Pitts, C.~Craddock, and G.~Myhre, ``Light-field aberration
  correction,'' U.S. Patent 9 628 684 B2, 18th April 2017.

\bibitem{Yu:2012}
Z.~Yu, J.~Yu, A.~Lumsdaine, and T.~Georgiev, ``An analysis of color demosaicing
  in plenoptic cameras,'' in \emph{IEEE International Conference on Computer
  Vision and Pattern Recognition (CVPR)}, 2012.

\end{thebibliography}
%
%
\vspace{-0.5cm}
\appendices%
\label{sec:app}%
\section{Scale Space Theorem}%
\label{subsec:app:scale_max}%
\vspace{-.2cm}
\textbf{Theorem} $M$ and $\sigma^\star$ are equal up to scale in \cref{eq:scale_space,eq:lap_scale}. %
\begin{proof}
	Let $G(\sigma, \mathbf{x})$ be a Gaussian function of $\mathbf{x}=\begin{bmatrix} k & l\end{bmatrix}^\intercal$
	and $\sigma$ be the scale in a Laplacian pyramid with blob radius $r$, then
	\begin{align*}
	G(\sigma, k, l) &=\text{exp}\left(-\frac{k^2+l^2}{2\sigma^2}\right) \\ 
	\partial_k G(\sigma, k, l) &= -\frac{k}{\sigma^2} \, \text{exp}\left(-\frac{k^2+l^2}{2\sigma^2}\right) \\
	\partial_{kk} G(\sigma, k, l)&=\left(-\frac{1}{\sigma^2} + \frac{k^2}{\sigma^4}\right) \, \text{exp}\left(-\frac{k^2+l^2}{2\sigma^2}\right) \\
	\nabla^2 G(\sigma, k, l) &=\partial_{kk} G(\sigma, k, l) + \partial_{ll} G(\sigma, k, l) \\
	&=\left(\frac{k^2+l^2}{\sigma^4}-\frac{2}{\sigma^2}\right) \, \text{exp}\left(-\frac{k^2+l^2}{2\sigma^2}\right) \\
	r^2 &= k^2+l^2 \\
	0&=\frac{r^2}{\sigma^4}-\frac{2}{\sigma^2} \\
	r&=\sqrt{2}\sigma \\
	M \approx 2r &= 2\sqrt{2}\sigma \qedhere
	\end{align*}
\end{proof}
{	
	\vspace{-1cm}
	\begin{IEEEbiography}[{\includegraphics[width=1in,height=1.25in,clip,keepaspectratio]{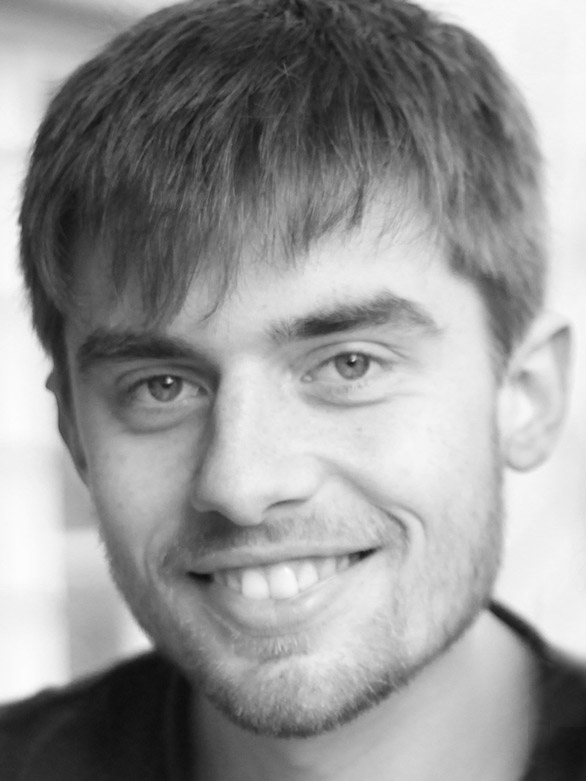}}]
		{Christopher Hahne} is a visiting scholar affiliated with the School of Mathematics and Computer Science at the University of Wolverhampton. He received the B.Sc. degree in Media Technology in 2012 from the University of Applied Sciences Hamburg while working at R\&D departments of Rohde~\&~Schwarz as well as the ARRI Group. 
		After becoming a visiting student at Brunel University in 2012, he started an M.Phil. and transferred to a bursary-funded Ph.D. programme at the University of Bedfordshire together with his supervisor Prof. Amar Aggoun. Upon completion of his doctoral studies, he published 10 high impact peer-reviewed journal and conference articles, most of them written as the main author. In his professional career, he has worked for trinamiX GmbH (BASF SE) and Pepperl+Fuchs SE, which led to 2 granted patents.
		His research interests concentrate on algorithm development in the fields of computer vision, audio processing and depth sensing. In 2021, he will join the Artificial Intelligence in Medical Imaging (AIMI) group as an advanced post-doctoral researcher at the ARTORG Center at the University of Bern.
	\end{IEEEbiography}

	\begin{IEEEbiography}[{\includegraphics[width=1in,height=1.25in,clip,keepaspectratio]{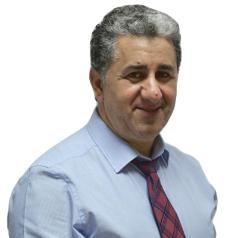}}]
		{Amar Aggoun} is currently the Head of School of Mathematics and Computer Science and Professor of Visual Computing at the University of Wolverhampton. He received the "Ingenieur d'\'etat" degree in Electronics Engineering in 1986 from Ecole Nationale Polytechnique d'Alger in (Algiers, Algeria) and a PhD degree in Electronic Engineering from the University of Nottingham. His academic career started at the University of Nottingham where he held the positions of research fellow in low power DSP architectures and visiting lecturer in Electronic Engineering and Mathematics. In 1993 he joined De Montfort University as a lecturer and progressed to the position of Principal Lecturer in 2000. In 2005 he joined Brunel University as Reader in Information and Communication Technologies. From 2013 to 2016, he was at the University of Bedfordshire as Head of School of Computer Science and Technology. He was also the director of the Institute for Research in Applicable Computing which oversees all the research within the School. His research is mainly focused on 3D Imaging and Immersive Technologies and he successfully secured and delivered research contracts worth in excess of \pounds6.9M, funded by the research councils UK, Innovate UK, the European commission and industry. Amongst the successful project, he was the initiator and the principal coordinator and manager of a project sponsored by the EU-FP7 ICT-4-1.5-Networked Media and 3D Internet, namely Live Immerse Video-Audio Interactive Multimedia (3D VIVANT). He holds three filed patents, published more than 200 peer-reviewed journals and conference publications and contributed to two white papers for the European Commission on the Future Internet. He also served as Associate Editor for the IEEE/OSA Journal of Display Technologies.
	\end{IEEEbiography}
\end{document}